\documentclass[prx,showpacs,aps,amssymb,amsmath,nofootinbib,notitlepage,superscriptaddress,twocolumn,longbibliography]{revtex4-1}

\usepackage{amsfonts}
\usepackage{graphicx,graphics,epsfig,times,bm,bbm,mathrsfs} 
\usepackage[normalem]{ulem}
\usepackage{subfigure}
\usepackage{upgreek}

\newcommand{\beq}{\begin{equation}}
\newcommand{\eeq}{\end{equation}}
\newcommand{\beqnn}{\begin{equation*}}
\newcommand{\eeqnn}{\end{equation*}}
\newcommand{\beann}{\begin{eqnarray*}}
\newcommand{\eeann}{\end{eqnarray*}}
\newcommand{\GPD}{\mathcal G \mathcal P \mathcal D}

\newcommand{\bes} {\begin{subequations}}
\newcommand{\ees} {\end{subequations}}
\newcommand{\bea} {\begin{eqnarray}}
\newcommand{\eea} {\end{eqnarray}}

\newcommand{\ignore}[1]{}

\newcommand{\run}{\mathrm{run}}

\begin{document}

\title{Optimally Stopped Optimization}

\author{Walter Vinci}
\affiliation{Department of Electrical Engineering, University of Southern California, Los Angeles, California 90089, USA}
\affiliation{Department of Physics and Astronomy, University of Southern California, Los Angeles, California 90089, USA}
\affiliation{Center for Quantum Information Science \& Technology, University of Southern California, Los Angeles, California 90089, USA}
\author{Daniel A. Lidar}
\affiliation{Department of Electrical Engineering, University of Southern California, Los Angeles, California 90089, USA}
\affiliation{Department of Physics and Astronomy, University of Southern California, Los Angeles, California 90089, USA}
\affiliation{Center for Quantum Information Science \& Technology, University of Southern California, Los Angeles, California 90089, USA}
\affiliation{Department of Chemistry, University of Southern California, Los Angeles, California 90089, USA}

\begin{abstract}
We combine the fields of heuristic optimization and optimal stopping. We propose a strategy for benchmarking randomized optimization algorithms that minimizes the expected total cost for obtaining a good solution with an optimal number of calls to the solver. 
To do so, rather than letting the objective function alone define a cost to be minimized, we introduce a further cost-per-call of the algorithm. 
We show that this problem can be formulated using optimal stopping theory. The expected cost is a flexible figure of merit for benchmarking probabilistic solvers 
that can be computed when the optimal solution is not known, and that avoids the biases and arbitrariness that affect other measures. The optimal stopping formulation of benchmarking directly leads to a real-time, optimal-utilization strategy for probabilistic optimizers with practical impact. We apply our formulation to benchmark simulated annealing on a class of MAX2SAT problems. We also compare the performance of a D-Wave 2X quantum annealer to the HFS solver, a specialized classical heuristic algorithm designed for low tree-width graphs. On a set of frustrated-loop instances with planted solutions defined on up to $N = 1098$ variables, the D-Wave device is two orders of magnitude faster than the HFS solver, and, modulo known caveats related to suboptimal annealing times, exhibits identical scaling with problem size. 

\end{abstract}
\maketitle

\section{Introduction} 
\label{sec:intro}

The performance of optimization algorithms is typically assessed in terms of either solution quality or computational effort~\cite{McGeoch:book,Barr1995,johnson2002theoretician}. In the case of randomized algorithms, these two quantities are random variables and their characterization is  usually performed via careful statistical analyses of finite samples~\cite{bartz2010experimental}.   A practical difficulty that one encounters in benchmarking optimization algorithms is that ``solution quality" and ``computational effort" are deeply intertwined. This difficulty seems to be one of the reasons why a consistent benchmarking methodology is lacking in the field of optimization algorithms \cite{hooker1995testing,bukh1992art,birattari2007assess,brownlee2007note}. 

The most common strategy for benchmarking is a ``quality first" approach where one fixes a minimum target for the quality of an acceptable solution. Performance is then measured in terms of the \emph{time-to-target}, i.e., the time to obtain an acceptable solution with a probability of, e.g., $99\%$. The target quality is usually specified as a percentage of the quality of the global optimum. When the target is the global optimum itself, the time-to-target reduces to the \emph{time-to-solution}. The time-to-target is a very useful measure of performance, but it has several drawbacks. First, if the global optimum is unknown, the approach described above cannot be used and choosing an appropriate target can become problematic. Common strategies such as setting the target as the best known solution, or as the best solution obtained by a competitive solver, involves an even larger number of arbitrary choices. Moreover, a time-to-target measure considers all the solutions whose cost is below that of the target as being equally good, thus ignoring precious information about the quality of solutions below the target.

In some cases it may be convenient to follow a ``time first" approach where one fixes the computation time. Performance is measured in terms of the \emph{target-in-time}, i.e., the (e.g., average) quality of the best solution found in the given computational window. This approach does not require knowledge of the optimal solution. The length of the window can, however, greatly bias the comparison between different solvers: some solvers may find good quality solutions more quickly than others, but may require more time to find solutions that are very close to the global minimum. 

The standard approaches described above are simple but also restrictive. As mentioned above, solution quality and computational effort are strictly interdependent quantities: intuitively, setting a more (less) ambitious target quality implies longer (shorter) computation times\footnote{Though this is not always the case; e.g., in the case of quantum adiabatic \textit{vs} diabatic optimization \cite{Somma:2012kx,crosson2014different,MAL:2016}.}. In many practical applications one is interested in exploiting this dependence to minimize both quantities at the same time. This may happen, e.g., when there is no reason to choose a specific target or a particular computation length. In other words, we would like to minimize the \emph{total cost} as a function of the computation time $t$:
\beq
C(t) = E(t) + T(t)\,,
\label{eq:gen}
\eeq
where $E(t)$ is the quality of the best solution found at time $t$ (without loss of generality we assume that $E(t)$ is the value of the objective function, which we may think of as an energy, that defines the optimization problem) and $T(t)$ is a measure of the computational effort. 
In practical applications, both $E(t)$ and $T(t)$ could represent, e.g., monetary costs. 

The time $t^*$ that minimizes the total cost $C(t)$ is the ``optimal stopping time", i.e., the time at which we should stop our computation and accept the best solution found. The total cost at the optimal stopping time is the ``optimal total cost" $C^* \equiv C(t^*)$. 
The total cost defined in Eq.~\eqref{eq:gen} expresses a natural tradeoff: an optimization algorithm can in principle achieve the same performance in terms of optimal total cost in a continuum of different ways: by taking more time to generate high quality solutions, or by taking less time to generate lower quality solutions. It is possible, of course, to define the total cost differently in order to express other variants of the tradeoff, but many such variants can be transformed into the form expressed in Eq.~\eqref{eq:gen} [e.g., by taking the logarithm of both sides in the case of total cost being defined instead as the product of $E(t)$ and $T(t)$], and moreover, as we shall see below, Eq.~\eqref{eq:gen} lends itself to an elegant analytical solution for $t^*$.

In the case of randomized optimizers, the minimization problem described above can be formulated in terms of  an \emph{optimal stopping problem}. Sequential statistical analysis, initiated by Wald \cite{Wald:1945bf,wald1973sequential} provides a general mathematical framework for finding analytical and approximate solutions to optimal stopping problems \cite{chow1971great,GOTTINGER:1976xe,Ferguson:book}. In order to keep our analysis as clear and simple as possible, we consider cost functions that are linear in time, i.e., $T(t) = c t$. This simple but practically relevant choice reduces our problem to the well-known optimal stopping problem known as ``house-selling" \cite{Ferguson:book,macqueen1960optimal,stigler1961economics}, which has an analytical solution. This will allow us to benchmark and compare optimization algorithms in terms of optimal total cost, rather than just solution quality or computational effort.

Using optimal total cost for benchmarking has several advantages. First, its definition only requires the specification of a cost function $T(t)$, as we have done, and optimal total cost can be easily used to directly compare widely different approaches to optimization. While its implementation is problem and solver independent, the optimal stopping approach naturally specifies a quality target and computational effort that do depend on the specific instance and solver used. In our approach, the choice of the cost function $T(t)$ plays a primary role in benchmarking using optimal total cost. The choice of the cost function is similar to the choice of the quality target when benchmarking via time-to-target measures \cite{Hoos:98,Aiex:2007ad,King:2015cs}.

The second advantage of optimal total cost is its flexibility: it can be computed without knowledge of the global minimum. Unbiased benchmarking of optimization algorithms can thus be performed rigorously on arbitrarily hard problems without prior knowledge of their solution, which is  the typical situation. Moreover, by choosing different cost functions $T(t)$ one can explore the performance of optimization algorithms in different utilization regimes. Large values of the cost function will result in less ambitious costs and faster computations, and vice versa. 
This flexibility is fully appreciated when considering that practical situations will essentially determine the value of the cost function, which then informs us about the quality target and computational time that are optimal for that application.

Optimal costs are also useful in determining the optimal amount of hardware resources required for certain computational tasks. This can be simply done by including the costs of using certain hardware into the cost function $T(t)$. Similarly, optimal total cost can help in assessing the practical viability of new technologies whose early adoption usually involve both a relevant improvement in computational performance and a relevant increase of utilization costs. This aspect is particularly delicate for quantum optimization, which promises a computational power (quantum speed-up \cite{speedup}) not achievable with classical computers. At the same time, the challenges in building a quantum computer will make the first prototypes very expensive. Indeed, commercially available quantum optimizers have been recently built \cite{Johnson:2010ys,Berkley:2010zr,Harris:2010kx,Bunyk:2014hb}. Remarkably, these prototype devices achieve a level of performance that is already comparable to modern CPUs \cite{speedup,Hen:2015rt,denchev2015computational}, despite several decades of continuous technological improvements in classical computation. It is likely that performance of quantum optimizers will increase with time, while their costs will drop. Optimal costs will thus be an important tool to determine the break-even point, i.e., the point where the optimal total cost obtained with quantum hardware will be smaller than that obtained with classical hardware.

This paper is organized as follows. In Section~\ref{sec:OPT} we formulate the optimal stopping problem mentioned above for randomized optimizers and linear cost functions. We also provide the analytical solution of the problem and discuss the connections between optimal total cost and other standard measures. In Section~\ref{sec:exp-det} we discuss how to experimentally determine optimal total cost. In Section~\ref{sec:BENCH} we present extensive numerical simulations in which we have benchmarked simulated annealing as a test-case randomized optimization algorithm. We discuss the use of optimal total cost to optimize the number of spin updates for simulated annealing and discuss the scaling of optimal total cost with problem size. In Section~\ref{sec:CvsQ} we compare the optimal total cost obtained with classical and quantum optimization hardware. In Section~\ref{sec:OPTUSE} we report on numerical experiments in which we show the feasibility of implementing optimal stopping when the behavior of a randomized solver on a given instance is learned during the computation. In Section~\ref{sec:PAR} we discuss how to use optimal total cost in the context  of parallel computation. We present our conclusions in Section~\ref{sec:CONC}.

\section{Benchmarking via Optimal Stopping}
\label{sec:OPT}

As explained above, we propose the optimal total cost as the appropriate measure of performance in the general case where both the solution quality $E(t)$ and the computation cost $T(t)$ play a role. In this section we explain how to find the stopping time $t^*$ and optimal total cost $C^*$ using the theory of optimal stopping. We then show how the optimal total cost measure can be reduced to standard quality-only measures such as time-to-target or time-to-solution. Using optimal total cost can thus be considered as a general framework for benchmarking that includes known benchmarking strategies as special cases.

\subsection{Optimal Total Cost}

We begin by formalizing our definition of optimal total cost in terms of a specific optimal stopping problem. Our fundamental assumption is that we can describe a randomized optimization algorithm in terms of an intrinsic ``quality distribution" $\mathcal P(e)$ of the qualities $e$ of the outcomes. From now on we will mostly use the term ``energy" to indicate the quality of a solution, with a smaller energy corresponding to a better quality. The distribution $\mathcal P(e)$ will depend on both the solver used and the optimization problem. We also assume that the runtime of the algorithm $t_{\run}$ is a constant when the same algorithm is repeatedly run on the same problem with the same parameters. In this scenario, the time dependence in Eq.~\eqref{eq:gen} is discretized in steps of $t_{\run}$ and can be rewritten as:
 \beq
 C_n =  \min \{e_1,\dots,e_n\} +T_n = E_n+T_n\,.
\label{eq:expense}
 \eeq
This equation is interpreted as follows. After the solver is run sequentially $n$ times, we have found the minimum energy $E_{n} = \min\{e_1,\dots,e_n  \}$ and spent an effort $T_n$ on the computation. At each step, we can either decide to perform more observations, thus trying to lower $E_{n}$ at the price of increasing the computational effort, or stop and accept the solution corresponding to  $E_{n}$.  The optimal strategy for this decision process is to minimize $C_n$ in Eq.~\eqref{eq:expense}. In the field of sequential statistical analysis, this decision-making problem is called optimal stopping \cite{Ferguson:book}. 
A brief introduction to the basics of optimal stopping theory is given in Appendix~\ref{sec:OS}. Using the \emph{principle of optimality} explained there, the optimal stopping rule calls for a stop as soon as one finds a solution with energy upper bounded by the optimal total cost $C^*$:\footnote{Henceforth an asterisk always denotes ``optimal".}
\beq
n^* = \min\{n \ge 1: e_n \le C^*\}\,.
\label{eq:opt_n}
\eeq
The principle of optimality elegantly encodes the optimal stopping rule into the knowledge of the optimal total cost $C^*$. Because the energies $e$ are i.i.d. random variables, the stopping step $n^*$, and thus all the terms in Eq.~\eqref{eq:expense}, can also be considered as random variables when the sequence of measurements is repeated. The optimal total cost is then by definition the average (expected) cost obtained when following the optimal stopping rule:
\beq
C^* \equiv \langle C_{n^*} \rangle = \langle E_{n^*} \rangle + \langle T_{n^*} \rangle \equiv E^*+ T^*\,,
\label{eq:split}
\eeq
where the average is taken over several repeated optimally stopped sequences, and where $E^*$ is the optimal energy and $T^*$ is the optimal computational effort. Note that the optimal stopping problem defined in Eq.~\eqref{eq:expense} is completely specified by the cost function $T_n$. Using a given solver for a particular application will result in a specific choice of the cost function and will thus specify the optimal stopping problem relevant for benchmarking. Optimal costs are in general very difficult to compute analytically, but a large and sophisticated set of tools has been developed to find approximate stopping rules \cite{chow1971great,GOTTINGER:1976xe,Ferguson:book}.

In order to study in detail the use of optimal total cost for benchmarking we consider a special, but practically relevant case that can be solved analytically. We assume that the cost function is linear in time:
\beq
 C_n =  \min \{e_1,\dots,e_n\} +n c t_{\run}\,,
\label{eq:linexp}
 \eeq
where the parameter  $c$ is interpreted as the cost per unit of time that specifies the computational effort. The quantity $c$ has thus units of energy per time, and the computation cost $T_n$ itself has units of energy. The optimal stopping problem defined above is then equivalent to the prototypical optimal stopping problem known as the ``house selling problem", which can be solved analytically since it is essentially a Markov model with translational invariance. In this case the optimal total cost $C^*$  is the solution of the following \emph{optimality equation} \cite{Ferguson:book}:
\beq
 C^*_c  \,\,:\,\, \int_{-\infty}^{C^*_c}(C^*_c-e)\mathcal P(e)de  =  c t_{\run}\,,
 \label{eq:opt_E}
\eeq
which is an implicit integral equation for  $C^*_c$ (see Appendix~\ref{sec:OS} for its derivation). Equation \eqref{eq:opt_E} involves the knowledge of of the probability distribution $\mathcal P(e)$, which is learned during benchmarking. As one intuitively expects, $C^*_c$ turns out to be a monotonically increasing function of $c$. We discuss the properties of Eq.~\eqref{eq:opt_E} in more detail in Appendix~\ref{sec:OS}. 

\subsection{Optimal Total Cost as an Energy Target}

Note that because the principle of optimality dictates that the sequence of observations in Eq.~\eqref{eq:linexp} stops as soon as one finds an energy $e$ below or equal to $C_c^*$, the optimal total cost can be regarded as an energy target.  

Pick an energy $e$ at random; the probability that it is at most $C_c^*$ is $p=\int_{-\infty}^{C^*_{c}} \mathcal P(e) de$. Since we stop when $e\leq C_c^*$, the probability of stopping after exactly $n$ attempts is $(1-p)^{n-1} p$. The mean stopping step $n_c^*$ is thus $\sum_{n=1}^\infty n (1-p)^{n-1} p = -p \frac{\partial}{\partial p} \sum_{n=0}^\infty (1-p)^n = -p \frac{\partial}{\partial p} \frac{1}{p} = \frac{1}{p}$, i.e.:
\beq
 n_c^*  = \left[ \int_{-\infty}^{C^*_{c}} \mathcal P(e) de \right]^{-1}\,.
\label{eq:meanstoppingtime}
\eeq
While the optimal stopping problem defined in Eq.~\eqref{eq:linexp} is problem and solver independent, the actual value of $C_c^*$ is not. The optimal stopping rule thus provides an energy target that is natural and appropriate for each solver and instance. Note, however, that $C_c^*$ is always larger than the optimal energy $E^*_c$, as long as $c>0$. This is because necessarily $n\geq 1$ (so that $T_n>0$) and due to the fact that stopping typically occurs when the last observed energy is strictly smaller than the target: $e_{n} <  C_c^*$. Thus, $C_c^*$ should not be confused with the optimal energy itself.

Using the optimal total cost as an energy target takes into account the occurrences and the values of energies below and above the target. This is an important difference between the total cost and time-to-target measures; the latter are binary in the sense they are only sensitive to whether energies are below or above the target, while the total cost measure is more general. With this in mind, as we show next, the total cost measure can be reduced to a time-to-target measure by an appropriate choice of $\mathcal P(e)$.

\subsection{Reduction to Time-to-Target}

The time-to-target is the total time required by a solver to reach the target energy at least once with a desired probability $p_d$, assuming each run takes a fixed time $t_{\run}$ \cite{King:2015cs}. Let $p$ be the $t_{\run}$-dependent probability that a single sample will reach the target energy (as estimated by the sample statistic), e.g., some percentage above the minimum energy. The probability of successively failing $k$ times to reach the target is $\left(1-p \right)^k$, so the probability of succeeding at least once after $k$ runs is $1-\left(1-p \right)^k$, which we set equal to the desired success probability $p_d$; from here one extracts the number of runs $k$ (approximated by a real number) and multiplies by $t_{\run}$ to get the time-to-target ${\rm TtT}$:
\beq
{\rm TtT} = t_{\run} \frac{\log(1-p_d)}{\log(1-p)} \sim  {t_{\run}}/{p}\,,
\eeq 
where the last relation holds for $p\ll 1$, and represents the mean time-to-target. With the appropriate choice of $\mathcal P(e)$, the optimal total cost is easily reduced to ${\rm TtT}$. Recalling that the time-to-target deals with a binary assignment (acceptable/unacceptable), we can assume that the energy distribution $\mathcal P(e)$ takes the following form:
\beq
\mathcal P(e) = p \delta(0) + (1-p) \delta(+\infty)\ ,
\eeq
where we have assigned a vanishing energy to acceptable solutions and and infinitely large energy to unacceptable solutions. For any finite value of $C_c^*$ Eq.~\eqref{eq:opt_E} then reduces to:
\beq
C_c^* p = c t_{\run} \Rightarrow C_c^*/c =  t_{\run}/p\,,
\eeq
which shows that the optimal total cost $ C_c^*$ is proportional to the computational time $t_{\run}/p$ required, on average, to hit the target for the first time, in agreement with ${\rm TtT}$ for small $p$.

\subsection{Reduction to Time-to-Solution}

The time-to-solution is a special case of the to time-to-target, with $p$ now being the probability of the solver finding the minimum energy \cite{speedup}. The reduction of optimal total cost to time-to-solution thus follows immediately from the previous subsection as the same special case.

It is instructive, however, to see how to extract the time-to-solution directly from the more general setting of Eqs.~\eqref{eq:linexp} and \eqref{eq:opt_E}.  In the case of binary optimization, the set of energies that can be observed is always discrete. In the limit $c\rightarrow 0$ the optimal total cost is smaller than the second best energy value $E_{0}<C_c^*< E_{1}$. Thus, for sufficiently small $c$ one stops when hitting the optimal solution. We then have from Eq.~\eqref{eq:linexp}:
\beq
 C_c^* =  E_0 + n^*  c t_{\run} \Rightarrow  (C_c^* -  E_0)/c =  t_{\run}/p\,,
 \eeq
where $p = 1/n^*$ is the probability to obtain the minimum energy. The difference between the optimal total cost and the minimum energy is thus proportional to the time-to-solution $t_{\run}/p$.

\subsection{Reduction to Average Energy}

In some cases, one may also be interested in the mean quality as another measure of performance. The mean quality is defined as the expected quality of the solution:
\beq
\bar E = \lim_{n \rightarrow \infty } {\rm mean} \{e_1,\dots, e_n\}\,.
\eeq
The optimal total cost can also be reduced to this quantity when the limit $c\rightarrow \infty$ is taken in Eq.~\eqref{eq:opt_E}. In this limit, in fact, the cost function is so large that it is optimal to stop after taking only one measurement. Assuming that  $C_c^*\gg e$, the optimality equation Eq.~\eqref{eq:opt_E} reduces to:
\beq
C_c^*-\bar E = c t_{\run} \Rightarrow \bar E  = C_c^*-c t_{\run} \,, 
\eeq
i.e., the mean energy is equal to the difference between the optimal total cost and the computational effort for running the solver once, in the limit of large $c$.

\subsection{Reduction to Target-in-Time}

The reduction of the optimal total cost to a target-in-time measure can be acheived by choosing an appropriate cost function $T_n$ as follows:
\beq
T_n = 
\left\{
\begin{array}{cc}
  0 &  n t_{\run} < T  \\
  +\infty &     n t_{\run} >T
\end{array}
\right.
\eeq
The optimal stopping rule is thus trivially to stop at $n^* = \lfloor T/t_{\run} \rfloor$, and there is no advantage to stopping the computation earlier. The optimal total cost is then the average best energy found in a time $T$, i.e., a target-in-time measure.

\section{Experimental Determination of Optimal Total Cost}
\label{sec:exp-det}

\begin{figure*}[t]
\subfigure[\, ]{\includegraphics[width=1\columnwidth]{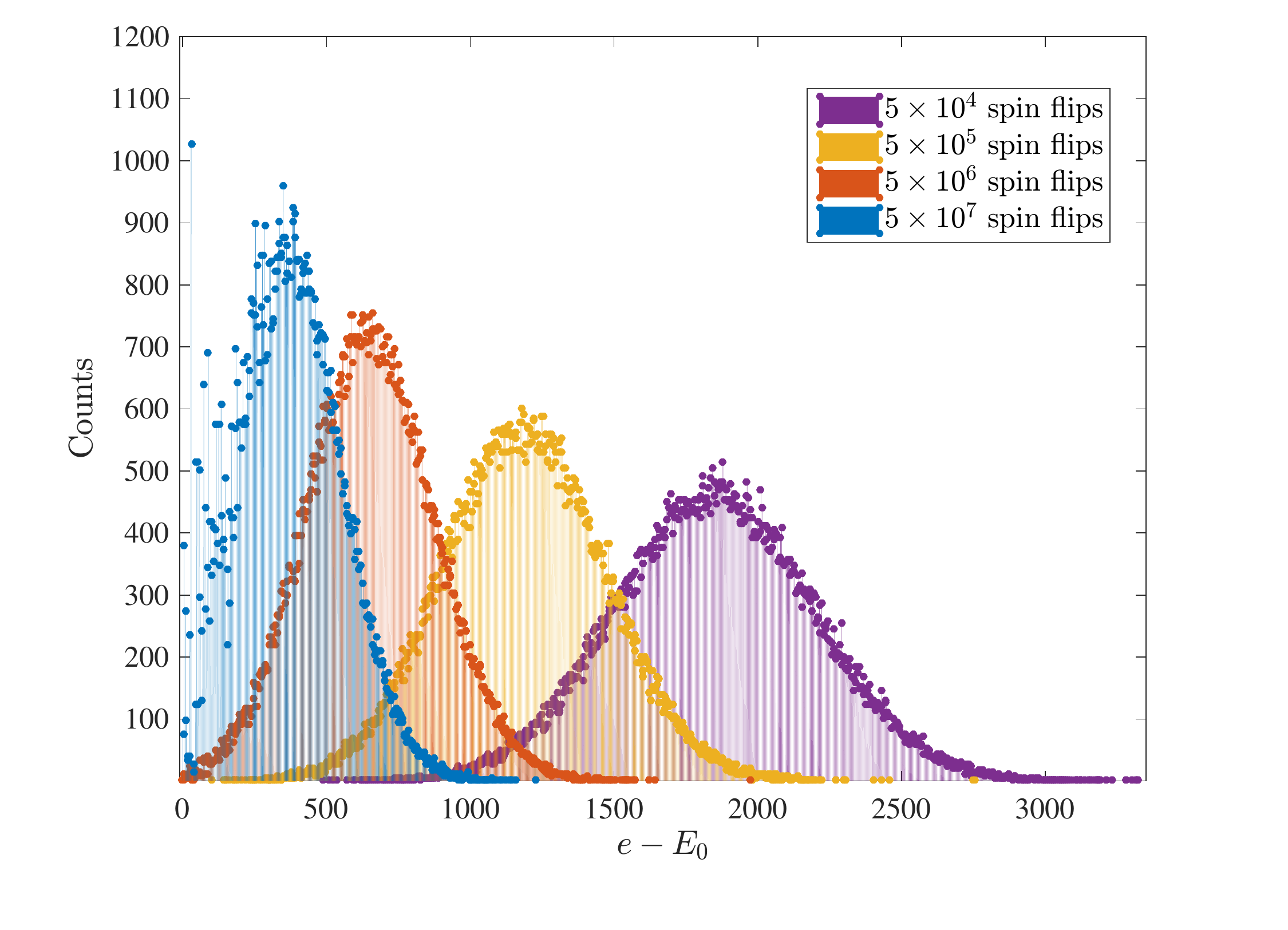} \label{fig:1a}}
\subfigure[\, ]{\includegraphics[width=1\columnwidth]{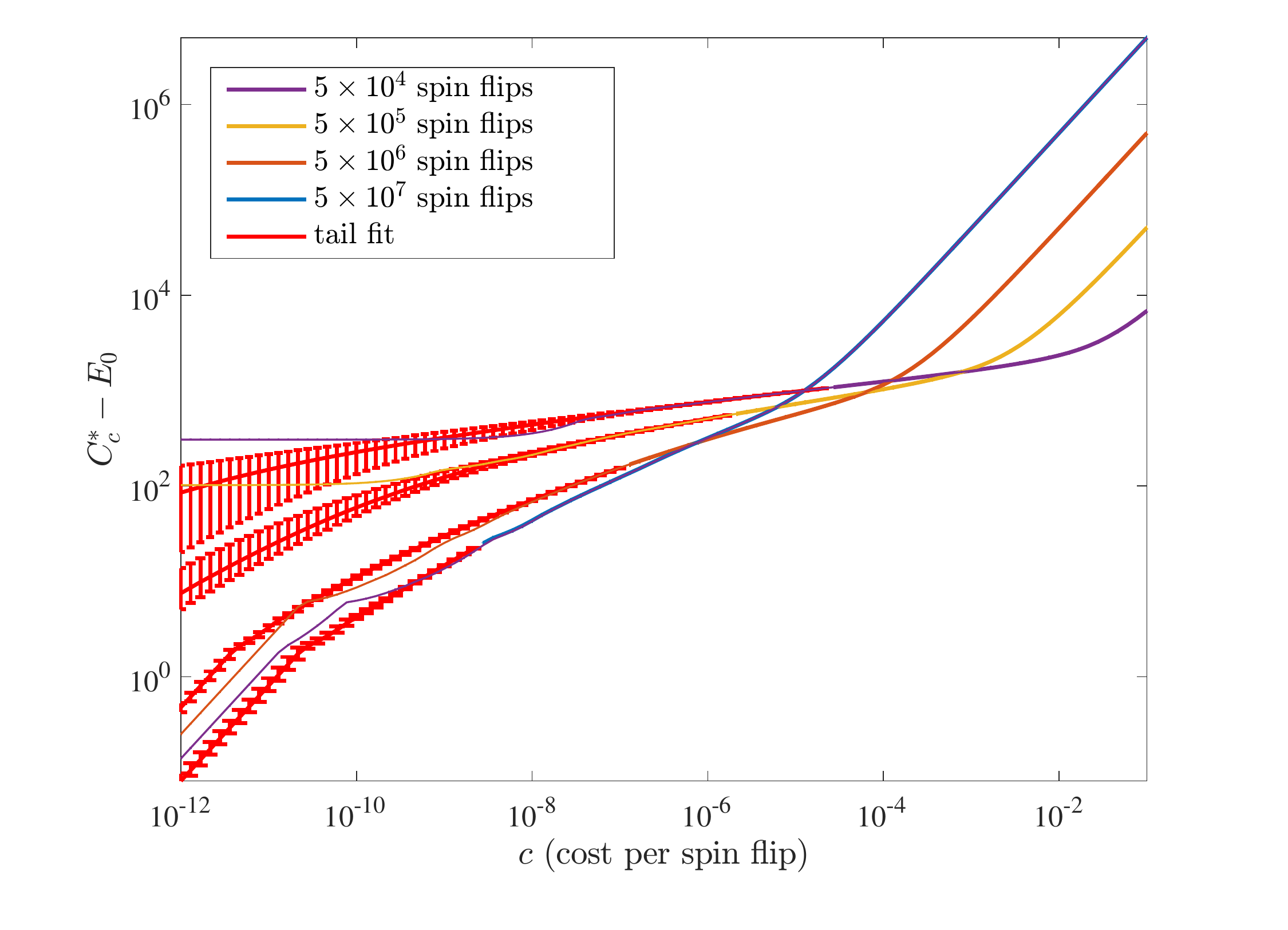} \label{fig:1b}}
\caption{Empirical quality functions (energy histograms) obtained running SA $10^5$ times on an instance of the optimization problem \eqref{eq:alltoall} with $N=1000$ variables. As expected, increasing the number of spin updates per run pushes the energy histogram towards lower energy values. (b) Optimal total cost $C^*_c$ computed from the empirical energy distributions $\mathcal P$ of Fig.~\ref{fig:1a}, after subtracting the ground state energy $E_0$ for the given problem instance. The lower envelope gives the optimal number of spin updates per run $n_{\mathrm{sf}}^*$ as a function of $c$. Error bars were computed via bootstrapping. The red parts correspond to values of the optimal total cost that fall into the lower tail (defined as the $0.1$th percentile) of the energy distributions. Note that large errors due to the under-sampling of the lower tail do not imply large errors in the large $c$ region.}
\label{fig:1} 
\end{figure*}

The energy distribution $\mathcal P(e)$ needs to be determined in order to compute the optimal total cost $C_c^*$ via the optimality equation~\eqref{eq:opt_E}. Because the integral function appearing in the optimality equation is  monotonic (see Appendix~\ref{sec:opteq}), Eq.~\eqref{eq:opt_E} can easily be solved using numerical methods. An empirical estimate of the energy distribution  $\mathcal P(e)$ can be obtained  by sampling a sufficiently large number of energies. The empirical energy distribution is always discrete, and can be written as a sum of Dirac deltas centered at the values of the observed energies: $\mathcal P(e) = \sum_{i} p_i \delta(e-e_i)$. The weights $p_i = n_i/N$ are the observed frequencies.
 
Experimental estimates are always performed on finite samples and statistical errors propagate non-trivially from the energy distribution to the calculation of the optimal total cost $C_c^*$.  This is due to the fact that rare or unobserved solutions may significantly contribute to the value of $C_c^*$.
In fact, the computation of $C^*_c$ involves an integration over the lower tail of $\mathcal P(e)$, which is typically under-sampled in the case of hard problems.  The statistical uncertainty $\delta \mathcal P(e)$ in the determination of the lower tail of the energy function corresponds to an error  $\delta C^*_c$ for the optimal total cost that can be estimated as follows (see Appendix~\ref{appsec:tail} for more details):
\beq
\delta C^*_c = \frac{\int_{\mathrm{tail}} e \delta \mathcal P(e)de}{\int_{-\infty}^{C^*_c} \mathcal P(e)de}\,.
\label{eq:errorbody}
\eeq
This expression shows that there are three main factors that contribute to the error in the estimate of $\delta C^*_c$. The first is, as usual, the sample size. A larger number of observations reduces the weight of the unobserved tail and lowers the value of the numerator. Another factor is the shape of the tail. The numerator can be large for heavy-tailed distribution even if the tail contains only very rare events. The third factor comes from the denominator and simply depends on the value of $C^*_c$. The denominator is smaller for smaller values of the optimal total cost, and thus the overall error is larger. This can be intuitively explained by the fact that smaller values of $C^*_c$ are more sensitive to the lower tail of the energy distribution.

In this work we assume that the energy distribution is not heavy-tailed. This is confirmed by all the numerical experiments we have performed, and ensures that the error in the determination of the optimal total cost is negligible when $C^*_c$ is larger than a purposely chosen percentile of $\mathcal P$.%
\footnote{Heavy tails are known to be an issue with simulated quantum annealing \cite{Steiger:2015fk}, not used in our work.} The possibility of reliably computing $C^*_c$ without good knowledge of the lower tail of the energy functions means that we can perform benchmarking without knowledge of the optimal solution. The advantage of this is that the time used for benchmarking is chosen by the experimenter, not imposed by the hardness of the problem. Benchmarking an optimization algorithm via optimal total cost on a set of arbitrarily hard instances will take more (or less) time depending of whether the experimenter needs to determine $C^*_c$ for smaller (or larger) values of $c$ with sufficient accuracy.

To illustrate the use of optimal total cost in a benchmarking study, we use simulated annealing (SA) \cite{kirkpatrick_optimization_1983} to study the following  optimization problem defined on $N=1000$ binary variables: 
\beq
H = \sum_{i < j }^{N} J_{ij}s_is_j, \quad s_i = \pm1\,,
\label{eq:alltoall}
\eeq 
where the $J_{ij}$ are integers randomly chosen uniformly from the set $\pm\{1,2,\dots,10\}$. This problem is equivalent to a weighted MAX2SAT with a number of clauses equal to the number of total variable pairs. It is also equivalent to finding the ground state of an Ising model defined on a complete graph $K_{1000}$. For all our numerical experiments we used the code provided in Ref.~\cite{Isakov:2015ao}.\footnote{We used the $\tt an\_ss\_ge\_fi\_vdeg$ solver. This optimized simulated annealing code achieves one spin update every about $10$ nanoseconds on modern CPUs (single core performance).} All our simulations were performed using a linear temperature schedule, with $T_{\mathrm{init}} = 10$ and $T_{\mathrm{fin}} = 1/3$. It is well known that the temperature schedule and the initial and final temperatures play a crucial role in determining the performance of SA, but here we chose to focus only on the number of spin updates as the central optimization parameter. In order to keep our discussion independent of the particular CPU used, we assume that the run-time $t_{\run}$ is proportional to the number of spin updates $ n_{\mathrm{sf}}$ performed during an SA run. We thus have $T_{n} = n c n_{\mathrm{sf}}$ for the computational cost function, with the constant $c$ now specifying the cost per spin update. The actual value of $c$ is, in practice, CPU-dependent.

Figure~\ref{fig:1a} shows the quality distribution $\mathcal P(e)$, or energy histogram, when $10^5$ samples are generated by running SA simulations with four different numbers of total spin updates, for a single randomly selected problem instance corresponding to Eq.~\eqref{eq:alltoall}. The number of samples collected was not sufficient to find the global optimal with only $5\times 10^4$ or $5\times 10^5$ updates. As expected, a larger number of spin updates pushes the distribution towards smaller energies.  Figure~\ref{fig:1b} shows the optimal total cost $C^*_c$ corresponding to the energy distributions of Fig.~\ref{fig:1a}. The optimal total cost is, as expected, a monotonic function of the cost per spin update $c$.  The optimal total cost is minimized by using a smaller (or larger) number of spin updates when $c$ is larger (or smaller). There is a simple intuitive explanation for this: expensive computations  (large $c$)  favor fast computations (small number of spin updates), while on the other hand cheap computations (small $c$)  favor long computations (large number of spin updates). At intermediate values of $c$, the SA algorithm gives the same performance (when two curves meet) using two different numbers of updates per run. This happens when the improvement in the average solution quality obtained by implementing a larger number of updates is exactly offset by the increase in the observational cost. 

\emph{The lower envelope is the optimal compromise between cost and efficacy of a single SA run}, which is determined by the number of spin updates per run, $n_{\mathrm{sf}}$. Namely, for any given cost one can infer the optimal value of $n_{\mathrm{sf}}$ by selecting the lowest of the curves at that value of $c$. This is a non-trivial conclusion obtained from our optimal stopping approach, that cannot be obtained within the traditional benchmarking framework that focuses entirely on minimizing the energy or the time to an energy target.

The thin colored lines in Fig.~\ref{fig:1b} are the optimal total cost computed by substituting the experimental energy distributions into the optimality equation. The bold lines are the mean and the standard errors of a sample of $1000$ values of  $C_c^*$ obtained as follows. We first generated $1000$ bootstrapped copies of each of the quality distributions of Fig.~\ref{fig:1a}. We modeled the lower tail (the first $0.1$th percentile) of each bootstrapped distribution with a maximum likelihood fit of a Generalized Pareto Distribution \cite{bartz2010experimental, coles2001introduction} (see Section~\ref{sec:asy} and Appendix~\ref{sec:GEV} for more details).  All the $C_c^*$ values were then computed using the bootstrapped distribution with the tail replaced by the Pareto fit. The bold red lines correspond to the values of the optimal total cost that fall into the fitted tail, while the bold colored lines are the values of $C_c^*$ that are outside the tail. Note that the error bars are relevant only when the optimal total cost falls inside the tail. \emph{We thus see that even an imprecise knowledge of the tail of the distribution does not affect the precise evaluation of $C_c^*$ for sufficiently large values of the cost $c$}, in agreement with Eq.~\eqref{eq:errorbody}.

\begin{figure*}[t]
\subfigure[\, ]{\includegraphics[width=1\columnwidth]{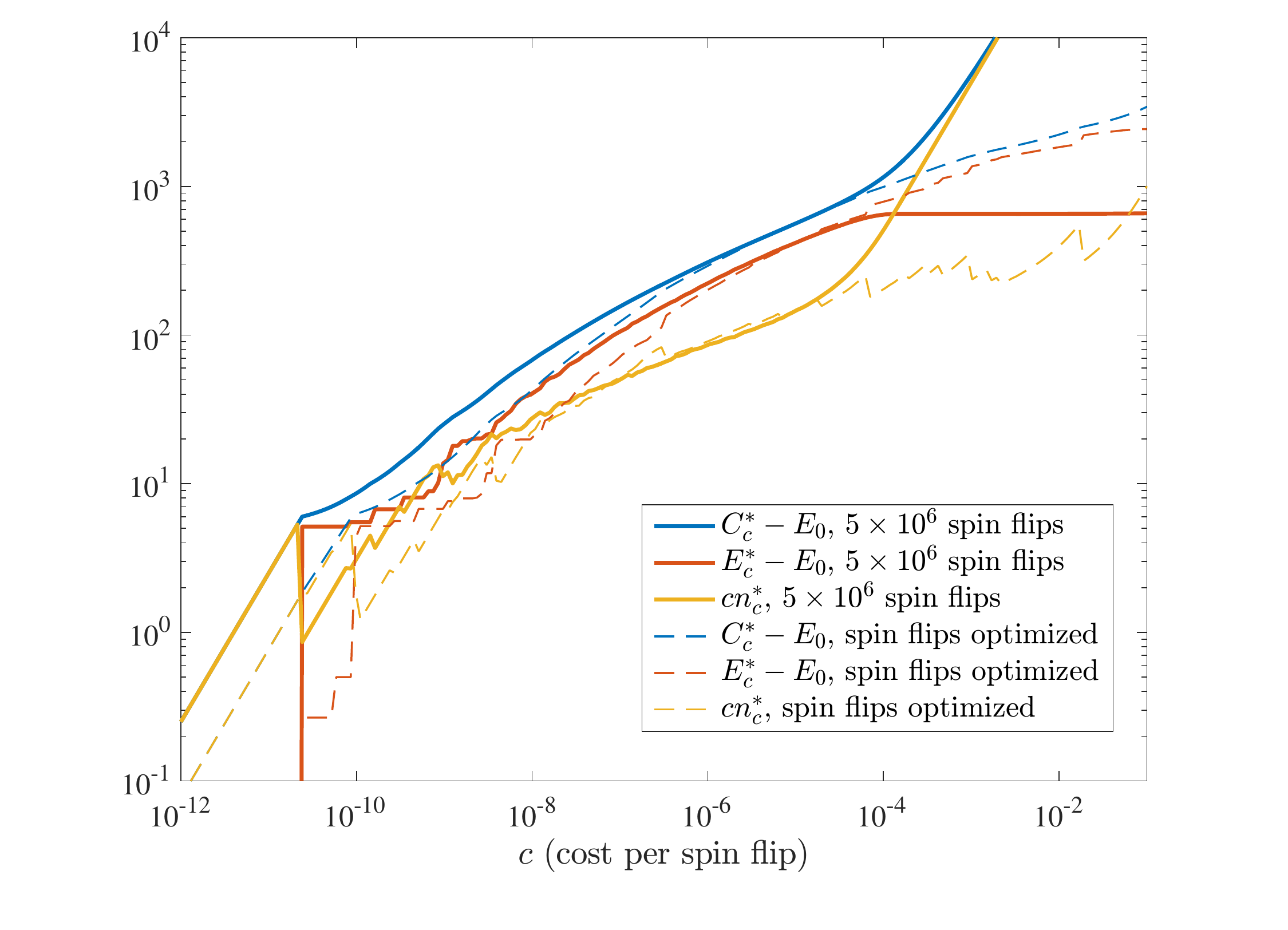} \label{fig:2a}}
\subfigure[\, ]{\includegraphics[width=1\columnwidth]{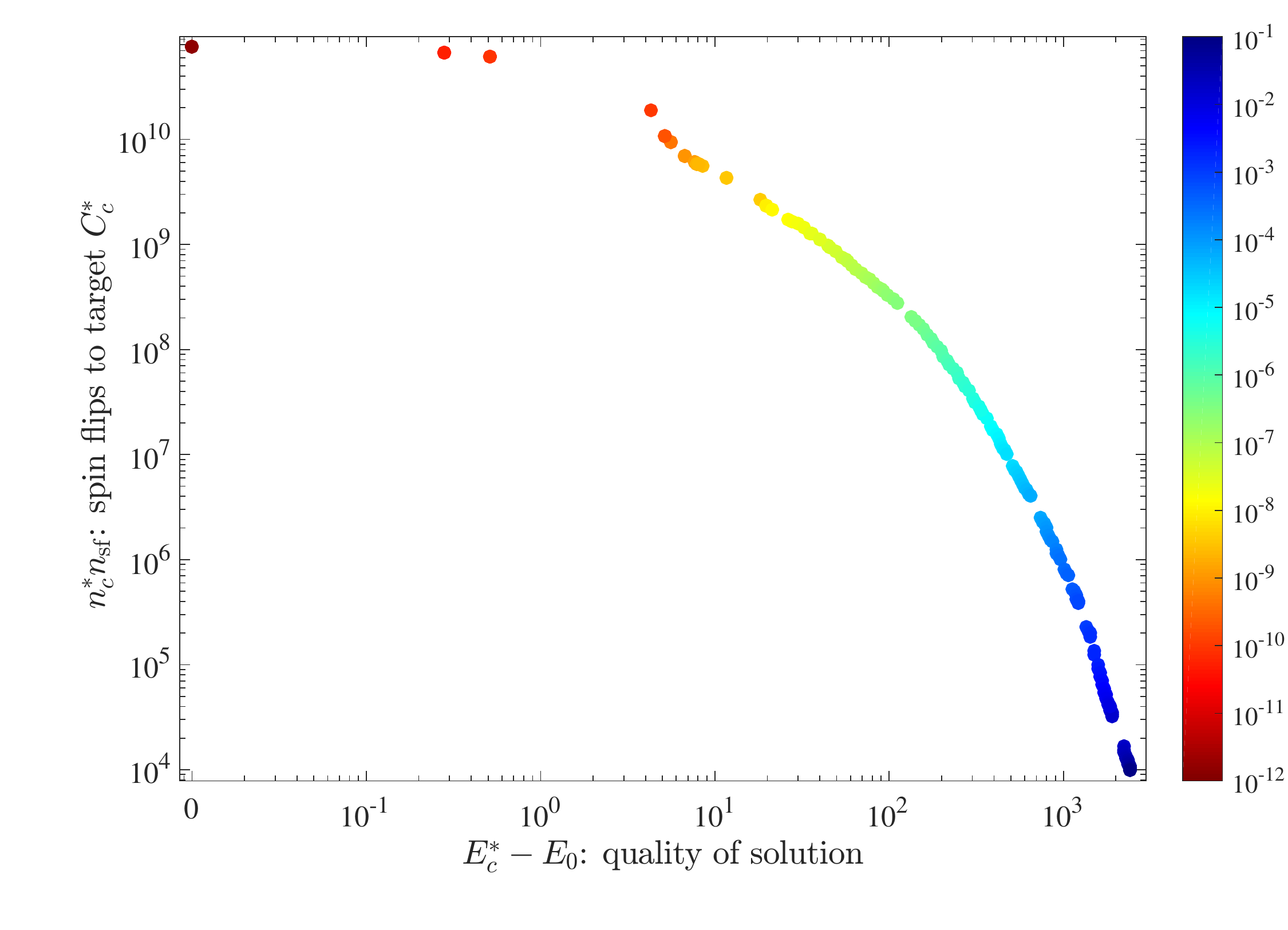} \label{fig:2b}}
\caption{(a) Optimal cost $C^*_c$ (blue), optimal energy $E^*_c$ (red), and optimal computational cost $T^*_c$ (yellow) computed by running SA with a fixed (solid lines) and optimized (dashed lines) number of spin updates (corresponding to the lower envelope of $C_c^*$ as described in Fig.~\ref{fig:1}). (b) Tradeoff between optimal stopping time $ n_c^* n_{\rm sf}$ (with optimized $n_{\rm sf}$) and solution quality $E^*_c$. The color code shows a monotonic inverse dependence of the stopping time on the unit cost $c$. Overlapping dots corresponding to very small $c$ (dark red colors) effectively have $E_c^* = E_0$; their common value on the vertical axis is the optimal time-to-solution. Data shown here corresponds to the same representative instance as in Fig.~\ref{fig:1}. }
\label{fig:2} 
\end{figure*}

We now illustrate how the average solution quality and computational effort contribute to the optimal total cost as a function of the cost $c$. To do so we write  Eq.~\eqref{eq:split} for our special case $T_n =  n c n_{\rm sf}$:

\beq
C_c^* =   E^*_c +  n_c^*c  n_{\rm sf}\,.
\eeq
We then  use Eq.~\eqref{eq:meanstoppingtime} to compute the average stopping step $  n_c^* $ as a function of the energy distribution. Figure~\ref{fig:2a} shows the values of the functions $C_c^*$, $E^*_c$ and $T^*_c$ for the same representative instance as in Fig.~\ref{fig:1}. Solid lines give the three functions computed for the case with $5\times 10^6$ spin updates, while dashed lines correspond to the case with the number of spin flip updates $n_{\rm sf}$ optimized to minimize the optimal total cost $C_c^*$, i.e., with $n_{\rm sf}$ chosen as a function of $c$ according to the lower envelope in Fig.~\ref{fig:1b}.%
\footnote{Note that there are two distinct notions of optimality at play at this point: optimal stopping, as dictated by the principle of optimality \eqref{eq:opt_n} (indicated by an asterisk superscript), and optimality of the number of spin updates, which is SA-specific and is found from the lower envelope of Fig.~\ref{fig:1b}. Other solvers, such as the D-Wave quantum annealer discussed below, are subject to an analogous ``optimal annealing time" notion. Also note that because the energy spectrum is discrete, $E^*_c$ and $T^*_c$ are not continuous functions of $c$, while the total cost $C_c^*$ is continuous.}

For the case with a fixed number of updates, we can identify three distinct regimes in Fig.~\ref{fig:2a}:\\
(i) In the small $c$ regime with $E_0 < C_c^*< E_1$ (optimal solution region) we have:
\beq
C_c^* = E_0 +  n_{0} c  n_{\rm sf},
\eeq
where $n_0$ is the  average number of samples needed to find the optimal solution. In this regime the optimal energy is equal to the minimum energy $E^*_c = E_0$, and the computation effort and  the optimal total cost grow linearly with $c$. The value of the optimal total cost in this regime is thus completely determined by the success probability $p_0$ of the solver in finding the optimal solution, i.e., $n_0 = 1/p_0$, and is determined (apart from the constant $E_0$) by the  computational cost  $T^*_c$;\\
(ii) In the intermediate $c$ regime the optimal total cost is given by a balance between solution quality and computational cost:
\beq
C_c^* = E^*_c + n_{C_c^*} c n_{\rm sf},
\eeq
where $n_{C_c^*}$ is the average number of samples needed to find a solution with quality at least equal to the optimal total cost. The optimal total cost in this regime depends non-trivially on the full energy distribution of the solver. Figure~\ref{fig:2a} shows that in this regime the value of the optimal total cost is dominated by the value of the optimal energy $E^*_c$. Moreover, from Fig.~\ref{fig:2a} we see that in this region $C_c^*$ appears to have a sub-polynomial dependence on $c$;\\
(iii) In the large $c$ regime the computational effort is so high that it is optimal to draw only one sample:
\beq
C_c^* = \bar E + c n_{\rm sf} \,,
\label{eq:high-c}
\eeq
where $\bar E$ is the mean energy obtained by running SA with a fixed number of spin updates ($5\times 10^6$). In this regime $C_c^*$ is dominated by the computational cost  $T^*_c$ and is again a linear function of $c$.

When the number of SA updates is optimized (dashed curves in Fig.~\ref{fig:1a}), we only distinguish two regions. The optimal number of spin updates decreases for larger $c$ to prevent the cost function $T_c^*$ from dominating the value of the optimal total cost. The third region thus disappears in favor of an extended intermediate region. Note that in this region minimizing the optimal total cost is not equivalent to optimizing the energy or the computational effort (the solid yellow and red lines can be below the corresponding dashed lines). 

It is also interesting to study the relation between the optimal stopping time $n_c^* n_{\rm sf}$ and the optimal energy $E^*_c$. Figure~\ref{fig:2b} shows the tradeoff between the two quantities when the number of spin updates $n_{\rm sf} $ has been optimized. A lower (better) energy solution requires, as intuitively expected, a longer stopping time. The color code in Fig.~\ref{fig:2b} represents the value of the unit cost $c$, which is inversely related to the stopping time.

\section{Benchmarking and Scaling via Optimal Total Cost}
\label{sec:BENCH}

\begin{figure*}[t]
\subfigure[\,]{\includegraphics[width=1\columnwidth]{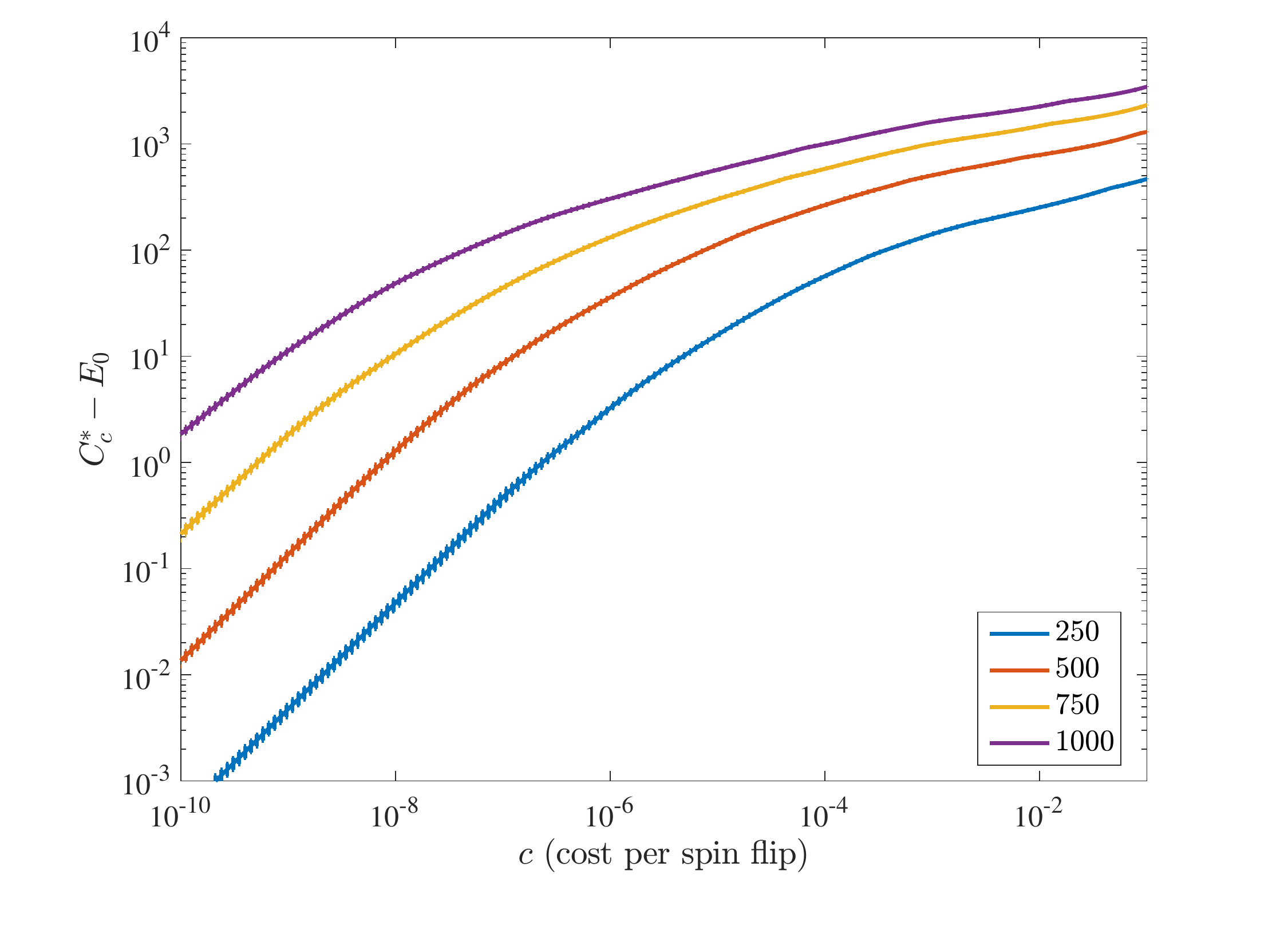} \label{fig:3a}}
\subfigure[\,]{\includegraphics[width=1\columnwidth]{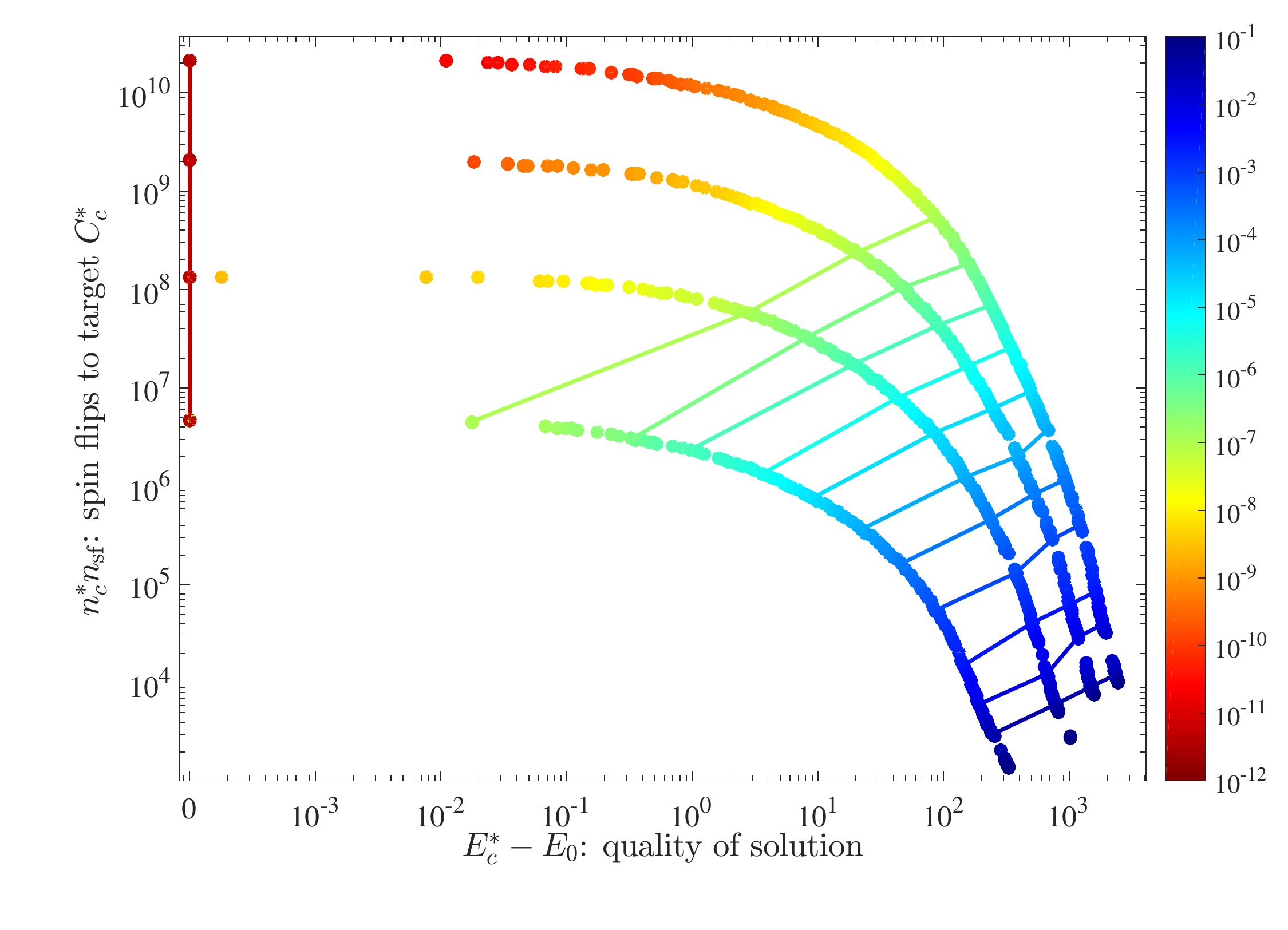} \label{fig:3b}}
\subfigure[\,]{\includegraphics[width=1\columnwidth]{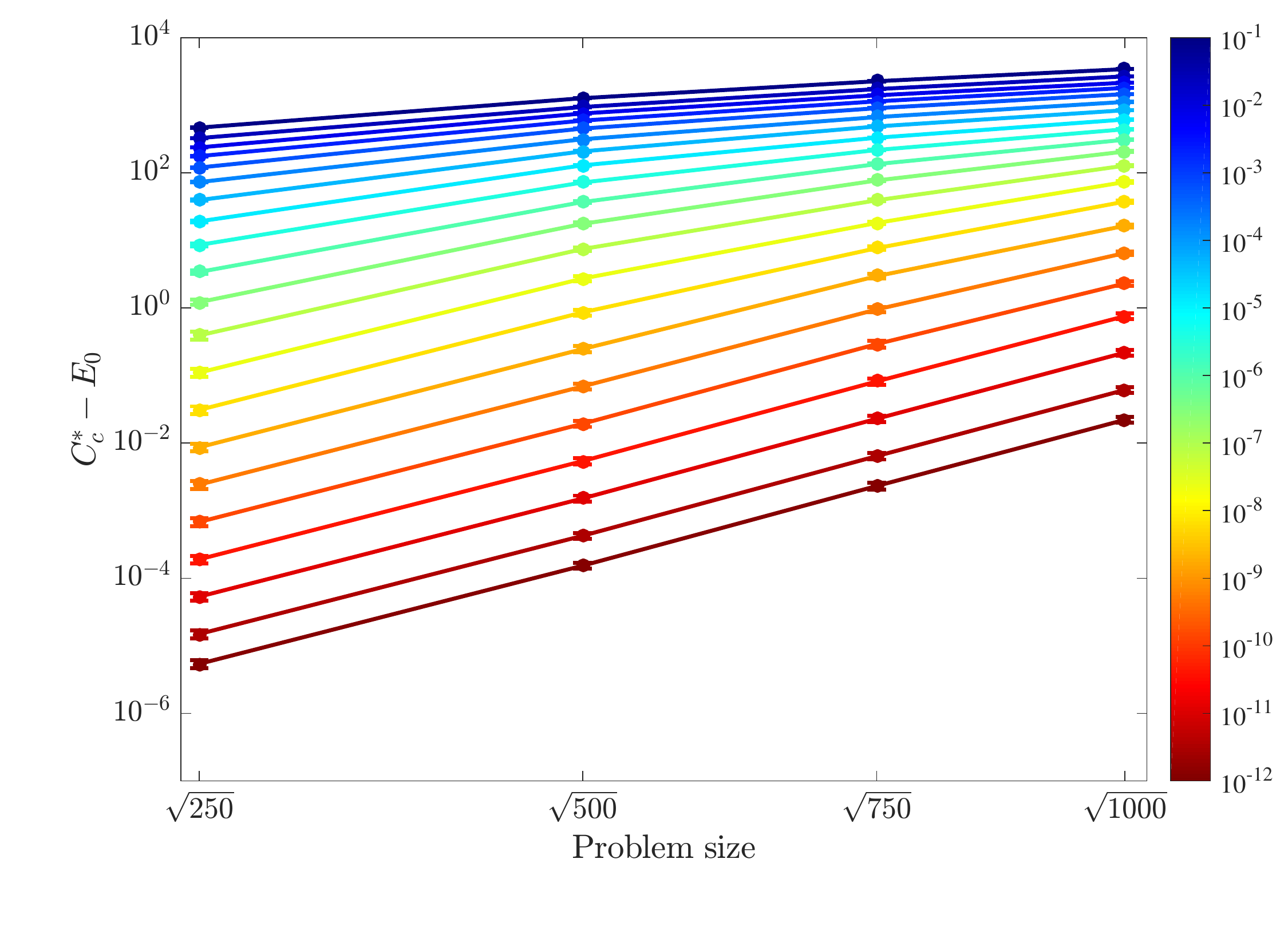} \label{fig:3c}}
\subfigure[\, ]{\includegraphics[width=1\columnwidth]{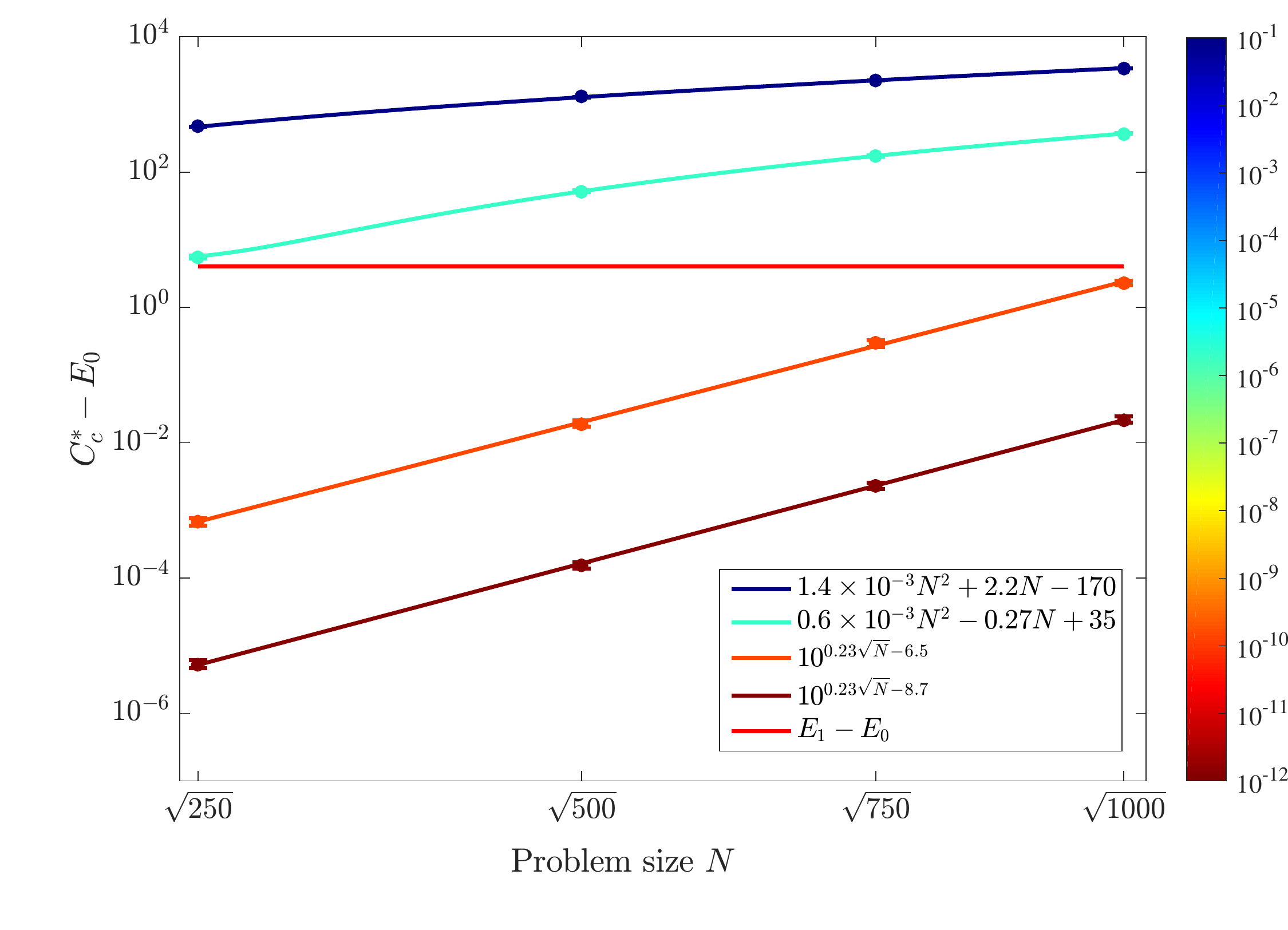} \label{fig:3d}}
\caption{(a)  Optimal total cost  $C_c^*$ for the four different problem sizes considered. Each point in the plot is the mean optimal total cost of the sample of $100$ instances. Error bars (small in the figure) are the standard deviation of the mean optimal total cost. (b) Tradeoff between mean stopping time $n_c^* n_{\rm sf}$ and solution quality $E^*_c$ averaged over the $100$-instance sample. The color code gives the $c$ value. The colored streaks are guides to the eye connecting dots with the same color (same value of $c$). The four curves are ordered by problem size with $N=1000$ at the top. (c) Scaling of the optimal total cost $C^*_c$ relative to the global optimum $E_0$. At fixed problem size the optimal total cost $C^*_c$ grows relative to $E_0$ as a function of the cost $c$. This is a necessary price to pay to keep the computational cost at the optimal level. (d) The red horizontal line is  $C_c^* = E_1$. Regression fits reveal a region of quadratic scaling (above the line) and a region of exponential scaling (below the line). Fit parameters are given in the legend.}
\label{fig:3} 
\end{figure*}

\begin{figure*}[ht]
\subfigure[\,]{\includegraphics[width=1\columnwidth]{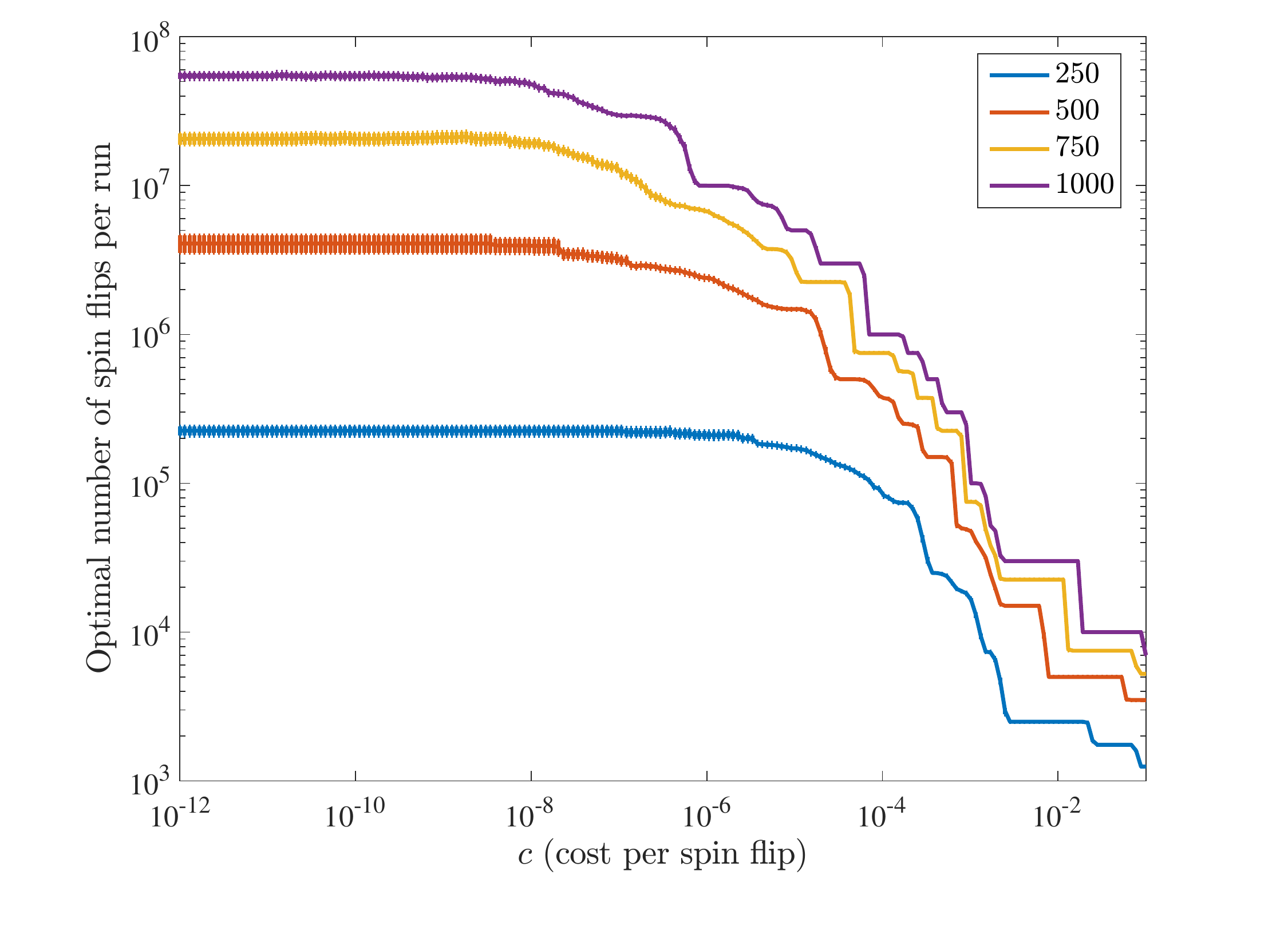} \label{fig:4a}}
\subfigure[\,]{\includegraphics[width=1\columnwidth]{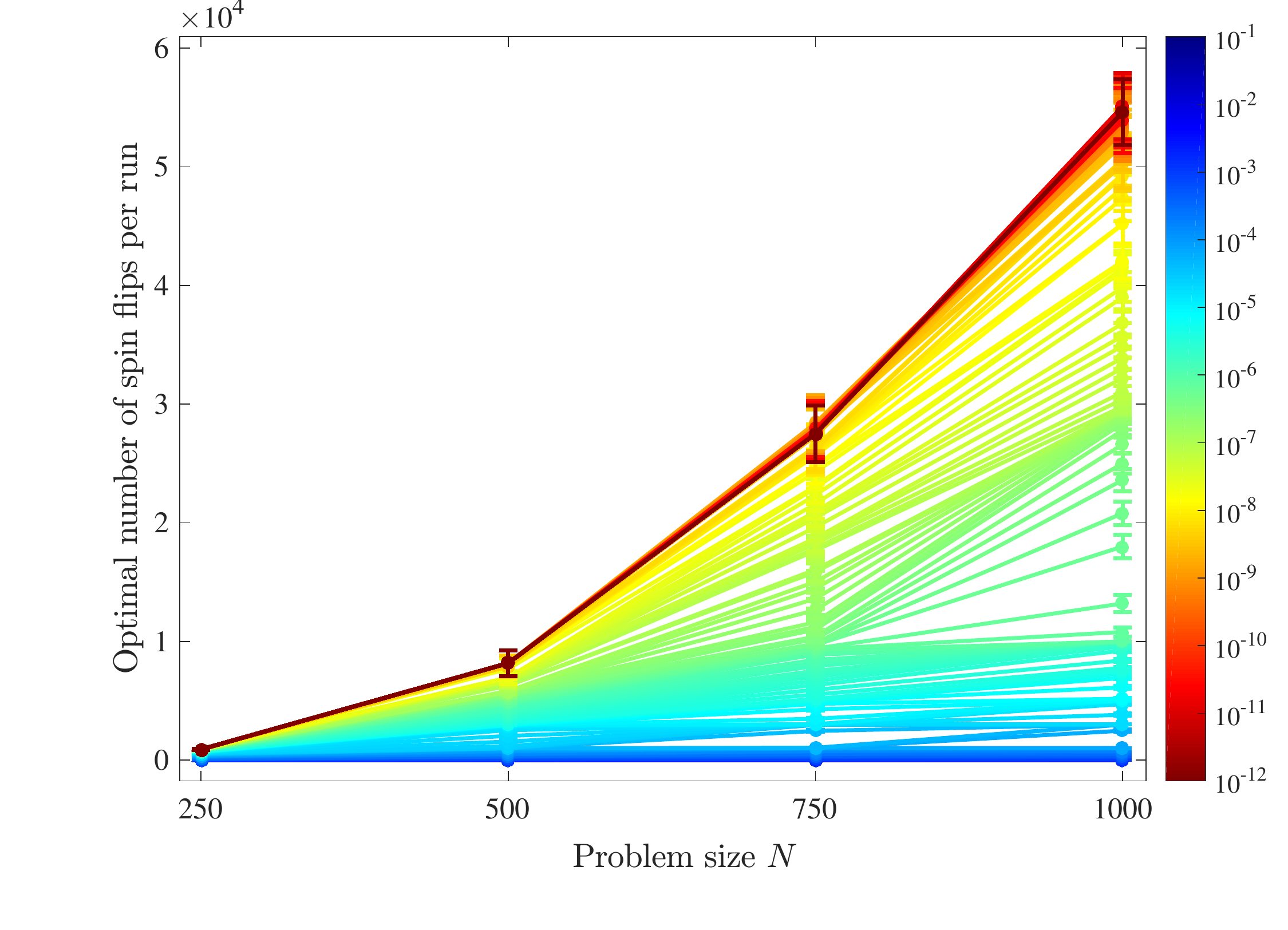} \label{fig:4b}}
\caption{(a) Optimal number of spin updates $n_{\rm sf}$ per SA run, as a function of the unit cost (per spin update) $c$. As expected, this number grows with both decreasing $c$ and increasing problem size $N$. The figure suggest a sub-polynomial scaling with $c$. (b) Optimal number of spin updates $n_{\rm sf}$ per SA run, as a function of the problem size $N$. While this number is expected to grow when the computational cost is negligible (small $c$, warm colors), it is seen to saturate for intermediate and large values of $c$ (cold colors).}
\label{fig:4} 
\end{figure*}

In this section we study scaling with problem size. Toward this end we generated $100$ Ising instances defined on complete graphs of sizes $N=250,500,750$ and $1000$. As in the example discussed above, each instance was randomly generated with integer coupling $J_{ij}$ randomly chosen uniformly from the set $\pm\{1,2,\dots,10\}$. The number of spin updates of the SA solver was optimized as before to minimize the optimal total cost $C_c^*$. For this optimization we  considered $21$ different values for the number of spin updates $n_{\mathrm{sf}} = N\times \{(1,3,5,7)\otimes(1,10,100,1000,10000),100000\}$, for each $N$ value. For each instance and number of spin updates, we  performed $10^5$ SA runs. An instance-by-instance estimation of the statistical error of $C^*_c$ could be performed as explained in the previous section. However, since we are interested in sample-wide properties, we computed the optimal total cost as if our experimental determination of $\mathcal P$ is exact. This approach is appropriate if the sample variability is larger than the single-instance errors in determining the optimal total cost (we checked that this is indeed the case). 

The results are summarized in Fig.~\ref{fig:3}. In Fig.~\ref{fig:3a} we show the optimized value of  $C^*_c$ as a function of $c$ for the four sample sizes of random instances. We again identify two regimes:  $C^*_c  < E_1$ (optimal stopping after finding the minimum energy) and $C^*_c > E_1$ (optimal stopping before finding the minimum energy). As expected, the transition between the two regimes happens at smaller values of the unit cost $c$ for harder problems defined on a larger number of variables $N$. That is, for fixed cost $c$, the larger is $N$ the sooner it is optimal to stop before finding the minimum energy. The difference $C^*_c -E_0$ also grows with the problem size at a fixed $c$: harder problems imply a larger distance of the stopping target  $C^*_c$ from the minimum energy.

Figure~\ref{fig:3b} shows the mean number of updates to stopping as a function of the expected solution quality $E^*_c$, for the four different problem sizes ($N=250$ is the bottom curve). If $c$ is sufficiently small (red vertical line), one always has $E^*_c = E_0$, and the optimal stopping time (in this case the time-to-solution) grows with the problem size. For larger $c$, however, in order to minimize $C_c^*$ it is optimal to both increase the computation time and reduce the expected quality solution $E^*_c$. This nontrivial dependence of $T^*_c$ and $E^*_c$ on the problem size (at fixed $c$; see the streaks in the figure) entails a very different scaling analysis than other typical approaches that keep a fixed target (or fix the target as a percentage of the optimal solution) \cite{King:2015cs}. 

The resulting scaling of the optimal total cost is shown in Fig.~\ref{fig:3c}. Note that for convenience in plotting the figures we always remove an overall constant (the energy of the global minimum $E_0$) from $C_c^*$. Such a constant is indeed irrelevant for benchmarking the performance of SA. While this is not obvious from Fig.~\ref{fig:3c}, the scaling behavior of $C_c^*$ depends crucially on which of the two regimes of  Fig.~\ref{fig:3a} we are probing.  This is shown in more detail in Fig.~\ref{fig:3d}, where the two regimes are separated by a red horizontal line ($C_c^* = E_1$). Below the horizontal line $C_c^*$ depends linearly on $c$, but at fixed $c$, it scales exponentially with the problem size. This corresponds to the exponential scaling of the time-to-solution. Above the horizontal line, the scaling of $C_c^*$  is quadratic in the problem size. Note that, at fixed $c$, $C_c^*$  crosses the horizontal line exponentially fast. This ``give-up" size corresponds to the point where it is no longer optimal to stop after finding the global solution. The optimal total cost will thus always have an initial exponential scaling followed by a quadratic scaling after a characteristic give-up size. The quadratic scaling for sufficiently large sizes can be explained by the fact that, for our choice of problems, the typical energy differences grow quadratically: $E_{2^N}-E_0 \sim N^2$. The scaling behavior of the optimal total cost can be summarized as follows
\beq
\label{eq:scaling}
C_c^*(N) \sim \left\{ \begin{array}{cc}
\alpha e^{\beta \sqrt{N}}& C_c^* < E_1 \\ 
\gamma N^2 +\delta N + \omega & C_c^* > E_1 
\end{array}
\right.\,,
\eeq
Using the relations above, the give-up size $N_{\text{gu}}$ at which scaling behavior changes can be estimated by:
\beq
\alpha e^{\beta \sqrt{N_{\text{gu}}}} \simeq E_1 \quad \Rightarrow \quad N_{\text{gu}} \simeq {\log^2(E_1/\alpha)}/{\beta^2},
\eeq
which means that the give-up size $N_{\text{gu}}$ decreases polynomially (quadratically in our case) with the exponential prefactor $\beta$ and decreases logarithmically with the prefactor $\alpha$. Thus, even obtaining a solver with a reduced exponential prefactor (the realistic goal for solvers tackling NP-hard problems, as opposed to an exponential speedup) would result in a modest polynomial increase in the give-up size $N_{\text{gu}}$. Quite remarkably, and importantly from a practical point of view, this means that \emph{the polynomial scaling regime is the relevant scaling regime for optimal total cost in scenarios with very hard optimization problems}

We conclude this section by analyzing how the optimal number of spin updates depends on $c$ and how it scales with problem size. The optimal number of spin updates, shown in Fig.~\ref{fig:4a} as a function of $c$,  is constant in the regime of small $c$ and monotonically decreasing in the large $c$ regime. The two regimes correspond again to the two described in Fig.~\ref{fig:3a}. Similarly, Fig.~\ref{fig:4b} shows the scaling of the optimal number of spin updates as a function of problem size. Although it is difficult to extract the scaling with problem size from the data presented, it is clear from Fig.~\ref{fig:4b} that the optimal number of spin updates grows more rapidly for smaller $c$. Interestingly, our results indicate that at larger values of $c$ the optimal number of spin updates stays almost constant. This very mild scaling is necessary to keep the computational cost under control when the problem size is increased.

To summarize, in this section we have shown how the optimal stopping approach leads to a non-trivial scaling of the optimal total cost $C_c^*$, and how this scaling is balanced between the optimal solution quality $E^*_c$ and optimal computational cost $T^*_c$. The only assumption we made is that it is possible to assign a meaningful cost per unit of time $c$, which we assume to be constant in the scaling analysis.

\section{Classical vs Quantum Optimization}
\label{sec:CvsQ}

\begin{figure*}[ht]
\subfigure[\,]{\includegraphics[width=1\columnwidth]{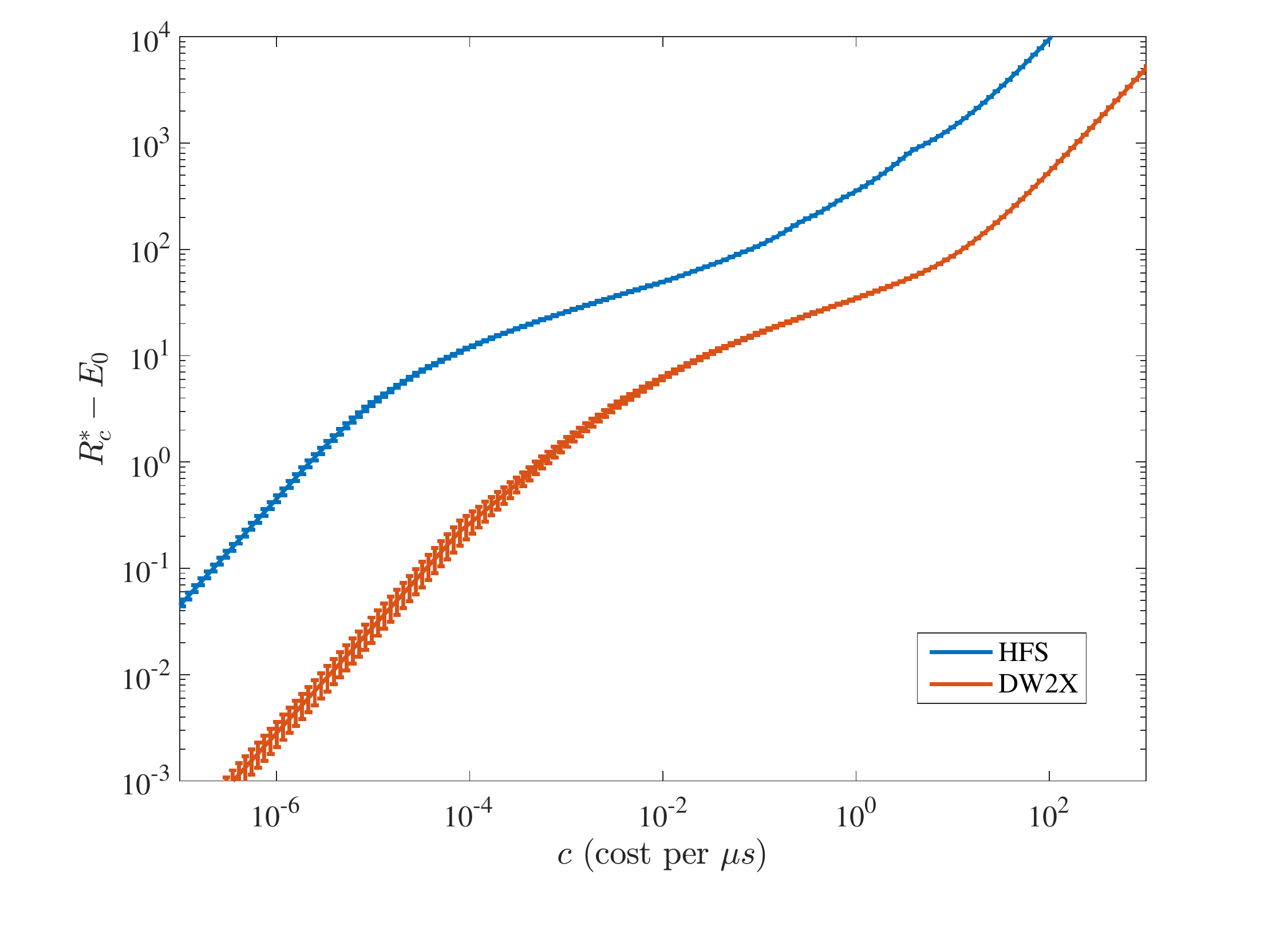} \label{fig:5abis}}
\subfigure[\,]{\includegraphics[width=1\columnwidth]{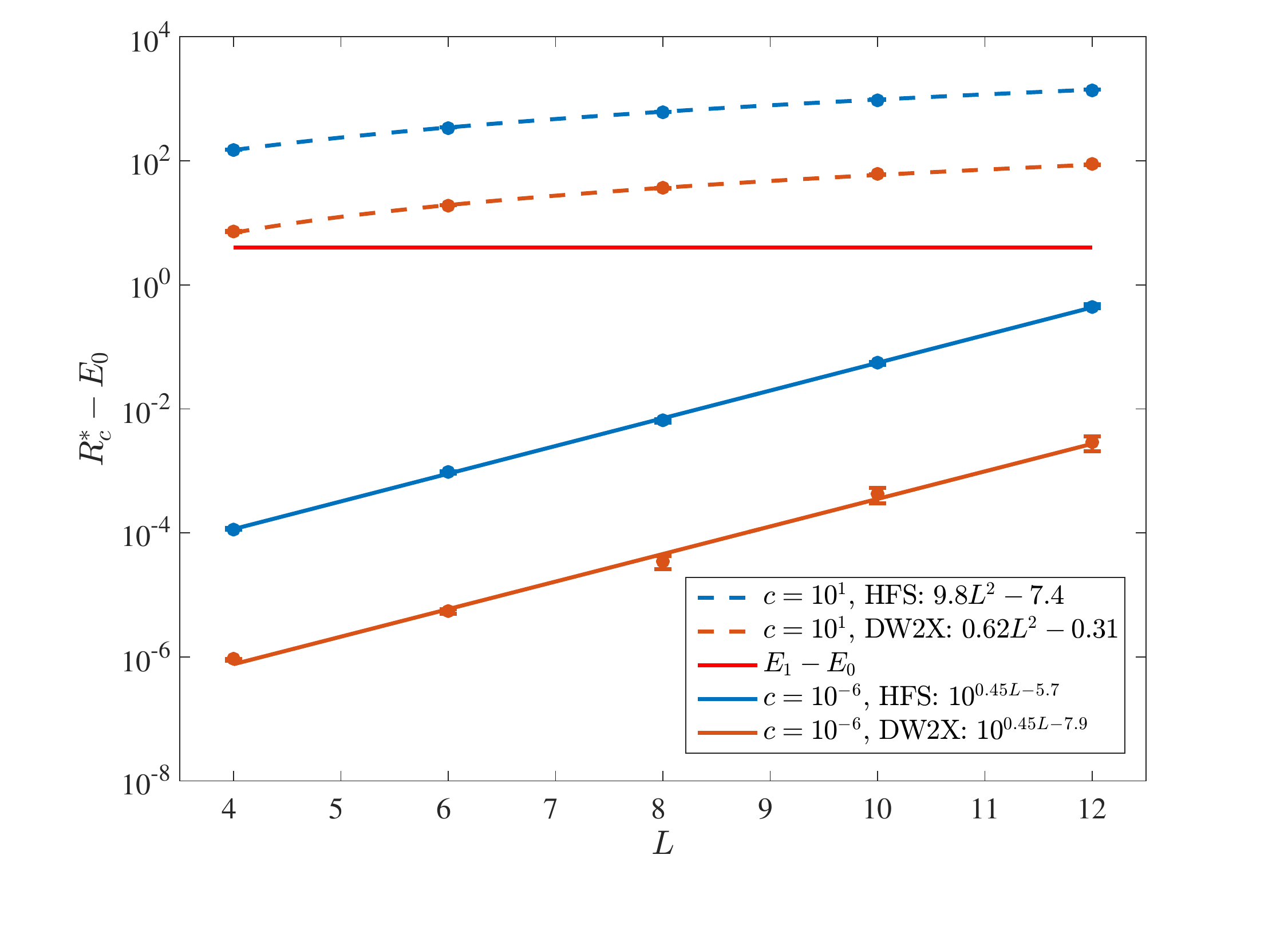} \label{fig:5bbis}}
\caption{(a) Optimal cost for HFS (blue) and DW2X (red), averaged over $100$ instances with planted solutions defined on the full Chimera graph. (b)  Scaling of the optimal total cost for HFS (blue) and DW2X (red) as a function of the problem size for $c= 10^{-6}$ (solid lines below the red horizontal line) and $c = 10^{1}$ (dashed lines above the horizontal line). The red horizontal line $C_c^* = E_1$ separates the regions of exponential (below the line) and linear (above the line) scaling. Note that DW2X and HFS have identical exponential scaling, but the DW2X prefactor is $2.2$ orders of magnitude smaller.}
\label{fig:5bis} 
\end{figure*}

In this section we use optimal total cost to compare the performance of a D-Wave quantum annealer (the DW2X ``Washington" chip installed at the University of Southern California)  to the Hamze-Freitas-Selby (HFS) classical optimization algorithm \cite{hamze:04,Selby:2014tx}. The HFS algorithm is optimized to take advantage of the specific low-treewidth structure of the Chimera graph  and is regarded as the state-of-the-art benchmark for classical optimization of Chimera-structured  problems.

Quantum annealers are analog quantum devices that exploit the physical process of quantum annealing to solve hard quadratic unconstrained optimization problems \cite{kadowaki_quantum_1998,Brooke1999,RevModPhys.80.1061,Dwave,q108,speedup,2014Katzgraber,Venturelli:2014nx,Boixo:2014yu}. The optimization problems that can be implemented on a DW2X processor are similar to those considered in the previous sections [see Eq.~\eqref{eq:alltoall}]:
\beq
H = \sum_{(i, j) \in {\rm  chimera} } J_{ij}s_is_j, \quad s_i = \pm1\,,
\label{eq:chimera}
\eeq 
with the important restriction that the sum now runs over the couplings that are physically implemented in the Chimera connectivity graph, defined on $12 \times 12$ unit cells of $8$ qubits each (see Fig.~\ref{fig:dw} in Appendix \ref{sec:DW}). 

To benchmark the three solvers, we considered sub-lattices of the Chimera graph defined on $L\times L$ unit cells, with $L = 4,6,8,10,12$ and $100$ instances per lattice size. Instances with $J_{ij} \in \pm\{1,2,3 \}$ were generated by constructing the problem Hamiltonian $H$ as a sum of frustrated loops with a known (planted) solution according to the technique developed in Ref.~\cite{Hen:2015rt} and further developed in Ref.~\cite{King:2015zr}.\footnote{Our instances were generated with a value of the clause density $\alpha$ equal to $0.35$.}

The DW2X processors has a tunable annealing time which we set to the smallest available value $t_{\rm DW} = 5\mu s$, which appeared to be optimal in the given range of annealing times for all values of $c$.\footnote{This $t_{\rm DW}$ value is unlikely to be the true optimum, i.e., a shorter annealing time is likely to result in improved performance \cite{Hen:2015rt}.} The cost of drawing a sample with the DW device is thus $c t_{\rm DW}$. 
The run-time of an the HFS algorithm can be considered to be  proportional to the number of  elementary operations $n_{\rm eo} $ performed per run. The parameter  $n_{\rm eo} $ can be optimized by the user.
On modern CPUs we typically have $t_{\rm HFS} = n_{\rm eo} \times 0.6 \mu s$ and the cost to draw an HFS sample is $c t_{\rm HFS}$.\footnote{The value chosen is representative and can change depending on the experimental setup.}

In Fig.~\ref{fig:5abis} we compare the averaged optimal total cost $C_c^*$ obtained by the two solvers on the set of $100$ instances defined on the full DW2X hardware graph ($L=12$). In this section we assume for simplicity that the unit cost $c$ is the same for all solvers (e.g., we do not account for the differences in the utilization costs). We discuss the practical necessity to include such costs at the end of this section.
With this in mind, we see that the DW2X quantum annealer significantly outperforms the HFS classical solver for all the $c$ values considered. For small $c$ values, DW2X is about 100 times faster than HFS. For large values of $c$  however, this advantage is reduced to about a factor of 10. This is due to the fact that the minimum run-time of  $t_{\rm DW} = 5\mu s$ for DW2X is highly suboptimal at large values of $c$. It is important to note that Fig.~\ref{fig:5abis}  compares the performance of a DW2X processor with the performance of the HFS algorithm run on a single core of a modern CPU; we address parallelization in Section~\ref{sec:PAR}.

In Fig.~\ref{fig:5bbis} we compare the scaling of $C_c^*$ for two values of the unit cost parameter $c$, probing the regimes $C_c^* < E_1 $  and $C_c^* > E_1$ (as defined in Sec.~\ref{sec:BENCH}). In the former regime the difference between the optimal total cost and the global minimum scales exponentially with the problem size. The scaling coefficients for DW2X and HFS are identical to within numerical precision. In the other regime the scaling is linear in the number of variables. This is due to the fact that, unlike the case of the previous section, the energy spread is linear in the problem size for the instances considered here: $E_{2^N}-E_0 \sim N \sim L^2$. Numerical fits give us the following scaling behavior for the optimal total cost:
\beq
C_c^*(N) \sim \left\{ \begin{array}{cc}
\alpha e^{\beta L}& C_c^* < E_1 \\ 
\gamma L^2 +\omega& C_c^* > E_1 
\end{array}
\right.\,,
\eeq
similar to the scaling result \eqref{eq:scaling} obtained for the class of instances considered in Sec.~\ref{sec:BENCH}. The numerical values of the corresponding fitting parameters are reported in the legend of Fig.~\ref{fig:5bbis}.\footnote{A well-known caveat is that the scaling of the DW2X processor can only be considered to be a lower bound for the true scaling. This is due to the fact that the annealing time $t_{\rm DW}$ is not optimized as a function of the problem size but rather fixed at its minimum value allowed by the current technology. It is expected that reducing the minimum annealing time on quantum annealers such as the DW2X will improve their prefactor $\alpha$ but worsen their scaling behavior (increasing the exponential prefactor $\beta$), since the latter is artificially boosted by taking too long to solve small-size problems \cite{speedup,Hen:2015rt}.}

We conclude this section by further commenting on the comparison between classical solvers, which run on standard CPUs, and quantum optimization, which requires dedicated hardware. Using optimal total cost naturally allows us to take into account the cost of the hardware which is necessary to utilize each optimizer to perform a fair practical comparison. As an illustrative example, we can assume that the cost of performing a classical computation is negligible while we have to pay an additional premium $c_{\text{qp}}$ to perform a computation on a quantum processor:  $c \mapsto  c+c_{\text{qp}}$. In our example, the parameter $c$ is solver independent (it depends on the application). A premium $c_{\text{qp}}$ has the effect of shifting the curves in Fig.~\ref{fig:5abis} to the left. A left shift for the DW2X processor, for example,  would make the HFS solver more competitive.

\section{Optimal Use of a Randomized Optimizer}
\label{sec:OPTUSE}

\begin{figure*}[ht]
\subfigure[\, ]{\includegraphics[width=1\columnwidth]{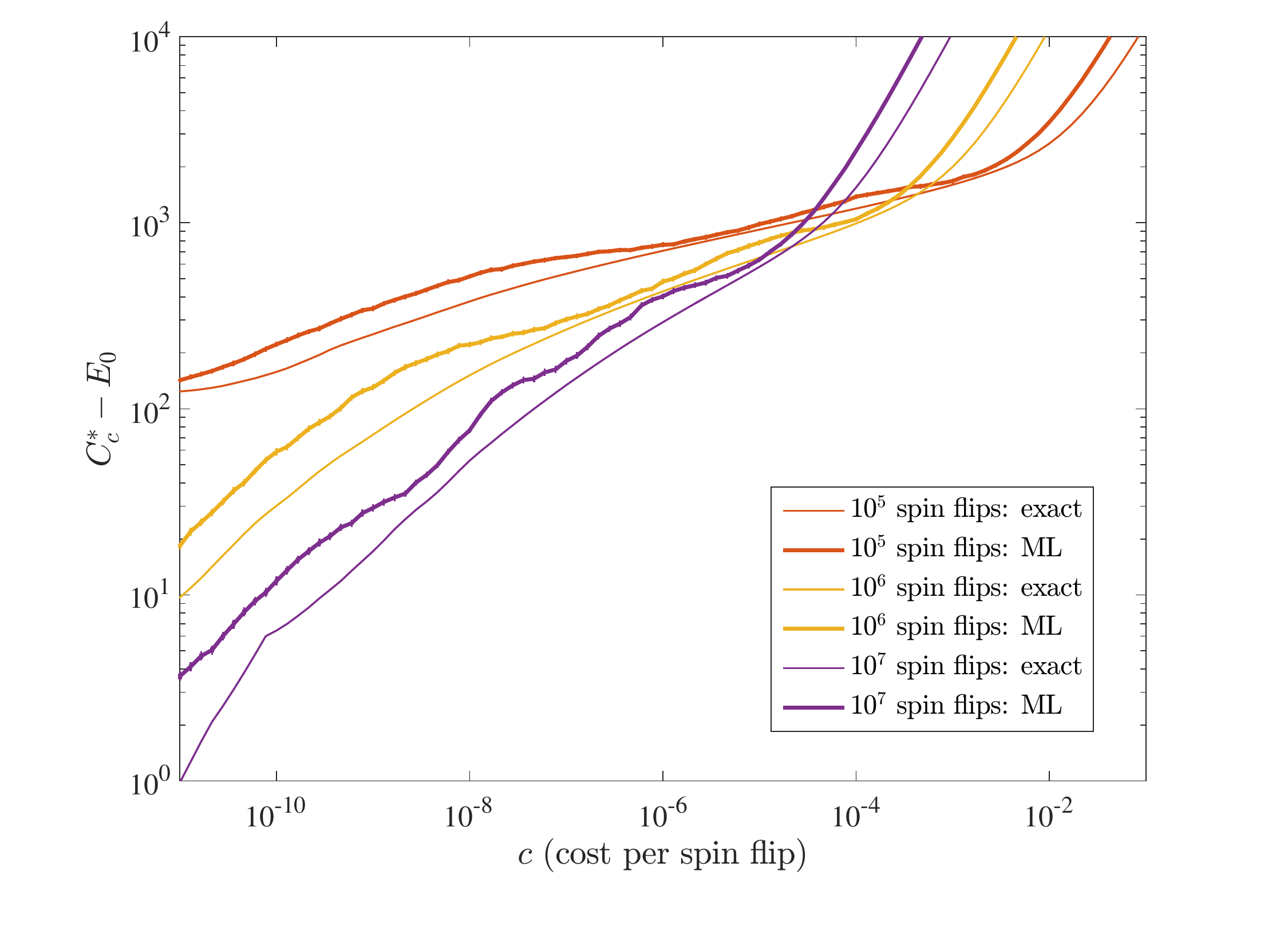} \label{fig:5a} }
\subfigure[\, ]{\includegraphics[width=1\columnwidth]{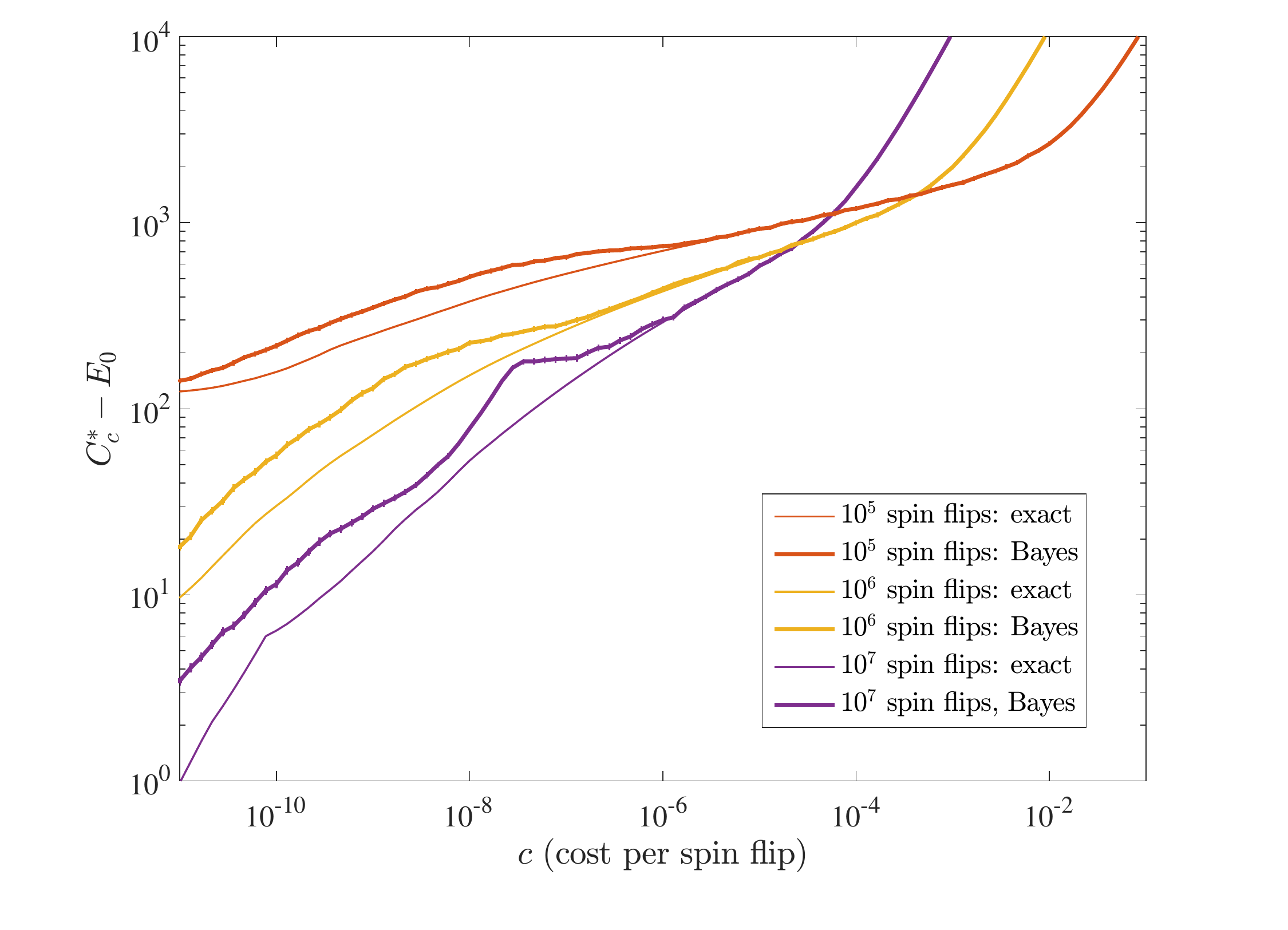} \label{fig:5b} }
\caption{Ideal (thin lines) and empirical (thick lines) optimal total cost for a representative instance on $N=1000$ variables and for different values of the number of spin updates $n_{\rm sf}$. The empirical optimal total costs are computed for the burn-in regime with (a) the maximum-likelihood fit method, and (b) the Bayesian method. For the class of problems studied, the prior knowledge obtained by knowing the behavior of other instances in the same family is good enough to make the Bayesian method almost exact in the large and intermediate $c$ regions.}
\label{fig:5} 
\end{figure*}

In the previous sections, we have interpreted benchmarking as a process to determine the energy distribution $\mathcal P(e)$ with the required confidence to compute $C^*_c$. In  practical applications, however, one must start the optimization process without a precise knowledge of the $\mathcal P(e)$. In the initial stages of the computation the determination of  $C_c^*$ will not be very precise. During the optimization process, as more more data is collected, one  updates the knowledge of $\mathcal P(e)$ to refine the calculation of the optimal total cost.

The calculation of the optimal total cost will depend on the previously observed energies and is updated every time a new solution is retrieved. Thus, Eq.~\eqref{eq:opt_E} is replaced by:
\beq
 C_{c,n}^* \,\,:\, \int_{-\infty}^{C^*_{c,n}}(C^*_{c,n}-e)\mathcal P_n(e) de =  c t_{\run}\,,
\eeq
where $\mathcal P_n(e)$ is our estimate of the quality distribution after $n$ energies have been retrieved from the optimizer. Note that a stopping decision based on incomplete knowledge of $\mathcal P(e)$ will necessarily be suboptimal. Any practical application of the optimal stopping approach will thus achieve an average cost larger than or equal to the ideal cost possible with full knowledge of $\mathcal P(e)$. 

It is important to note that at every stage of the computation, the learning process involves inferring the probability to observe solutions with energy smaller than all other previously observed solutions. It is easy to see why this is the case. A stopping rule solely based on the so-far observed empirical distribution $\mathcal P_n(e)$ necessarily calls for stopping at every stage. This is because the so-far observed lowest energy is empirically interpreted as the global solution, suggesting (incorrectly) that there is no advantage in continuing with the optimization process.

As we have shown above, for large values of $c$ the value of the optimal total cost $C^*_c$ is determined by the mean of the energy distribution [recall Eq.~\eqref{eq:high-c}]. We expect this quantity to be estimated fairly well even with a small number of observations. This is important because for large values of $c$ it is optimal to stop after a small number of observations. On the other hand, for small $c$ the optimal total cost is dominated by the behavior of the lower tail of the energy distribution. The experimental determination of the tail of a distribution requires, by definition, the collection of a large sample of energies. Fortunately, for small $c$ we expect that it is  optimal to stop only after a typically large number of energies is observed. This allows us to perform a tail inference analysis of $\mathcal P(e)$.

The observations above suggest the implementation of the following strategy for the optimal use of a randomized optimizer. During an initial  ``burn-in" regime, when the number of energies collected is small, $C^*_c$ can be obtained by inferring the general shape of the energy distribution $\mathcal P(e)$. We have considered two approaches. In one case, at each step of the burn-in regime a parametric maximum-likelihood fit is used to estimate $\mathcal P(e)$. In the other case, knowledge of $\mathcal P(e)$ is updated in a Bayesian manner. The latter case would be more appropriate when some prior knowledge of the energy distribution is available. In the ``asymptotic" regime a large number of energy samples is available. In this regime the optimal total cost can be estimated with an inference of the lower tail of $\mathcal P$. Tail inference could be performed, e.g., via a maximum-likelihood fit with Pareto distributions.

\subsection{Burn-in regime}

\subsubsection{Maximum-likelihood fits}

This approach to the burn-in regime requires the choice of a parametric representation $\mathcal P(e|\alpha_i)$ of the quality distribution, where the parameters $\alpha_i$ are determined via a maximum-likelihood fit. We thus determine the time-dependent optimal total cost $C_{c,n}^*$ using the energy distribution:
\beq
\mathcal P_n(e) = \mathcal P(e|\bar \alpha_{i,n})\,
\eeq
where $\bar \alpha_{i,n}$ are the maximum-likelihood estimations of the parameters $\alpha_{i}$ after $n$ observations. In general we expect that in the burn-in regime $C^*_{c}$ is mostly determined by the first moments of the energy distribution. Simple choices  should thus work well (for example parametric distributions with only the first few non-vanishing moments). Note that we must assume that the discrete distribution $\mathcal P$ is well-behaved, i.e., it can can be well-described by discretizing a smooth, continuous probability distribution.

\subsubsection{Bayesian updates}

A Bayesian approach to the burn-in regime can be preferable when one has prior knowledge of $\mathcal P(e)$. This prior  knowledge could have been acquired, e.g., by previously solving other similar optimization problems.  Following the Bayesian approach the energy distribution is now expressed in terms of a parametric function $\mathcal P(e|\alpha_i)$, with the parameters $\alpha_i$  also distributed according to a prior distribution $\mathcal D (\alpha_i|\nu_\alpha)$, which encodes our prior knowledge of $\mathcal P(e)$ and depends parametrically on a further set of  hyperparameters $\nu_\alpha$. The prior is updated as usual via the Bayes formula
\beq
\mathcal D ( \alpha_i|e_1,\dots,e_n,\nu_\alpha) = \frac{\mathcal P(e_1|\alpha_i)\dots \mathcal P(e_n|\alpha_i)\mathcal  D(\alpha_i|\nu_\alpha)}{\int \mathcal P(e_1|\alpha_i)\dots \mathcal P(e_n|\alpha_i) d\alpha_i}\,.
\eeq
After each observation, our best guess for $\mathcal P_{n}(e)$ is obtained by marginalizing over the parameters:
\beq
\mathcal P_{n}(e) = \int P(e|\alpha_i) \mathcal D ( \alpha_i|e_1,\dots,e_n,\nu_\alpha) d\alpha_i\,.
\eeq

An additional advantage of the Bayesian approach is the possibility to work with fully discrete distributions $\mathcal P(e)$. In many practical applications one can regard  $\mathcal P(e)$ as a multinomial distribution $P(e|\alpha_i)$, with the parameters $\alpha_i =p_i$ identifying the probability to obtain the energy $e_i$. A convenient choice for the prior distribution $\mathcal D (\alpha_i|\nu_\alpha)$ is the Dirichlet distribution, which is the conjugate prior of the multinomial distribution. In this convenient setup, Bayesian updates can be easily performed using only algebraic operations (see Appendix~\ref{sec:bayes} for more details).

\subsection{Asymptotic regime: tail inference with Pareto distributions}
\label{sec:asy}

In one of its possible formulations, the theorem of extreme values states that, under very general assumptions, the tail of a given distribution is well-approximated by an element of a three-parameter family of distributions known as Generalized Pareto Distributions: $\GPD(e,\lambda,k,\mu)$ \cite{coles2001introduction} (see Appendix~\ref{sec:GEV} for more details). In the asymptotic regime we define  $\mathcal P_{n}(e)$ as a  the following piece-wise probability distribution:
\beq
\mathcal P_{n}(e) = 
\left\{
\begin{array}{cc}
   \GPD(e,\bar \lambda_n,\bar k_n,\mu_{n}) & e\le \mu_{n}  \\
     \mathcal P_{\mathrm{emp},n}(e)     &  e >  \mu_{n} 
\end{array}\,.
\right.
\label{eq:pareto}
\eeq
The parameters $\bar \lambda_n$ and $\bar k_n$ are estimated via maximum-likelihood after each observation, while $\mu_{n}$ is a conveniently chosen threshold that defines the tail of the empirical distribution $P_{\mathrm{emp},n}(e)$ that is substituted by the $\GPD$ fit. The choice of the threshold $\mu_{n}$ is crucial to obtain a good tail inference via $\GPD$ \cite{coles2001introduction}. The threshold has to be as small as possible in order for the $\GPD$ to model the tail of $\mathcal P(e)$  with the smallest possible bias. On the other hand, $\mu_{n}$ should also be large enough so that  a sufficient number of observations is used for the maximum-likelihood fit of the parameters of  $\GPD$.

\subsection{Numerical Experiments with Optimal Stopping}

We performed optimal stopping experiments to study how close to the ideal optimal total cost is the average cost obtained by implementing the strategy described in the previous subsections. We assume that the empirical distribution obtained after $10^7$ SA runs is the exact distribution $\mathcal P(e) = \mathcal P_{\mathrm{emp},10^7}(e)$. Under this assumption, each of the outcomes of independent SA runs can be reproduced by a random sampling of $\mathcal P_{\mathrm{emp},10^7}(e)$. This approximation is crucial to keep the computational time manageable.

\subsubsection{Maximum-likelihood and Pareto fits}

Each optimal stopping experiment is performed as follows. We build a sequence of observations $\{ e_1,\dots, e_n \}$ via random sampling. We define a burn-in regime $1\le n \le 500$ and an asymptotic regime $n > 500$.   In the burn-in regime, $\mathcal P_n(e)$ is determined by fitting a Gaussian distribution via maximum-likelihood. In the asymptotic regime, we use the distribution defined in Eq.~\eqref{eq:pareto}, with the parameters $\bar \lambda_n$ and $\bar k_n$ obtained via a maximum-likelihood fit. The parameter $\mu_n$ is chosen to be the $2\times 10^4/n$-th percentile of the empirical distribution $P_{\mathrm{emp},n}(e)$. This means that $100$ observations are always used in the fit. At each step, $C_{n,c}^*$ is computed using the estimated distribution $\mathcal P_n(e)$ in the optimality equation Eq.~\eqref{eq:opt_E} and the principle of optimality Eq.~\eqref{eq:opt_n} is used to determine whether to stop or continue the sequence of observations. While the support of the intrinsic and empirical quality distributions is always compact, the support of the $\GPD$ tail can extend to $-\infty$. This may result in an estimate of the optimal total cost that is smaller than the global minimum ($C^*_{c,n}<E_0$) and as a consequence the principle of optimality never calls for stopping. To avoid this situation, we override the principle of optimality with an additional stopping rule, i.e., we stop as soon as the number of observations is large enough that there is a $99\%$ probability that we should already have observed a stopping value.  Once stopped, the final cost $C_n = \min(e_1,\dots,e_n) +n c t_{\run} $ is recorded. We repeated this process $1000$ times to determine the average cost.

Numerical results are shown in Fig.~\ref{fig:5a} for the same random instance as in Figs.~\ref{fig:1} and~\ref{fig:2} defined on $N=1000$ variables and for several values of the number of spin updates. The thin lines are the exact optimal total cost computed using $\mathcal P_{\mathrm{emp},10^7}(e)$ as the energy distribution. The thick lines are the average empirical costs obtained using the method described above.  As expected, the empirical costs are larger than the exact costs. Even with our simple approach, however, we were able to obtain empirical costs that satisfactorily reproduce the values of the ideal costs. This demonstrates the viability of the optimal stopping approach for the optimal use of randomized optimizers. 

We again identify three regions in Fig.~\ref{fig:5a}. In the large $c$ region, where it is optimal to stop after one observation, the empirical cost is systematically higher because we need at least two observation to perform a maximum-likelihood fit. 
In the intermediate $c$ region, the optimal total cost is mainly determined by the overall shape of the energy distribution. In this region it is important that the parametric distribution used for the fit accurately reproduces the quality distribution. We see that even the simple choice of a Gaussian distribution gives reasonably good results. In the small $c$ region the empirical optimal total cost are most likely dominated by the $\GPD$ tail fit. In this regime, however, it is more difficult to assess the effectiveness of our approach, because the tail of the empirical distribution $\mathcal P_{\mathrm{emp},10^7}(e)$ is not a statistically good approximation of the intrinsic energy distribution $\mathcal P(e)$.

\subsubsection{Bayesian updates}

We have also performed similar experiments where the energy distribution $\mathcal P_n(e)$ is obtained, in the burn-in regime, via Bayesian updates. As prior knowledge for the energy distribution, we used a family distribution $\mathcal P_{\mathrm{fam}}(e)$ obtained by including in the same energy distribution all the energies of all $100$ instances in the same family ($10^7$ energies in total). The idea is to use information about a family of similar instances to obtain a better guess of $C_c^*$ for a new optimization problem in the same family. The prior we choose is the Dirichlet distribution $\mathcal D(\alpha_i|500 p_{\mathrm{fam},i})$, where $p_{\mathrm{fam},i}$ is the probability to observe the energy $e_i$ from the family distribution  $\mathcal P_{\mathrm{fam}}(e)$. Intuitively, this function corresponds to a prior knowledge of $500$ ``virtual" observations distributed according to the family distribution. Results are shown in Fig.~\ref{fig:5b}. We see that in the regions of intermediate and large $c$ values the empirical cost almost exactly matches the exact cost. This is due to the fact that the Bayesian update can be performed after one observation (optimal in the large $c$ region), and that the family distribution used to build the prior is an excellent representation of the exact quality function $\mathcal P(e)$ for the class of problems considered in this study. Note that the maximum-likelihood and the Bayesian approaches give the same result in the small $c$ region. This is simply due to the fact that in this region stopping occurs after entering the asymptotic regime, which is the same in both cases.

\section{Parallelization}
\label{sec:PAR}
\subsection{Optimal Parallelization}

\begin{figure*}[ht]
\subfigure[\,]{\includegraphics[width=1\columnwidth]{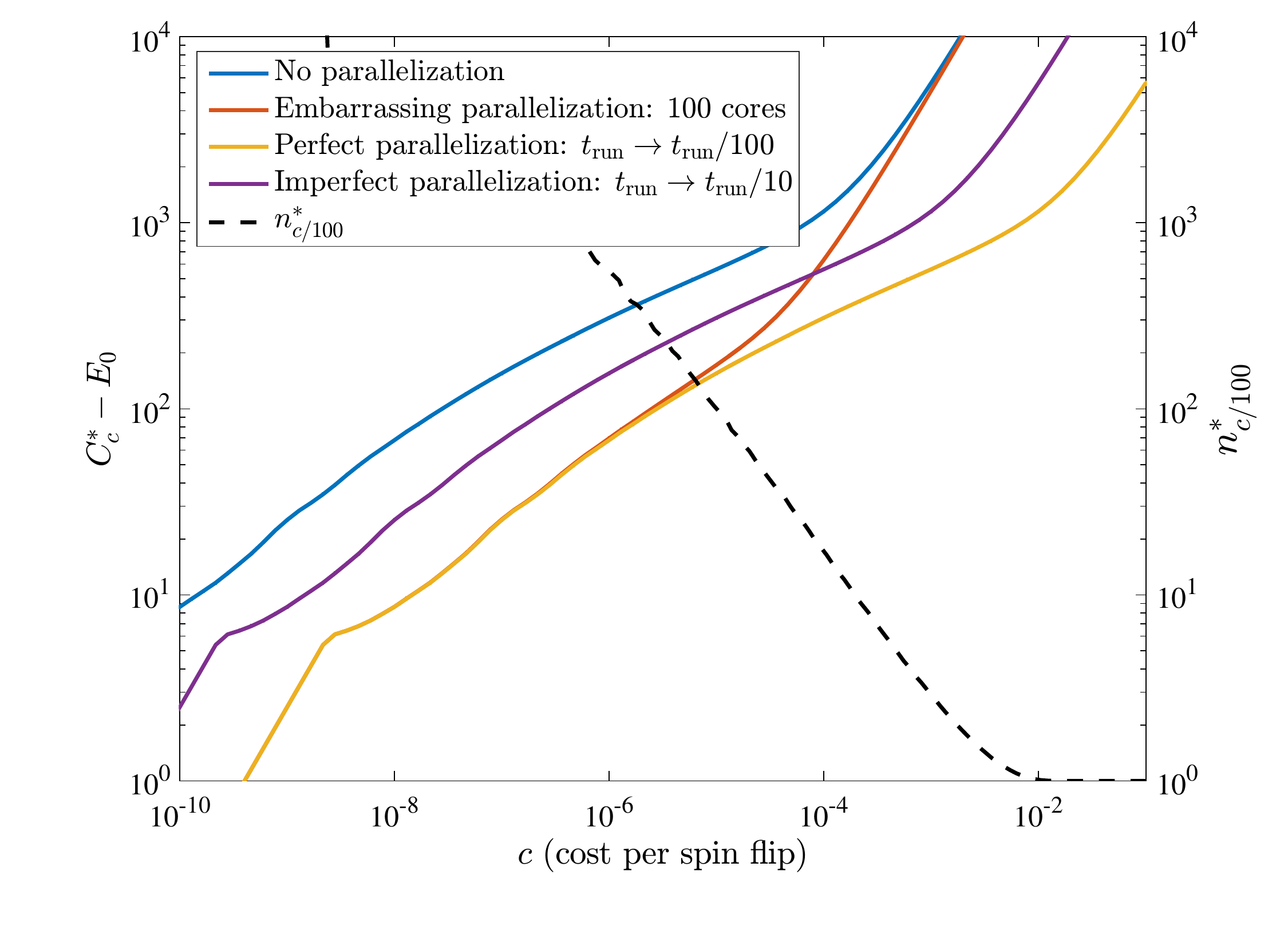} \label{fig:6a}}
\subfigure[\,]{\includegraphics[width=1\columnwidth]{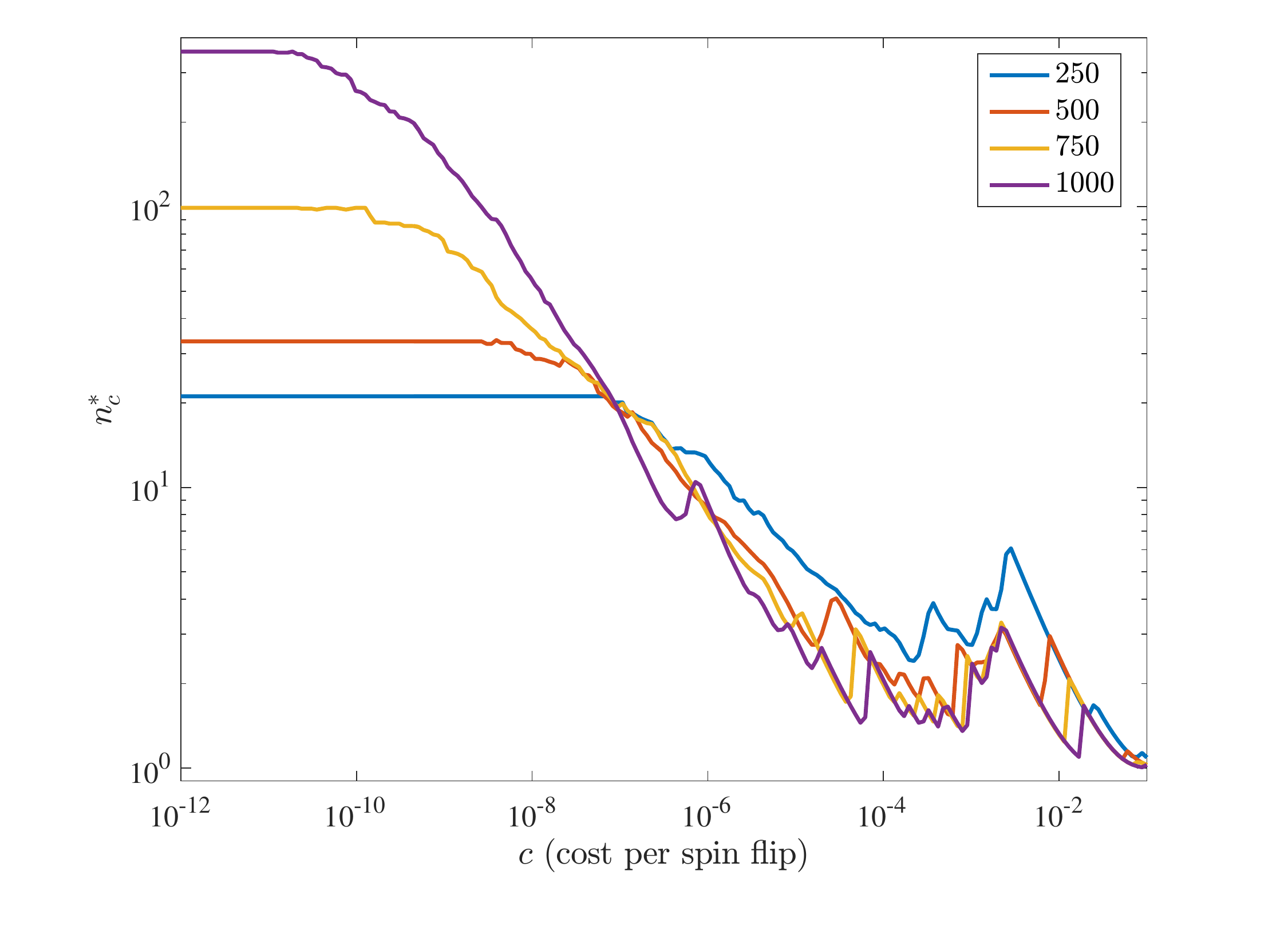} \label{fig:6b}}
\caption{(a) Optimal cost $C_c^*$ for the same instance as in Fig.~\ref{fig:1}, computed running SA with $5\times 10^6$ spin updates and different parallelization strategies. The black dashed line is the optimal stopping step $n_{c/100}^*$ corresponding to perfect parallelization. Embarrassing parallelization is notably worse than perfect parallelization when $n_{c/100}^* < 100$. (b) Stopping step $n_c^*$ computed after optimizing the number of spin updates and averaged over the $100$ instances of each size $N$ indicated in the legend. According to the discussion in the main text,  large values of $n_c^*$ allow for efficient embarrassing parallelization.}
\label{fig:6} 
\end{figure*}

The optimal stopping approach allows us to carefully address the question of parallelization. SA is an example of a solver that is ``embarrassingly" parallelizable, which means that multiple independent runs of the same algorithm can be executed in parallel to speed up the computation. When embarrassing parallelization is implemented, we can assume that  the $n$-th energy observation is given as follows:
\beq
 e_{n}(n_{\mathrm{cpu}}) = \min \{e_{n,1}, e_{n,2},\dots,e_{n,n_{\mathrm{cpu}}} \}\,,
 \label{eq:parallel_e}
\eeq
 where $n_{\mathrm{cpu}}$ is the number of cores, or processes used in the parallelization of the algorithm. A less trivial parallelization can also be implemented in many randomized optimization algorithms to shorten the length $t_{\run}$ of a computation. ``Perfect" parallelization is achieved when $t_{\run}\rightarrow t_{\run}/n_{\mathrm{cpu}}$. In most cases algorithms cannot be perfectly parallelized. Even when perfect parallelization is possible in theory, practical limitations may only allow for ``imperfect" parallelization. We have imperfect parallelization when  $t_{\run}\rightarrow t_{\run}/n_{\mathrm{imp}}$ with $1 < n_{\mathrm{imp}} < n_{\mathrm{cpu}}$.

What is the best parallelization strategy to minimize the total cost? Let us assume for the time being  that running multiple parallel processes does not increase the unit cost  $c$. Figure~\ref{fig:6a} shows $C_c^*$ under this assumption, for the same random instance as in Fig.~\ref{fig:1}, obtained by running SA using $5\times 10^6$ spin updates. The blue line is the optimal total cost without parallelization. The purple line corresponds to a situation where only imperfect parallelization is achievable. Perfect parallelization (yellow line) achieved with $100$ cores is equivalent to rescaling $c \rightarrow c/100$ and represents the ideal way of using the available computational resources. We find that embarrassing parallelization (red line) performs as well as perfect parallelization in the small $c$ regime.
This is an important practical observation: \emph{trivial embarrassing parallelization performs as well as ideal but practically unachievable perfect parallelization, in the low unit cost regime.} 

For large $c$, embarrassing parallelization becomes less effective and approaches the performance of the no parallelization result, but stays below it. This is because in the limit of large $c$, optimal total cost is dominated by the cost of drawing single observations, which is not reduced by embarrassing parallelization. Still, each of the parallel processes returns a single energy before stopping, so embarrassing parallelization yields the minimum energy over the set of ${\mathrm{cpu}}$ samples, while the no parallelization case draws a single sample from from the same distribution. Intuitively, embarrassing parallelization becomes essentially equivalent to perfect parallelization when the perfectly parallelized optimal stopping step $n_{c/n_{\mathrm{cpu}}}^*$ is larger or comparable to the number of parallel processes, i.e., $n_{\mathrm{cpu}} <  n_{c/n_{\mathrm{cpu}}}^*$, which requires $c$ to be sufficiently small. This observation is confirmed in  Fig.~\ref{fig:6a} which shows how embarrassing parallelization performed with $100$ processes starts to perform  notably worse than perfect parallelization when $ n_{c/100}^* < 100$. Thus, a rule of thumb for an optimal parallelization strategy is to use all parallel processes to perform embarrassing parallelization in the limit of small $c$, when  $n_{\mathrm{cpu}} <  n_{c/n_{\mathrm{cpu}}}^*$, or imperfect parallelization in the large $c$ limit, when  $1 \simeq  n_{c/n_\mathrm{imp}}^* $. In the intermediate regime parallel processes should be optimally distributed between embarrassing and imperfect parallelization. Embarrassing parallelization can thus be efficiently exploited only if there is a regimes where the optimal stopping step $n_c^*$ is large.  

Figure~\ref{fig:6b} shows $n_c^*$ (averaged over $100$ instances) for different problem sizes obtained after optimization of the number of spin updates. The optimal stopping step $n_c^*$ grows larger than $1$ for larger problem size and small $c$. As expected, $n_c^*$  is close to $1$ in the limit of large $c$. 
Figure~\ref{fig:6b} shows, e.g., that embarrassing parallelization can be efficiently implemented (i.e., it is equivalent to perfect parallelization) in solving the $N=1000$ class of problems when the number of parallel processes used is $\sim 300$ (and $\sim 20$ for the $N=250$ case). It should then be more effective to use a larger amount of computational resources to implement a certain degree of imperfect parallelization. 

We expect these observations to guide future benchmarking studies in giving useful information about parallelization optimization. 

\subsection{Optimal Number of Cores}

So far we assigned a cost $c$ to the flow of time, but have not taken into account the cost of the hardware resources necessary for parallelization. More generally, we must include the cost of implementing a certain amount of  computational resources into the cost function $T(t)$. A simple practical approach is to use the following type of cost function:
\beq
T(t, n_{\mathrm{cpu}}) = (c_t + c_{\mathrm{cpu}} n_{\mathrm{cpu}}) t_{\run}\,.
\label{eq:parallel_c}
\eeq
The cost $c$ now comprises two contributions. The term $c_t$ measures a cost that is simply due to the flow of time, and is solver and hardware independent.  This term could, e.g., be related to a loss in revenue for a certain business to simply idling while the optimizer is running. The term $c_{\mathrm{cpu}} n_{\mathrm{cpu}}$ depends on the hardware used and is proportional to the number of cores used. This term may include, for example, utility bills, maintenance or rent costs that scale linearly with the size of the cluster. 
Equation~\eqref{eq:parallel_c} can be viewed as defining the cost $c$ for a parallelized version of a given solver, with the parameter $n_{\mathrm{cpu}}$ being interpreted as an additional parameter to be optimized in order to minimize the total cost $C_{c_t,c_{\mathrm{cpu}}}^*(n_{\mathrm{cpu}})$, which is now considered a  function of  $n_{\mathrm{cpu}}$. 

In the previous subsection we showed that, for sufficiently small $c = c_t + c_{\mathrm{cpu}} n_{\mathrm{cpu}}$, embarrassing parallelization is practically equivalent to perfect parallelization.  We thus have that for sufficiently small $c_t + c_{\mathrm{cpu}} n_{\mathrm{cpu}}$ parallelization is equivalent to the rescaling $ c_t + c_{\mathrm{cpu}} n_{\mathrm{cpu}} \rightarrow  c_t/n_{\mathrm{cpu}} + c_{\mathrm{cpu}} $. This cost is monotonically decreasing in $n_{\mathrm{cpu}}$. Because the optimal total cost is a monotonic function of the unit cost $c$, it follows that it is optimal to increase $n_{\mathrm{cpu}}$ at least as long as the embarrassing parallelization process is effective, i.e., as long as $n_{\mathrm{cpu}} \sim n^*_{c_t/n_{\mathrm{cpu}} +c_{\mathrm{cpu}}}$.  As explained in the previous subsection, a larger number of cores can be optimal if a certain degree of imperfect parallelization is possible. An important observation is that when the cost per CPU $c_\mathrm{cpu}$ is included, even the efficacy of perfect parallelization is greatly reduced  when  $n_{\mathrm{cpu}}$ is large. The effective unit cost converges to $ c_t/n_\mathrm{cpu} + c_{\mathrm{cpu}}  \rightarrow  c_{\mathrm{cpu}} $, with the optimal total cost converging to $C^*_{c_{\mathrm{cpu}}}$ with no further improvements. This limiting value is uniquely determined by the unit hardware cost $c_{\mathrm{cpu}}$ and has a simple intuitive explanation: including the cost of the hardware in the unit cost places a practical limit on the amount of hardware resources that should be implemented in solving an optimization problem. In general, the optimal number of cores $n^*_{\mathrm{cpu}}(c_t, c_{\mathrm{cpu}})$ will depend on the specific solver, instances, and cost function considered, and should be estimated by performing benchmarking studies similar to those presented here.

\section{Conclusions}
\label{sec:CONC}

We have presented an optimal stopping approach to benchmarking randomized optimization algorithms. Rather than focusing, as is customary, on optimizing solution quality (e.g., minimizing the energy) alone, we considered the more general problem of optimizing solution quality along with the associated cost of obtaining samples of the fitness function. This approach is natural given that the cost of more samples grows with time, so that the total cost should account for the fitness function along with the latter cost. We have shown that this problem lends itself naturally to an analytical solution within the framework of optimal stopping theory, and can be recast as the well-known ``house-selling" problem, under the assumption that samples are statistically independent random variables. This approach yields both the optimal stopping time and the optimal total cost [Eqs.~\eqref{eq:opt_n} and \eqref{eq:opt_E}, respectively]. Moreover, our optimal stopping approach includes as special cases all the standard variants of randomized benchmarking, including time-to-solution, time-to-target, average energy, and target-in-time.

To find the optimal stopping time, at which a round of randomized benchmarking concludes and one settles for the lowest energy solution obtained thus far, requires knowledge of the energy distribution function. We have shown how this distribution can be found on the fly using either maximum likelihood fits or Bayesian updates, and by inferring the occurrence of rare samples using the theory of generalized Pareto distributions. This demonstrates that optimal stopping criteria lead to an optimal-utilization strategy of randomized optimization algorithms that can be successfully used in practical scenarios.

Our key findings, based on studying MAX2SAT problems over complete graphs and frustrated-loop problems with planted solutions on the Chimera graph (using  the D-Wave 2X quantum annealing device and the classical HFS algorithm) can be summarized as follows:

\begin{itemize}
\item As the unit cost (the cost per operation) increases, the run-time must be lowered, i.e., the optimal run-time depends on the unit cost, as exemplified in Fig.~\ref{fig:1b}. After optimization of the run-time, optimal total cost give the optimal trade-off between computational time and solution quality, as shown in Fig.~\ref{fig:2b}.

\item Hard optimization problems exhibit exponential scaling with problem size of the optimal total cost only in the regime of small unit cost. When the unit cost is sufficiently high, the total cost for such optimization problems instead exhibits polynomial scaling [see Fig.~\ref{fig:3d}]. This is a consequence of the optimal stopping criterion, which favors stopping before the lowest energy is found, when the unit cost is high.

\item Assuming equal unit cost, we find that the D-Wave 2X quantum annealer outperforms the HFS algorithm run on a single CPU core by a factor of $\sim 100$ in terms of the total cost [see Fig.~\ref{fig:5bbis}]. 
This should not be seen as a claim of quantum speedup (e.g., because it is difficult to assign consistent unit costs across different technologies), but rather as an encouraging sign for quantum annealing that it can be competitive with the best classical optimization heuristics.

\item Optimal total cost provides a precise criterion for optimal parallelization strategies. We found that, in the low unit cost regime, ``embarrassing parallelization" performs as well as ideal, but practically unachievable, perfect parallelization. In the large unit cost regime, on the other hand, even imperfect parallelization (to reduce the run-time of the algorithm) is preferable. In the intermediate unit cost regime, embarrassing and imperfect parallelization should be optimally balanced to minimize the optimal total costs [see Fig.~\ref{fig:6a}]. 

\end{itemize}

We hope that our approach to benchmarking randomized optimization algorithms, which balances optimizing the objective function with the computational cost of optimization, will inspire future investigations into this important tradeoff, and will result in an appreciation of the useful role optimal stopping theory can play in heuristic optimization.

\begin{acknowledgments}
We thank Tameem Albash and Itay Hen for useful discussions during the preparation of this work. We also thank Tameem Albash for providing the set of instances with planted solutions used in  section \ref{sec:CvsQ}. This work was supported under ARO MURI Grant Nos. W911NF-11-1-0268 and W911NF-15-1-0582, and NSF grant number INSPIRE-1551064.
\end{acknowledgments}

 \appendix

\section{Optimal Stopping}
\label{sec:OS}

In this section we give a one-page introduction to the theory of optimal stopping. Our presentation closely follows the lecture notes \cite{Ferguson:book}, which can be consulted for more details on the subject. 

Quoting Ferguson: ``the theory of optimal stopping is concerned with the problem of choosing a time to take a given action based on sequentially observed random variables in order to maximize an expected payoff or to minimize an expected cost". In the most general set-up, an optimal stopping problem is defined by two objects:
\begin{enumerate}
\item A sequence of random variables, $X_1,X_2,\dots$, whose joint distribution is assumed known;
\item A sequence of real-valued reward functions $y_0,y_1(x_1),y_2(x_1,x_2),\dots,y_{\infty}(x_1,x_2,\dots)$.
\end{enumerate}
The optimal stopping problem is then defined as follows. One sequentially observes the values $x_1,x_2,\dots, x_n$. At each step $n=1,2,\dots$ one decides whether to stop and obtain a reward, $y_{n}(x_1,x_2,\dots,x_n)$ or to continue. If no observations are taken, one obtain the reward $y_0$, while never stopping results in the reward $y_{\infty}(x_1,x_2,\dots)$. The goal is to find a time to stop (stopping rule) that maximizes the expected reward.  A stopping rule is a sequence of functions  $0 \le \phi_n(x_1,\dots,x_n) \le 1$ defining the probability to stop at each step $n$.  The expected reward associated with the stopping rule $\phi$ (i.e., the family of functions $\{ \phi_n \}$) and the sequence of observations $x_1,x_2,\dots$ is then
\begin{align}
V_\phi(x_1,x_2,\dots) &= \sum_{n=0}^\infty \prod_{j=1}^{n-1}\left[1-\phi_j(x_1,\dots,x_j)\right] \\
&\quad \times\phi_n(x_1,\dots,x_n)y_n(x_1,\dots,x_n) \notag \ .
\label{eq:optrew}
\end{align}
The product is the probability of not stopping for the first $n-1$ steps, followed by a stop at step $n$; multiplied by the reward at step $n$ this gives the expected reward at that step, and summed over all step values this gives the expected reward associated with the rule $\phi$ for a given sequence of observations. The reward associated with the stopping rule $\phi$ is obtained by averaging over all possible sequences of observations $V_\phi = E\{V_\phi(X_1,X_2,\dots)\}$. The optimal stopping rule  is thus given by:
\beq
\phi^* \,\, : \,\, V^* \equiv V_{\phi^*} = \sup_\phi\{ V_{\phi} \}\,.
\eeq
A special case is where $\phi_n(x_1,\dots,x_n)=0,1$ are binary valued  functions, i.e., instead of a randomized stopping rule, at each step one takes a deterministic stopping decision. In this case we simply have $V_\phi(x_1,x_2,\dots) = y_n(x_1,x_2,\dots,x_n)$, with $n$ being the stopping step, and we can also write  $V_\phi = E\{Y_{N_\phi}\}$. We have used capital letters to stress that the rewards $Y_{N_\phi}$ and the stopping time $N_\phi$, being functions of the random variables $X_n$, are themselves random variables.

A central result in the theory of optimal stopping regards the existence of optimal stopping rules. In particular, it is possible to prove that an optimal stopping rule exists if the following two conditions are satisfied:
\bes
\begin{align}
&E\{\sup_n Y_n\} < \infty\,, \\
& \limsup_{n \rightarrow \infty} Y_n  \le Y_\infty\,.  
\end{align}
\ees
These conditions have a very simple and intuitive explanation. The first means that even a prophet that has knowledge of the whole sequence $y_1,y_2,\dots$, and thus knows in advance when it is optimal to stop, can only obtain a finite reward. The second is simply an asymptotic regularity condition. Under these conditions one can also show that the optimal stopping rule is given by the \emph{principle of optimality}. Let us define the optimal reward $V^*_n(x_1,\dots,x_n)$ conditioned on the observation of $n$ values
\beq
V^*_n(x_1,\dots,x_n) = {\rm ess \, sup}_{N_\phi \ge n}E \{ Y_{N_\phi} | x_1,\dots,x_n \}\,,
\eeq
where the essential supremum is taken over all the stopping rules that call for drawing at least $n$ observations. The principle of optimality then states that it is optimal to stop as soon as $y_n(x_1,\dots,x_n) = V^*_n(x_1,\dots,x_n)$:
\beq
n^* = \min \{ n\, | \, y_n(x_1,\dots,x_n) \ge V^*_n(x_1,\dots,x_n) \}\,.
\label{eq:optprinc}
\eeq
The principle of optimality is also very intuitive: at any stage $n$, it is optimal to stop if the reward $y_n(x_1,\dots,x_n)$ obtained in case of stopping  is at least as large as the optimal reward $V^*_n(x_1,\dots,x_n)$ that one may expect if a decision to continue is taken. Finally, it is possible to prove the following \emph{optimality equation}: 
\beq
V^*_n = E\{\max \{Y_n, E\{V^*_{n+1}(x_1,\dots,x_n, X_{n+1})   \}  \}\}\,.
\label{eq:opteq}
\eeq
This equation plays a central role in dynamic programming. It is a recursive equation between $V^*_n$ and $V^*_{n+1}$: it states, again quite intuitively, that the optimal reward that can be obtained at stage $n$ is the maximum between the expected reward $Y_n$ obtained from stopping exactly at the stage $n$ and the reward $V^*_{n+1}(x_1,\dots,x_n, X_{n+1})$ expected if one were to use the optimal among all stopping rules that call for at least one other draw (from the distribution $X_{n+1}$).

Equation~\eqref{eq:opteq} is, in most practical situations, difficult to solve. Therefore, several techniques have been developed to find approximate, or near-optimal, stopping rules. An important class of problems, which allows for powerful analytical results, is that of Markov models, in which the distribution $X_{n+1}$ at stage $n+1$ does not depend on the previous $n$ observations, but only on the distribution $X_n$ at stage $n$:   $X_{n+1}(x_1,\dots,x_n) = X_{n+1}(X_n)$. Moreover, the rewards are functions of the last observation only: $Y_n = y_n(X_n)$. An additional simplification arises when some symmetries are present. In the next section we solve a Markov optimal stopping problem with translational symmetry that was used in the main text.

\subsection{The House Selling Problem}

A prototypical optimal stopping problem is the so-called house selling problem. The problem is formulated as follows. An asset is on sale, and offers are presented daily to the seller. On day $n$, an offer $x_n$ is made which one assumes is an observation of a random variable $X$ extracted from a distribution $\mathcal P(x)$ which does not depend on the day. One also assumes that each additional day of waiting implies a cost $c$. Accepting an offer on day $n$ thus gives the reward
 \beq
y_n = x_n-c n\quad (y_0 = 0, y_\infty = -\infty)\,
\eeq
if recalling previous offers is not possible, or 
\beq
y_n = \max\{x_1,\dots,x_n\}-c n\,,
\eeq
if recall is allowed. In this problem, we recognize the structure of a Markov model with additional translational invariance due to the invariance over time of the distribution of offers $X$ and the linear dependence on the cost. We can exploit these properties to analytically solve the optimality equation~\eqref{eq:opteq}. Because of the above-mentioned translational invariance, the problem at stage $n$ is equivalent to the problem at stage $0$, with the only difference being that at stage $n$ a price $cn$ has been paid that cannot be recovered. This allows us to write the following for the expected reward:
\beq
V^*_{n+1} = V^*_{n}-c = V_1^*-c n\equiv V^*-c n\,.
\eeq
With the condition above, the optimality equation~\eqref{eq:opteq} can be written as follows:
\begin{align}
V^*_{n} & =   E\{\max \{X_n-cn, V^*_{n+1}  \} \}\notag \\
&= E\{\max \{X_n-cn, V^*_{n}-cn  \}\} \ ,
\end{align}
which implies
\bea
V^*   =   E\{\max \{X_n, V^*  \} \}-c \ .
\eea
This can be rewritten as $\int_{-\infty}^{+\infty}V^* \mathcal P(x) dx = \int_{-\infty}^{V^*}V^* \mathcal P(x) dx + \int_{V^*}^{+\infty} x \mathcal P(x) dx -c$, which yields our final form:
\bea
\int_{V^*}^{+\infty} (x-V^*)  \mathcal P(x) dx = c\, ,
\label{eq:eopder}
\eea
an equation we can solve for the optimal reward $V^*$. The principle of optimality then instructs to stop as soon as an offer larger than or equal to the optimal reward has been received:
\beq
n^* = \min\{n \ge 1: x_n \ge V^*\}\,.
\eeq
Note that since $V^*$ is $n$-independent, it makes no difference whether we are allowed to recall previous offers. With recall, we would stop as soon as $\max\{x_1,\dots,x_n\} = x_n \ge V^*$, and both the expected reward and the optimal stopping rule are equivalent in the cases with or without recall. In this specific problem, therefore, the possibility to recall previous offers does not translate into a practical advantage. With this comment in mind, and by simply changing the sign of our quantities (minimizing the costs rather than maximizing the rewards), we can trivially put Eq.~\eqref{eq:eopder} into the form of Eq.~\eqref{eq:opt_E}.

\subsection{The Optimality Equation}
\label{sec:opteq}
In this section we give more details about the optimality equation~\eqref{eq:eopder}. Let us rewrite it as follows:
\beq
\int_{V^*}^{+\infty} (x-V^*)  \mathcal P(x) dx  \equiv I(V^*) = c\,.
\label{eq:I}
\eeq
It is easy to show that
\beq
\partial_{V^*} I(V^*) =  - \int_{V^*}^{+\infty}  \mathcal P(x) dx = {\rm CDF}(\mathcal P(x), V^*)-1\,.
\eeq
The integral $I(V^*)$ is thus a positive, monotonically decreasing function of $V^*$. The optimality equation $I(V^*)=c$ thus has a unique solution and can easily be solved numerically. Moreover, letting $f(V^*,c)\equiv I(V^*) - c$, this implies that the equation $f(V_c^*,c)=0$  is an implicit definition of $V^*_c$ as a function of $c$. From the theorem of implicit functions\footnote{The condition  $f(V^*,c) = 0$ is an implicit definition of $V^*$ as a function of $c$, which we denote by $V_c^*$. Differentiating  $f(V_c^*,c) = 0$ with respect to $c$ gives $\partial_{c}f(V_c^*,c) +\partial_{V_c^*}f(V_c^*,c)\partial_cV_c^* = 0$, which is the first equivalence in Eq.~\eqref{eq:impl}.}
we have:
\begin{align}
\partial_c V^*_c  &= 
- \left. \frac{\partial_c f(V^*,c)}{\partial_{V^*}f(V^*,c)}\right|_{V_c^*} =
-  \left.\frac{\partial_c( -c)}{\partial_{V^*} I(V^*)}\right|_{V_c^*} \notag \\
&= \left({\rm CDF}(\mathcal P(x), V_c^*)-1\right)^{-1}\,.
\label{eq:impl}
\end{align}
Thus, the dependence of the expected reward $V^*$ on the cost $c$ can be inferred from the CDF. The slope of $V^*_c$ as a function of $c$ is an increasing function of $c$, going from $-\infty$ to $-1$. As a consequence,  $V^*_c$ has a left vertical asymptote at $c= 0$  and an right oblique asymptote at $c \rightarrow +\infty$.
 The optimality equation can be analytically solved on the right oblique asymptote where $|V^*_c|$ is very large. The integrals above can then be approximated by $\int_{-\infty}^{+\infty}$ and the optimality equation reduces to $E(X)-V^*_c = c$. In the vicinity of the vertical asymptote $V^*_c$ depends on the specific form of the upper tail of the distribution $\mathcal P(x)$.

\subsection{Tail Contribution to Optimal Rewards}
\label{appsec:tail}

When applying optimal stopping ideas to benchmarking probabilistic optimizers, we use an experimental estimate for $\mathcal P(x)$ in order to determine the experimental optimal total cost $V^*_c$. An important subtlety in this regard is understanding the influence of rare events (or the contribution of the upper tail of $\mathcal P(x)$) in determining $V^*_c$. This is important because rare events corresponding to very large values of $|x|$ 
can have a non-negligible contribution to determining the optimal  reward. An estimate of the contribution of statistical errors can be done as follows. Let us assume that the true distribution $\mathcal P(x)$ is approximated by a parametric fit $\mathcal P_{\rm{fit}}(x|\alpha)$ depending on the parameter $\alpha$. The error $\delta\alpha$ now encodes the statistical error in the  knowledge of the distribution $\mathcal P(x)$. We can then estimate  $\delta_\alpha V^*_c$, the uncertainty of $V^*_c$ given the uncertainty of  $\mathcal P(x)$ encoded in $\delta \alpha$, as follows :
\begin{align}
\delta_\alpha V^*_c  &= (\partial_\alpha V^*_c) \delta \alpha = -\frac{\partial_\alpha I(V_c^*)}{{\rm CDF}(\mathcal P(x), V_c^*)-1}\delta \alpha \notag \\
&= -\frac{\delta_\alpha I(V_c^*)}{{\rm CDF}(\mathcal P(x), V_c^*)-1},
\label{eq:impl2}
\end{align}
where in the second equality we have used the theorem of implicit functions again.\footnote{We use $f(V^*_c(\alpha),c) = I(V_c^*(\alpha)-c = 0$. Differentiating with respect to $\alpha$ gives $\partial_{\alpha}f(V_c^*,c) +\partial_{V_c^*}f(V_c^*,c)\partial_\alpha V_c^*(\alpha) = 0$ (we suppressed the dependence on $\alpha$ where possible). Using Eq.~\eqref{eq:impl} gives ${\partial_{V^*_c} I(V^*_c)} = {\rm CDF}(\mathcal P(x), V_c^*)-1$. Together, this yields the second equality in Eq.~\eqref{eq:impl2}.}
We can now write for the numerator:
\begin{align}
\delta_\alpha I (V_c^*) &= \int_{V_c^*}^{+\infty} (x-V_c^*)  \delta_\alpha \mathcal P_{\rm{fit}}(x|\alpha)dx  \notag \\
&= \int_{\mathrm{tail}} (x-V_c^*)  \delta_\alpha \mathcal P_{\rm{fit}}(x|\alpha)dx \notag \\
&=  \int_{\mathrm{tail}}  x   \delta_\alpha \mathcal P_{\rm{fit}}(x|\alpha)dx ,
\label{eq:A19}
\end{align}
where in the first equality we used the definition of $I (V_c^*)$ in Eq.~\eqref{eq:I} but replaced $\mathcal P(x)$ by $\mathcal P_{\rm{fit}}(x|\alpha)$, and remembered to hold $V_c^*(\alpha)$ constant. In the second equality we assumed that $\delta_\alpha \mathcal P_{\rm{fit}}(x|\alpha)$ is non-negligible only in the upper tail of $\mathcal P_{\rm{fit}}(x|\alpha)$, and in the third equality we assumed $x \gg V_c^*$, i.e., $V_c^*$ is not in the upper tail of $\mathcal P_{\rm{fit}}(x|\alpha)$. Combining the two equations above and removing the label $\alpha$ from the equations yields the estimate  [Eq.~\eqref{eq:errorbody}] reported in the main text for the error $\delta V^*_c$, where we now replaced by $\mathcal P_{\rm{fit}}(x|\alpha)$ by $\mathcal P(x)$:
\beq
\delta V^*_c \sim \frac{\int_{\mathrm{tail}} x  \delta \mathcal P(x) dx}{\int_{V^*}^{\infty}  \mathcal P(x) dx}\,.
\eeq
The error $\delta V^*_c$ will be thus negligible if the denominator is large enough, which can be achieved by considering sufficiently small values of $V^*$, or a sufficiently small numerator, which can be achieved by collecting enough measurements to reduce the weight of the unobserved tail. Of course, we must assume that the intrinsic probability distribution $\mathcal P(x)$ has a well-behaved tail, i.e., $\left| \int_{\mathrm{tail}} x  \mathcal P(x) dx\right|$ should be small as long as the tail weight $\left| \int_{\mathrm{tail}}  \mathcal P(x) dx \right|$ is small.

\section{Bayesian Updates of the Multinomial Distribution}
\label{sec:bayes}

In the case of discrete optimization the quality distribution $\mathcal P$ is always a multinomial distribution  $ \mathcal P(e | p_i \equiv \alpha_i  )$, where the parameters $\alpha_i \equiv p_i$ are the probabilities to obtain the corresponding energy values $e_i$.\footnote{Note that the set of observed energies is always finite, thus discrete. The quality distribution can thus be described by a multinomial distribution even in the case of continuous optimization.} In the Bayesian approach, a previous knowledge, or ``best guess", of the parameters $p_i$ is described by a prior, i.e., a probability distribution $\mathcal D(p_i|\nu_\alpha)$ for the parameters $p_i$ that depends on a set of additional hyper-parameters $\nu_\alpha$.

A convenient choice for the prior is the Dirichlet distribution, which we denote by $\mathcal D(p_i|\nu_i)$ \cite{gershman2012tutorial}. The hyper-parameters $\nu_i$ are positive numbers and the support of the Dirichlet distribution is a set of probabilities $p_i$ (i.e., $\sum_i p_i = 1$). We use the same Latin index $i$ to indicate that the number of hyper-parameters $\nu_i$ is the same as the number of parameters $p_i$, i.e., equal to the number of non-equal energy values $e_i$. The Dirichlet distribution is the conjugate prior of the multinomial distribution, which means that the posterior distribution is itself a Dirichlet distribution. In particular, one can show that the Bayesian update of the Dirichlet distribution after $n$ observations
\beq
\mathcal D ( p_i |e_1,\dots,e_n, \nu_i) = \frac{\mathcal P(e_1|p_i)\dots \mathcal P(e_n|p_i)\mathcal  D(p_i|\nu_i)}{\int \mathcal P(e_1|p_i)\dots \mathcal P(e_n|p_i) dp_i}\ ,
\eeq
is given by:
\beq
\mathcal D ( p_i |e_1,\dots,e_n, \nu_i) = \mathcal D ( p_i | \nu_i+n_i)\,,
\eeq
where $n_i$ is the number of observations equal to $e_i$ and $n = \sum_i n_i$.  The expression above shows that the hyper-parameters $\nu_i$ can indeed be interpreted as a set of virtual observations that encode our expectation before any observation is performed. More precisely, the average of a Dirichlet distribution is given by:
\beq
\bar p_i =   \int p_i\mathcal  D ( p_i | \nu_i+n_i)dp_i = (\nu_i+n_i)/\sum_i (n_i+\nu_i)\,.
\eeq
As expected, when the number of real observations $n_i$ is larger than the number of virtual observations $\nu_i$, the average probabilities $\bar p_i$ are equal to the observed probabilities $p_i = n_i/\sum_i n_i$.

We finally mention the following very useful property: marginalizing a multinomial distribution over a set of hyper-parameters distributed according to a Dirichlet distribution still gives a multinomial distribution with probabilities given by the average probabilities $\bar p_i$.  In other words, the posterior predictive distribution $\mathcal P(e | e_1,\dots, e_n,\nu_i)$ is given by

\begin{align}
\mathcal P(e | e_1,\dots, e_n,\nu_i) &= \int \mathcal P(e | p_i) \mathcal D ( p_i | e_1,\dots,e_n, \nu_i) dp_i \notag \\
& = \mathcal P(e | \bar p_i)\,.
\end{align}

\begin{figure*}[t]
\includegraphics[width=1.84\columnwidth]{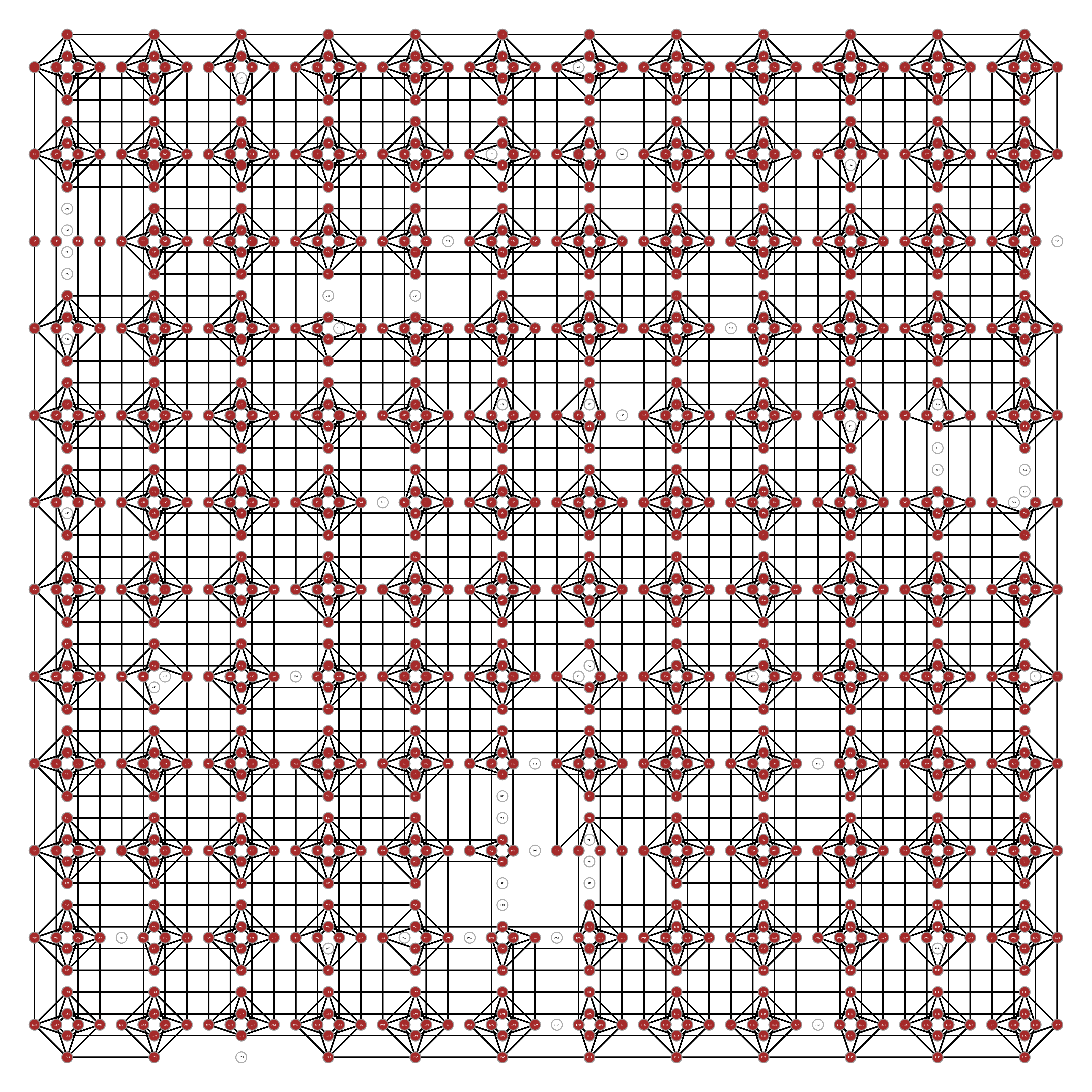}
\caption{Hardware connectivity of the DW2X device installed at the University of Southern California. Red circles represent usable qubits, while white circles represent the $54$ deactivated qubits. Black lines represent all the available couplings between active qubits.}
\label{fig:dw}
\end{figure*}

\section{Generalized Pareto Distribution}
\label{sec:GEV}

Extreme Value Theory provides a general framework to model the statistical behavior of extreme, i.e., very rare, events \cite{coles2001introduction}. The theorem of extreme values is valid under some very general assumptions on the regularity of the unknown distribution $\mathcal P(x)$. It states that the distribution of threshold exceedances, i.e., the conditional probability
\beq
\Pr \{ X > \mu  + x | X > \mu \} = \frac{1-{\rm{CDF}}(\mathcal P, x+\mu)}{1-\rm{CDF}(\mathcal P,\mu)} 
\eeq
is approximated, for very large $\mu$, by a Generalized Pareto Distribution $\GPD(x,\lambda,k,\mu)$
\beq
\frac{1-{\rm CDF}(\mathcal P, x+\mu)}{1-{\rm{CDF}}(\mathcal P,\mu)}  \sim {\rm{CDF}}(\GPD(\lambda,k,\mu),x)\,,
\eeq
where $\lambda$ and $k$ are positive parameters.   $\GPD(x,\lambda,k,\mu)$ is thus a three-parameter family of distributions that can be used to model the tail $x > \mu$ of an unknown distribution $\mathcal P(x)$, and is given by
$ \frac{1}{\lambda} \left( 1+k\frac{x-\mu}{\lambda}  \right)^{-1-1/k}$
for $(\mu < x\,\, {\rm if}\,\, k>0)\, {\rm or}\, (\mu<x<\mu-\lambda/k \, {\rm if}\,  k<0)$, or by
$\frac{1}{\lambda} e^{ -\frac{x-\mu}{\lambda}}$
for $(\mu< x,k = 0)$.
The distribution $\GPD$ takes three basic forms according to the sign of the parameter $k$. Exponentially decaying tails are described by an element of the family $\GPD(x,\lambda,k,\mu)$ with $k=0$. The parameter $k$ is positive for polynomial tails, while it is negative for finite tails. The theorem of extreme values can be seen as the equivalent of the central limit theorem for the mean of random samples drawn from an unknown distribution.

\section{DW2X quantum annealer}
\label{sec:DW}

We used the D-Wave 2X ``Washington" chip installed at the University of Southern California's Information Sciences Institute. Some qubits are deactivated for technical reasons, leaving a total of $1098$ working qubits (out of $1152$) and $3049$ tunable couplers $J_{ij}$ (out of $3360$). In using different sublattices of the Chimera graph defined on $L\times L$ unit cells, with $L = 4,6,8,10,12$, as stated in the main text, we always started from the bottom right of Fig.~\ref{fig:dw}. The D-Wave devices have been described in great detail before and we refer to the reader to Refs.~\cite{Johnson:2010ys,Harris:2010kx,Berkley:2010zr,Bunyk:2014hb,speedup,q-sig2} for more information.

\bibliography{refs}

\begin{thebibliography}{45}%
\makeatletter
\providecommand \@ifxundefined [1]{%
 \@ifx{#1\undefined}
}%
\providecommand \@ifnum [1]{%
 \ifnum #1\expandafter \@firstoftwo
 \else \expandafter \@secondoftwo
 \fi
}%
\providecommand \@ifx [1]{%
 \ifx #1\expandafter \@firstoftwo
 \else \expandafter \@secondoftwo
 \fi
}%
\providecommand \natexlab [1]{#1}%
\providecommand \enquote  [1]{``#1''}%
\providecommand \bibnamefont  [1]{#1}%
\providecommand \bibfnamefont [1]{#1}%
\providecommand \citenamefont [1]{#1}%
\providecommand \href@noop [0]{\@secondoftwo}%
\providecommand \href [0]{\begingroup \@sanitize@url \@href}%
\providecommand \@href[1]{\@@startlink{#1}\@@href}%
\providecommand \@@href[1]{\endgroup#1\@@endlink}%
\providecommand \@sanitize@url [0]{\catcode `\\12\catcode `\$12\catcode
  `\&12\catcode `\#12\catcode `\^12\catcode `\_12\catcode `\%12\relax}%
\providecommand \@@startlink[1]{}%
\providecommand \@@endlink[0]{}%
\providecommand \url  [0]{\begingroup\@sanitize@url \@url }%
\providecommand \@url [1]{\endgroup\@href {#1}{\urlprefix }}%
\providecommand \urlprefix  [0]{URL }%
\providecommand \Eprint [0]{\href }%
\providecommand \doibase [0]{http://dx.doi.org/}%
\providecommand \selectlanguage [0]{\@gobble}%
\providecommand \bibinfo  [0]{\@secondoftwo}%
\providecommand \bibfield  [0]{\@secondoftwo}%
\providecommand \translation [1]{[#1]}%
\providecommand \BibitemOpen [0]{}%
\providecommand \bibitemStop [0]{}%
\providecommand \bibitemNoStop [0]{.\EOS\space}%
\providecommand \EOS [0]{\spacefactor3000\relax}%
\providecommand \BibitemShut  [1]{\csname bibitem#1\endcsname}%
\let\auto@bib@innerbib\@empty
\bibitem [{\citenamefont {{Catherine C. McGeoch}}(2012)}]{McGeoch:book}%
  \BibitemOpen
  \bibfield  {author} {\bibinfo {author} {\bibnamefont {{Catherine C.
  McGeoch}}},\ }\href
  {http://www.cambridge.org/us/academic/subjects/computer-science/algorithmics-complexity-computer-algebra-and-computational-g/guide-experimental-algorithmics?format=PB&isbn=9780521173018#contentsTabAnchor}
  {\emph {\bibinfo {title} {{A Guide to Experimental Algorithmics}}}}\
  (\bibinfo  {publisher} {{Cambridge University Press}},\ \bibinfo {address}
  {{Cambride, UK}},\ \bibinfo {year} {2012})\BibitemShut {NoStop}%
\bibitem [{\citenamefont {Barr}\ \emph {et~al.}(1995)\citenamefont {Barr},
  \citenamefont {Golden}, \citenamefont {Kelly}, \citenamefont {Resende},\ and\
  \citenamefont {Stewart}}]{Barr1995}%
  \BibitemOpen
  \bibfield  {author} {\bibinfo {author} {\bibfnamefont {Richard~S.}\
  \bibnamefont {Barr}}, \bibinfo {author} {\bibfnamefont {Bruce~L.}\
  \bibnamefont {Golden}}, \bibinfo {author} {\bibfnamefont {James~P.}\
  \bibnamefont {Kelly}}, \bibinfo {author} {\bibfnamefont {Mauricio G.~C.}\
  \bibnamefont {Resende}}, \ and\ \bibinfo {author} {\bibfnamefont
  {William~R.}\ \bibnamefont {Stewart}},\ }\bibfield  {title} {\enquote
  {\bibinfo {title} {Designing and reporting on computational experiments with
  heuristic methods},}\ }\href {\doibase 10.1007/BF02430363} {\bibfield
  {journal} {\bibinfo  {journal} {Journal of Heuristics}\ }\textbf {\bibinfo
  {volume} {1}},\ \bibinfo {pages} {9--32} (\bibinfo {year}
  {1995})}\BibitemShut {NoStop}%
\bibitem [{\citenamefont {Johnson}(2002)}]{johnson2002theoretician}%
  \BibitemOpen
  \bibfield  {author} {\bibinfo {author} {\bibfnamefont {David~S}\ \bibnamefont
  {Johnson}},\ }\enquote {\bibinfo {title} {{Data Structures, Near Neighbor
  Searches, and Methodology: Fifth and Sixth DIMACS Implementation
  Challenges}},}\ \ (\bibinfo  {publisher} {{AMS}},\ \bibinfo {year} {2002})\
  pp.\ \bibinfo {pages} {215--250}\BibitemShut {NoStop}%
\bibitem [{\citenamefont {Bartz-Beielstein}\ \emph {et~al.}(2010)\citenamefont
  {Bartz-Beielstein}, \citenamefont {Chiarandini}, \citenamefont {Paquete},\
  and\ \citenamefont {Preuss}}]{bartz2010experimental}%
  \BibitemOpen
  \bibfield  {author} {\bibinfo {author} {\bibfnamefont {Thomas}\ \bibnamefont
  {Bartz-Beielstein}}, \bibinfo {author} {\bibfnamefont {Marco}\ \bibnamefont
  {Chiarandini}}, \bibinfo {author} {\bibfnamefont {Luis}\ \bibnamefont
  {Paquete}}, \ and\ \bibinfo {author} {\bibfnamefont {Mike}\ \bibnamefont
  {Preuss}},\ }\href {http://link.springer.com/book/10.1007/978-3-642-02538-9}
  {\emph {\bibinfo {title} {Experimental methods for the analysis of
  optimization algorithms}}}\ (\bibinfo  {publisher} {Springer},\ \bibinfo
  {year} {2010})\BibitemShut {NoStop}%
\bibitem [{\citenamefont {Hooker}(1995)}]{hooker1995testing}%
  \BibitemOpen
  \bibfield  {author} {\bibinfo {author} {\bibfnamefont {J.~N.}\ \bibnamefont
  {Hooker}},\ }\bibfield  {title} {\enquote {\bibinfo {title} {Testing
  heuristics: We have it all wrong},}\ }\href {\doibase 10.1007/BF02430364}
  {\bibfield  {journal} {\bibinfo  {journal} {Journal of Heuristics}\ }\textbf
  {\bibinfo {volume} {1}},\ \bibinfo {pages} {33--42} (\bibinfo {year}
  {1995})}\BibitemShut {NoStop}%
\bibitem [{\citenamefont {Jain}(1991)}]{bukh1992art}%
  \BibitemOpen
  \bibfield  {author} {\bibinfo {author} {\bibfnamefont {Raj}\ \bibnamefont
  {Jain}},\ }\href
  {http://www.eletrica.ufpr.br/pedroso/2014/TE816/Art_Of_Computer_Systems_Performance_Analysis_Techniques_For_Experimental_Measurements_Simulation_And_Modeling-Raj_Jain.pdf}
  {\emph {\bibinfo {title} {The art of computer systems performance analysis,
  techniques for experimental design, measurement, simulation and modeling}}}\
  (\bibinfo  {publisher} {{Wiley Computer Publishing}},\ \bibinfo {year}
  {1991})\BibitemShut {NoStop}%
\bibitem [{\citenamefont {Birattari}\ and\ \citenamefont
  {Dorigo}(2007)}]{birattari2007assess}%
  \BibitemOpen
  \bibfield  {author} {\bibinfo {author} {\bibfnamefont {Mauro}\ \bibnamefont
  {Birattari}}\ and\ \bibinfo {author} {\bibfnamefont {Marco}\ \bibnamefont
  {Dorigo}},\ }\bibfield  {title} {\enquote {\bibinfo {title} {How to assess
  and report the performance of a stochastic algorithm on a benchmark problem:
  mean or best result on a number of runs?}}\ }\href {\doibase
  10.1007/s11590-006-0011-8} {\bibfield  {journal} {\bibinfo  {journal}
  {Optimization Letters}\ }\textbf {\bibinfo {volume} {1}},\ \bibinfo {pages}
  {309--311} (\bibinfo {year} {2007})}\BibitemShut {NoStop}%
\bibitem [{\citenamefont {Brownlee}(2007)}]{brownlee2007note}%
  \BibitemOpen
  \bibfield  {author} {\bibinfo {author} {\bibfnamefont {Jason}\ \bibnamefont
  {Brownlee}},\ }\href
  {http://citeseerx.ist.psu.edu/viewdoc/summary?doi=10.1.1.73.6198} {\emph
  {\bibinfo {title} {A note on research methodology and benchmarking
  optimization algorithms}}},\ \bibinfo {type} {Tech. Rep.}\ (\bibinfo
  {institution} {Complex Intelligent Systems Laboratory (CIS), Centre for
  Information Technology Research (CITR), Faculty of Information and
  Communication Technologies (ICT), Swinburne University of Technology,
  Victoria, Australia, Technical Report ID},\ \bibinfo {year}
  {2007})\BibitemShut {NoStop}%
\bibitem [{\citenamefont {Somma}\ \emph {et~al.}(2012)\citenamefont {Somma},
  \citenamefont {Nagaj},\ and\ \citenamefont {Kieferov{\'a}}}]{Somma:2012kx}%
  \BibitemOpen
  \bibfield  {author} {\bibinfo {author} {\bibfnamefont {Rolando~D.}\
  \bibnamefont {Somma}}, \bibinfo {author} {\bibfnamefont {Daniel}\
  \bibnamefont {Nagaj}}, \ and\ \bibinfo {author} {\bibfnamefont {M{\'a}ria}\
  \bibnamefont {Kieferov{\'a}}},\ }\bibfield  {title} {\enquote {\bibinfo
  {title} {Quantum speedup by quantum annealing},}\ }\href
  {http://link.aps.org/doi/10.1103/PhysRevLett.109.050501} {\bibfield
  {journal} {\bibinfo  {journal} {Phys. Rev. Lett.}\ }\textbf {\bibinfo
  {volume} {109}},\ \bibinfo {pages} {050501--} (\bibinfo {year}
  {2012})}\BibitemShut {NoStop}%
\bibitem [{\citenamefont {Crosson}\ \emph {et~al.}(2014)\citenamefont
  {Crosson}, \citenamefont {Farhi}, \citenamefont {Lin}, \citenamefont {Lin},\
  and\ \citenamefont {Shor}}]{crosson2014different}%
  \BibitemOpen
  \bibfield  {author} {\bibinfo {author} {\bibfnamefont {Elizabeth}\
  \bibnamefont {Crosson}}, \bibinfo {author} {\bibfnamefont {Edward}\
  \bibnamefont {Farhi}}, \bibinfo {author} {\bibfnamefont {Cedric Yen-Yu}\
  \bibnamefont {Lin}}, \bibinfo {author} {\bibfnamefont {Han-Hsuan}\
  \bibnamefont {Lin}}, \ and\ \bibinfo {author} {\bibfnamefont {Peter}\
  \bibnamefont {Shor}},\ }\bibfield  {title} {\enquote {\bibinfo {title}
  {Different strategies for optimization using the quantum adiabatic
  algorithm},}\ }\href {http://arxiv.org/abs/1401.7320} {\bibfield  {journal}
  {\bibinfo  {journal} {arXiv preprint arXiv:1401.7320}\ } (\bibinfo {year}
  {2014})}\BibitemShut {NoStop}%
\bibitem [{\citenamefont {Muthukrishnan}\ \emph {et~al.}(2016)\citenamefont
  {Muthukrishnan}, \citenamefont {Albash},\ and\ \citenamefont
  {Lidar}}]{MAL:2016}%
  \BibitemOpen
  \bibfield  {author} {\bibinfo {author} {\bibfnamefont {Siddharth}\
  \bibnamefont {Muthukrishnan}}, \bibinfo {author} {\bibfnamefont {Tameem}\
  \bibnamefont {Albash}}, \ and\ \bibinfo {author} {\bibfnamefont {Daniel~A.}\
  \bibnamefont {Lidar}},\ }\bibfield  {title} {\enquote {\bibinfo {title}
  {Tunneling and speedup in quantum optimization for permutation-symmetric
  problems},}\ }\href {http://arXiv.org/abs/1511.03910} {\bibfield  {journal}
  {\bibinfo  {journal} {{Phys. Rev. X}}\ }\textbf {\bibinfo {volume} {?}},\
  \bibinfo {pages} {?} (\bibinfo {year} {2016})}\BibitemShut {NoStop}%
\bibitem [{\citenamefont {Wald}(1945)}]{Wald:1945bf}%
  \BibitemOpen
  \bibfield  {author} {\bibinfo {author} {\bibfnamefont {A.}~\bibnamefont
  {Wald}},\ }\bibfield  {title} {\enquote {\bibinfo {title} {Sequential tests
  of statistical hypotheses},}\ }\href {\doibase 10.1214/aoms/1177731118}
  {\bibfield  {journal} {\bibinfo  {journal} {Ann. Math. Statist.}\ }\textbf
  {\bibinfo {volume} {16}},\ \bibinfo {pages} {117--186} (\bibinfo {year}
  {1945})}\BibitemShut {NoStop}%
\bibitem [{\citenamefont {Wald}(1973)}]{wald1973sequential}%
  \BibitemOpen
  \bibfield  {author} {\bibinfo {author} {\bibfnamefont {Abraham}\ \bibnamefont
  {Wald}},\ }\href@noop {} {\emph {\bibinfo {title} {Sequential analysis}}}\
  (\bibinfo  {publisher} {Courier Corporation},\ \bibinfo {year}
  {1973})\BibitemShut {NoStop}%
\bibitem [{\citenamefont {Chow}\ \emph {et~al.}(1971)\citenamefont {Chow},
  \citenamefont {Robbins},\ and\ \citenamefont {Siegmund}}]{chow1971great}%
  \BibitemOpen
  \bibfield  {author} {\bibinfo {author} {\bibfnamefont {Yuan~Shih}\
  \bibnamefont {Chow}}, \bibinfo {author} {\bibfnamefont {Herbert}\
  \bibnamefont {Robbins}}, \ and\ \bibinfo {author} {\bibfnamefont {David}\
  \bibnamefont {Siegmund}},\ }\href@noop {} {\emph {\bibinfo {title} {Great
  expectations: The theory of optimal stopping}}}\ (\bibinfo  {publisher}
  {Houghton Mifflin},\ \bibinfo {year} {1971})\BibitemShut {NoStop}%
\bibitem [{\citenamefont {Gottinger}(1976)}]{GOTTINGER:1976xe}%
  \BibitemOpen
  \bibfield  {author} {\bibinfo {author} {\bibfnamefont {Hans~W.}\ \bibnamefont
  {Gottinger}},\ }\bibfield  {title} {\enquote {\bibinfo {title} {Sequential
  analysis and optimal stopping},}\ }\href
  {http://www.jstor.org/stable/40749860} {\bibfield  {journal} {\bibinfo
  {journal} {Journal of Institutional and Theoretical Economics}\ }\textbf
  {\bibinfo {volume} {132}},\ \bibinfo {pages} {41--62} (\bibinfo {year}
  {1976})}\BibitemShut {NoStop}%
\bibitem [{\citenamefont {Ferguson}(2008)}]{Ferguson:book}%
  \BibitemOpen
  \bibfield  {author} {\bibinfo {author} {\bibfnamefont {Thomas~S.}\
  \bibnamefont {Ferguson}},\ }\href
  {http://www.e-booksdirectory.com/details.php?ebook=5651} {\enquote {\bibinfo
  {title} {{Optimal Stopping and Applications}},}\ } (\bibinfo {year}
  {2008})\BibitemShut {NoStop}%
\bibitem [{\citenamefont {MacQueen}\ and\ \citenamefont
  {Miller~Jr}(1960)}]{macqueen1960optimal}%
  \BibitemOpen
  \bibfield  {author} {\bibinfo {author} {\bibfnamefont {James}\ \bibnamefont
  {MacQueen}}\ and\ \bibinfo {author} {\bibfnamefont {RG}~\bibnamefont
  {Miller~Jr}},\ }\bibfield  {title} {\enquote {\bibinfo {title} {Optimal
  persistence policies},}\ }\href@noop {} {\bibfield  {journal} {\bibinfo
  {journal} {Operations Research}\ }\textbf {\bibinfo {volume} {8}},\ \bibinfo
  {pages} {362--380} (\bibinfo {year} {1960})}\BibitemShut {NoStop}%
\bibitem [{\citenamefont {Stigler}(1961)}]{stigler1961economics}%
  \BibitemOpen
  \bibfield  {author} {\bibinfo {author} {\bibfnamefont {George~J}\
  \bibnamefont {Stigler}},\ }\bibfield  {title} {\enquote {\bibinfo {title}
  {The economics of information},}\ }\href@noop {} {\bibfield  {journal}
  {\bibinfo  {journal} {The journal of political economy}\ ,\ \bibinfo {pages}
  {213--225}} (\bibinfo {year} {1961})}\BibitemShut {NoStop}%
\bibitem [{\citenamefont {Hoos}\ and\ \citenamefont
  {St\"{u}tzle}(1998)}]{Hoos:98}%
  \BibitemOpen
  \bibfield  {author} {\bibinfo {author} {\bibfnamefont {Holger~H.}\
  \bibnamefont {Hoos}}\ and\ \bibinfo {author} {\bibfnamefont {Thomas}\
  \bibnamefont {St\"{u}tzle}},\ }\bibfield  {title} {\enquote {\bibinfo {title}
  {{Evaluating Las Vegas Algorithms: Pitfalls and Remedies}},}\ }in\ \href@noop
  {} {\emph {\bibinfo {booktitle} {{Proceedings of the 14th conference on
  uncertainty in artificial intelligence}}}}\ (\bibinfo {year}
  {1998})\BibitemShut {NoStop}%
\bibitem [{\citenamefont {Aiex}\ \emph {et~al.}(2007)\citenamefont {Aiex},
  \citenamefont {Resende},\ and\ \citenamefont {Ribeiro}}]{Aiex:2007ad}%
  \BibitemOpen
  \bibfield  {author} {\bibinfo {author} {\bibfnamefont {Renata~M.}\
  \bibnamefont {Aiex}}, \bibinfo {author} {\bibfnamefont {Mauricio G.~C.}\
  \bibnamefont {Resende}}, \ and\ \bibinfo {author} {\bibfnamefont {Celso~C.}\
  \bibnamefont {Ribeiro}},\ }\bibfield  {title} {\enquote {\bibinfo {title}
  {Ttt plots: a perl program to create time-to-target plots},}\ }\href
  {\doibase 10.1007/s11590-006-0031-4} {\bibfield  {journal} {\bibinfo
  {journal} {Optimization Letters}\ }\textbf {\bibinfo {volume} {1}},\ \bibinfo
  {pages} {355--366} (\bibinfo {year} {2007})}\BibitemShut {NoStop}%
\bibitem [{\citenamefont {King}\ \emph
  {et~al.}(2015{\natexlab{a}})\citenamefont {King}, \citenamefont {Yarkoni},
  \citenamefont {Nevisi}, \citenamefont {Hilton},\ and\ \citenamefont
  {McGeoch}}]{King:2015cs}%
  \BibitemOpen
  \bibfield  {author} {\bibinfo {author} {\bibfnamefont {James}\ \bibnamefont
  {King}}, \bibinfo {author} {\bibfnamefont {Sheir}\ \bibnamefont {Yarkoni}},
  \bibinfo {author} {\bibfnamefont {Mayssam~M.}\ \bibnamefont {Nevisi}},
  \bibinfo {author} {\bibfnamefont {Jeremy~P.}\ \bibnamefont {Hilton}}, \ and\
  \bibinfo {author} {\bibfnamefont {Catherine~C.}\ \bibnamefont {McGeoch}},\
  }\bibfield  {title} {\enquote {\bibinfo {title} {Benchmarking a quantum
  annealing processor with the time-to-target metric},}\ }\href
  {http://arXiv.org/abs/1508.05087} {\bibfield  {journal} {\bibinfo  {journal}
  {arXiv:1508.05087}\ } (\bibinfo {year} {2015}{\natexlab{a}})}\BibitemShut
  {NoStop}%
\bibitem [{\citenamefont {R{\o}nnow}\ \emph {et~al.}(2014)\citenamefont
  {R{\o}nnow}, \citenamefont {Wang}, \citenamefont {Job}, \citenamefont
  {Boixo}, \citenamefont {Isakov}, \citenamefont {Wecker}, \citenamefont
  {Martinis}, \citenamefont {Lidar},\ and\ \citenamefont {Troyer}}]{speedup}%
  \BibitemOpen
  \bibfield  {author} {\bibinfo {author} {\bibfnamefont {Troels~F.}\
  \bibnamefont {R{\o}nnow}}, \bibinfo {author} {\bibfnamefont {Zhihui}\
  \bibnamefont {Wang}}, \bibinfo {author} {\bibfnamefont {Joshua}\ \bibnamefont
  {Job}}, \bibinfo {author} {\bibfnamefont {Sergio}\ \bibnamefont {Boixo}},
  \bibinfo {author} {\bibfnamefont {Sergei~V.}\ \bibnamefont {Isakov}},
  \bibinfo {author} {\bibfnamefont {David}\ \bibnamefont {Wecker}}, \bibinfo
  {author} {\bibfnamefont {John~M.}\ \bibnamefont {Martinis}}, \bibinfo
  {author} {\bibfnamefont {Daniel~A.}\ \bibnamefont {Lidar}}, \ and\ \bibinfo
  {author} {\bibfnamefont {Matthias}\ \bibnamefont {Troyer}},\ }\bibfield
  {title} {\enquote {\bibinfo {title} {Defining and detecting quantum
  speedup},}\ }\href {\doibase 10.1126/science.1252319} {\bibfield  {journal}
  {\bibinfo  {journal} {Science}\ }\textbf {\bibinfo {volume} {345}},\ \bibinfo
  {pages} {420--424} (\bibinfo {year} {2014})}\BibitemShut {NoStop}%
\bibitem [{\citenamefont {Johnson}\ \emph {et~al.}(2010)\citenamefont
  {Johnson}, \citenamefont {Bunyk}, \citenamefont {Maibaum}, \citenamefont
  {Tolkacheva}, \citenamefont {Berkley}, \citenamefont {Chapple}, \citenamefont
  {Harris}, \citenamefont {Johansson}, \citenamefont {Lanting}, \citenamefont
  {Perminov}, \citenamefont {Ladizinsky}, \citenamefont {Oh},\ and\
  \citenamefont {Rose}}]{Johnson:2010ys}%
  \BibitemOpen
  \bibfield  {author} {\bibinfo {author} {\bibfnamefont {M~W}\ \bibnamefont
  {Johnson}}, \bibinfo {author} {\bibfnamefont {P}~\bibnamefont {Bunyk}},
  \bibinfo {author} {\bibfnamefont {F}~\bibnamefont {Maibaum}}, \bibinfo
  {author} {\bibfnamefont {E}~\bibnamefont {Tolkacheva}}, \bibinfo {author}
  {\bibfnamefont {A~J}\ \bibnamefont {Berkley}}, \bibinfo {author}
  {\bibfnamefont {E~M}\ \bibnamefont {Chapple}}, \bibinfo {author}
  {\bibfnamefont {R}~\bibnamefont {Harris}}, \bibinfo {author} {\bibfnamefont
  {J}~\bibnamefont {Johansson}}, \bibinfo {author} {\bibfnamefont
  {T}~\bibnamefont {Lanting}}, \bibinfo {author} {\bibfnamefont
  {I}~\bibnamefont {Perminov}}, \bibinfo {author} {\bibfnamefont
  {E}~\bibnamefont {Ladizinsky}}, \bibinfo {author} {\bibfnamefont
  {T}~\bibnamefont {Oh}}, \ and\ \bibinfo {author} {\bibfnamefont
  {G}~\bibnamefont {Rose}},\ }\bibfield  {title} {\enquote {\bibinfo {title} {A
  scalable control system for a superconducting adiabatic quantum optimization
  processor},}\ }\href {http://stacks.iop.org/0953-2048/23/i=6/a=065004}
  {\bibfield  {journal} {\bibinfo  {journal} {Superconductor Science and
  Technology}\ }\textbf {\bibinfo {volume} {23}},\ \bibinfo {pages} {065004}
  (\bibinfo {year} {2010})}\BibitemShut {NoStop}%
\bibitem [{\citenamefont {Berkley}\ \emph {et~al.}(2010)\citenamefont
  {Berkley}, \citenamefont {Johnson}, \citenamefont {Bunyk}, \citenamefont
  {Harris}, \citenamefont {Johansson}, \citenamefont {Lanting}, \citenamefont
  {Ladizinsky}, \citenamefont {Tolkacheva}, \citenamefont {Amin},\ and\
  \citenamefont {Rose}}]{Berkley:2010zr}%
  \BibitemOpen
  \bibfield  {author} {\bibinfo {author} {\bibfnamefont {A~J}\ \bibnamefont
  {Berkley}}, \bibinfo {author} {\bibfnamefont {M~W}\ \bibnamefont {Johnson}},
  \bibinfo {author} {\bibfnamefont {P}~\bibnamefont {Bunyk}}, \bibinfo {author}
  {\bibfnamefont {R}~\bibnamefont {Harris}}, \bibinfo {author} {\bibfnamefont
  {J}~\bibnamefont {Johansson}}, \bibinfo {author} {\bibfnamefont
  {T}~\bibnamefont {Lanting}}, \bibinfo {author} {\bibfnamefont
  {E}~\bibnamefont {Ladizinsky}}, \bibinfo {author} {\bibfnamefont
  {E}~\bibnamefont {Tolkacheva}}, \bibinfo {author} {\bibfnamefont {M~H~S}\
  \bibnamefont {Amin}}, \ and\ \bibinfo {author} {\bibfnamefont
  {G}~\bibnamefont {Rose}},\ }\bibfield  {title} {\enquote {\bibinfo {title} {A
  scalable readout system for a superconducting adiabatic quantum optimization
  system},}\ }\href {http://stacks.iop.org/0953-2048/23/i=10/a=105014}
  {\bibfield  {journal} {\bibinfo  {journal} {Superconductor Science and
  Technology}\ }\textbf {\bibinfo {volume} {23}},\ \bibinfo {pages} {105014}
  (\bibinfo {year} {2010})}\BibitemShut {NoStop}%
\bibitem [{\citenamefont {Harris}\ \emph {et~al.}(2010)\citenamefont {Harris},
  \citenamefont {Johnson}, \citenamefont {Lanting}, \citenamefont {Berkley},
  \citenamefont {Johansson}, \citenamefont {Bunyk}, \citenamefont {Tolkacheva},
  \citenamefont {Ladizinsky}, \citenamefont {Ladizinsky}, \citenamefont {Oh},
  \citenamefont {Cioata}, \citenamefont {Perminov}, \citenamefont {Spear},
  \citenamefont {Enderud}, \citenamefont {Rich}, \citenamefont {Uchaikin},
  \citenamefont {Thom}, \citenamefont {Chapple}, \citenamefont {Wang},
  \citenamefont {Wilson}, \citenamefont {Amin}, \citenamefont {Dickson},
  \citenamefont {Karimi}, \citenamefont {Macready}, \citenamefont {Truncik},\
  and\ \citenamefont {Rose}}]{Harris:2010kx}%
  \BibitemOpen
  \bibfield  {author} {\bibinfo {author} {\bibfnamefont {R.}~\bibnamefont
  {Harris}}, \bibinfo {author} {\bibfnamefont {M.~W.}\ \bibnamefont {Johnson}},
  \bibinfo {author} {\bibfnamefont {T.}~\bibnamefont {Lanting}}, \bibinfo
  {author} {\bibfnamefont {A.~J.}\ \bibnamefont {Berkley}}, \bibinfo {author}
  {\bibfnamefont {J.}~\bibnamefont {Johansson}}, \bibinfo {author}
  {\bibfnamefont {P.}~\bibnamefont {Bunyk}}, \bibinfo {author} {\bibfnamefont
  {E.}~\bibnamefont {Tolkacheva}}, \bibinfo {author} {\bibfnamefont
  {E.}~\bibnamefont {Ladizinsky}}, \bibinfo {author} {\bibfnamefont
  {N.}~\bibnamefont {Ladizinsky}}, \bibinfo {author} {\bibfnamefont
  {T.}~\bibnamefont {Oh}}, \bibinfo {author} {\bibfnamefont {F.}~\bibnamefont
  {Cioata}}, \bibinfo {author} {\bibfnamefont {I.}~\bibnamefont {Perminov}},
  \bibinfo {author} {\bibfnamefont {P.}~\bibnamefont {Spear}}, \bibinfo
  {author} {\bibfnamefont {C.}~\bibnamefont {Enderud}}, \bibinfo {author}
  {\bibfnamefont {C.}~\bibnamefont {Rich}}, \bibinfo {author} {\bibfnamefont
  {S.}~\bibnamefont {Uchaikin}}, \bibinfo {author} {\bibfnamefont {M.~C.}\
  \bibnamefont {Thom}}, \bibinfo {author} {\bibfnamefont {E.~M.}\ \bibnamefont
  {Chapple}}, \bibinfo {author} {\bibfnamefont {J.}~\bibnamefont {Wang}},
  \bibinfo {author} {\bibfnamefont {B.}~\bibnamefont {Wilson}}, \bibinfo
  {author} {\bibfnamefont {M.~H.~S.}\ \bibnamefont {Amin}}, \bibinfo {author}
  {\bibfnamefont {N.}~\bibnamefont {Dickson}}, \bibinfo {author} {\bibfnamefont
  {K.}~\bibnamefont {Karimi}}, \bibinfo {author} {\bibfnamefont
  {B.}~\bibnamefont {Macready}}, \bibinfo {author} {\bibfnamefont {C.~J.~S.}\
  \bibnamefont {Truncik}}, \ and\ \bibinfo {author} {\bibfnamefont
  {G.}~\bibnamefont {Rose}},\ }\bibfield  {title} {\enquote {\bibinfo {title}
  {Experimental investigation of an eight-qubit unit cell in a superconducting
  optimization processor},}\ }\href {\doibase 10.1103/PhysRevB.82.024511}
  {\bibfield  {journal} {\bibinfo  {journal} {Phys. Rev. B}\ }\textbf {\bibinfo
  {volume} {82}},\ \bibinfo {pages} {024511} (\bibinfo {year}
  {2010})}\BibitemShut {NoStop}%
\bibitem [{\citenamefont {Bunyk}\ \emph {et~al.}(Aug. 2014)\citenamefont
  {Bunyk}, \citenamefont {Hoskinson}, \citenamefont {Johnson}, \citenamefont
  {Tolkacheva}, \citenamefont {Altomare}, \citenamefont {Berkley},
  \citenamefont {Harris}, \citenamefont {Hilton}, \citenamefont {Lanting},
  \citenamefont {Przybysz},\ and\ \citenamefont {Whittaker}}]{Bunyk:2014hb}%
  \BibitemOpen
  \bibfield  {author} {\bibinfo {author} {\bibfnamefont {P.~I}\ \bibnamefont
  {Bunyk}}, \bibinfo {author} {\bibfnamefont {E.~M.}\ \bibnamefont
  {Hoskinson}}, \bibinfo {author} {\bibfnamefont {M.~W.}\ \bibnamefont
  {Johnson}}, \bibinfo {author} {\bibfnamefont {E.}~\bibnamefont {Tolkacheva}},
  \bibinfo {author} {\bibfnamefont {F.}~\bibnamefont {Altomare}}, \bibinfo
  {author} {\bibfnamefont {AJ.}\ \bibnamefont {Berkley}}, \bibinfo {author}
  {\bibfnamefont {R.}~\bibnamefont {Harris}}, \bibinfo {author} {\bibfnamefont
  {J.~P.}\ \bibnamefont {Hilton}}, \bibinfo {author} {\bibfnamefont
  {T.}~\bibnamefont {Lanting}}, \bibinfo {author} {\bibfnamefont {AJ.}\
  \bibnamefont {Przybysz}}, \ and\ \bibinfo {author} {\bibfnamefont
  {J.}~\bibnamefont {Whittaker}},\ }\bibfield  {title} {\enquote {\bibinfo
  {title} {Architectural considerations in the design of a superconducting
  quantum annealing processor},}\ }\href {\doibase 10.1109/TASC.2014.2318294}
  {\bibfield  {journal} {\bibinfo  {journal} {IEEE Transactions on Applied
  Superconductivity}\ }\textbf {\bibinfo {volume} {24}},\ \bibinfo {pages}
  {1--10} (\bibinfo {year} {Aug. 2014})}\BibitemShut {NoStop}%
\bibitem [{\citenamefont {Hen}\ \emph {et~al.}(2015)\citenamefont {Hen},
  \citenamefont {Job}, \citenamefont {Albash}, \citenamefont {R{\o}nnow},
  \citenamefont {Troyer},\ and\ \citenamefont {Lidar}}]{Hen:2015rt}%
  \BibitemOpen
  \bibfield  {author} {\bibinfo {author} {\bibfnamefont {Itay}\ \bibnamefont
  {Hen}}, \bibinfo {author} {\bibfnamefont {Joshua}\ \bibnamefont {Job}},
  \bibinfo {author} {\bibfnamefont {Tameem}\ \bibnamefont {Albash}}, \bibinfo
  {author} {\bibfnamefont {Troels~F.}\ \bibnamefont {R{\o}nnow}}, \bibinfo
  {author} {\bibfnamefont {Matthias}\ \bibnamefont {Troyer}}, \ and\ \bibinfo
  {author} {\bibfnamefont {Daniel~A.}\ \bibnamefont {Lidar}},\ }\bibfield
  {title} {\enquote {\bibinfo {title} {Probing for quantum speedup in
  spin-glass problems with planted solutions},}\ }\href
  {http://link.aps.org/doi/10.1103/PhysRevA.92.042325} {\bibfield  {journal}
  {\bibinfo  {journal} {{Phys. Rev. A}}\ }\textbf {\bibinfo {volume} {92}},\
  \bibinfo {pages} {042325--} (\bibinfo {year} {2015})}\BibitemShut {NoStop}%
\bibitem [{\citenamefont {Denchev}\ \emph {et~al.}(2015)\citenamefont
  {Denchev}, \citenamefont {Boixo}, \citenamefont {Isakov}, \citenamefont
  {Ding}, \citenamefont {Babbush}, \citenamefont {Smelyanskiy}, \citenamefont
  {Martinis},\ and\ \citenamefont {Neven}}]{denchev2015computational}%
  \BibitemOpen
  \bibfield  {author} {\bibinfo {author} {\bibfnamefont {Vasil~S.}\
  \bibnamefont {Denchev}}, \bibinfo {author} {\bibfnamefont {Sergio}\
  \bibnamefont {Boixo}}, \bibinfo {author} {\bibfnamefont {Sergei~V.}\
  \bibnamefont {Isakov}}, \bibinfo {author} {\bibfnamefont {Nan}\ \bibnamefont
  {Ding}}, \bibinfo {author} {\bibfnamefont {Ryan}\ \bibnamefont {Babbush}},
  \bibinfo {author} {\bibfnamefont {Vadim}\ \bibnamefont {Smelyanskiy}},
  \bibinfo {author} {\bibfnamefont {John}\ \bibnamefont {Martinis}}, \ and\
  \bibinfo {author} {\bibfnamefont {Hartmut}\ \bibnamefont {Neven}},\
  }\bibfield  {title} {\enquote {\bibinfo {title} {What is the computational
  value of finite range tunneling?}}\ }\href {http://arXiv.org/abs/1512.02206}
  {\bibfield  {journal} {\bibinfo  {journal} {arXiv:1512.02206}\ } (\bibinfo
  {year} {2015})}\BibitemShut {NoStop}%
\bibitem [{\citenamefont {Steiger}\ \emph {et~al.}(2015)\citenamefont
  {Steiger}, \citenamefont {R{\o}nnow},\ and\ \citenamefont
  {Troyer}}]{Steiger:2015fk}%
  \BibitemOpen
  \bibfield  {author} {\bibinfo {author} {\bibfnamefont {Damian~S.}\
  \bibnamefont {Steiger}}, \bibinfo {author} {\bibfnamefont {Troels~F.}\
  \bibnamefont {R{\o}nnow}}, \ and\ \bibinfo {author} {\bibfnamefont
  {Matthias}\ \bibnamefont {Troyer}},\ }\bibfield  {title} {\enquote {\bibinfo
  {title} {Heavy tails in the distribution of time to solution for classical
  and quantum annealing*},}\ }\href
  {http://link.aps.org/doi/10.1103/PhysRevLett.115.230501} {\bibfield
  {journal} {\bibinfo  {journal} {Physical Review Letters}\ }\textbf {\bibinfo
  {volume} {115}},\ \bibinfo {pages} {230501--} (\bibinfo {year}
  {2015})}\BibitemShut {NoStop}%
\bibitem [{\citenamefont {Kirkpatrick}\ \emph {et~al.}(1983)\citenamefont
  {Kirkpatrick}, \citenamefont {Gelatt},\ and\ \citenamefont
  {Vecchi}}]{kirkpatrick_optimization_1983}%
  \BibitemOpen
  \bibfield  {author} {\bibinfo {author} {\bibfnamefont {S.}~\bibnamefont
  {Kirkpatrick}}, \bibinfo {author} {\bibfnamefont {C.~D.}\ \bibnamefont
  {Gelatt}}, \ and\ \bibinfo {author} {\bibfnamefont {M.~P.}\ \bibnamefont
  {Vecchi}},\ }\bibfield  {title} {\enquote {\bibinfo {title} {Optimization by
  simulated annealing},}\ }\href {\doibase 10.1126/science.220.4598.671}
  {\bibfield  {journal} {\bibinfo  {journal} {Science}\ }\textbf {\bibinfo
  {volume} {220}},\ \bibinfo {pages} {671--680} (\bibinfo {year}
  {1983})}\BibitemShut {NoStop}%
\bibitem [{\citenamefont {Isakov}\ \emph {et~al.}(2015)\citenamefont {Isakov},
  \citenamefont {Zintchenko}, \citenamefont {R{\o}nnow},\ and\ \citenamefont
  {Troyer}}]{Isakov:2015ao}%
  \BibitemOpen
  \bibfield  {author} {\bibinfo {author} {\bibfnamefont {S.~V.}\ \bibnamefont
  {Isakov}}, \bibinfo {author} {\bibfnamefont {I.~N.}\ \bibnamefont
  {Zintchenko}}, \bibinfo {author} {\bibfnamefont {T.~F.}\ \bibnamefont
  {R{\o}nnow}}, \ and\ \bibinfo {author} {\bibfnamefont {M.}~\bibnamefont
  {Troyer}},\ }\bibfield  {title} {\enquote {\bibinfo {title} {Optimised
  simulated annealing for ising spin glasses},}\ }\href {\doibase
  http://dx.doi.org/10.1016/j.cpc.2015.02.015} {\bibfield  {journal} {\bibinfo
  {journal} {Computer Physics Communications}\ }\textbf {\bibinfo {volume}
  {192}},\ \bibinfo {pages} {265--271} (\bibinfo {year} {2015})}\BibitemShut
  {NoStop}%
\bibitem [{\citenamefont {Coles}\ \emph {et~al.}(2001)\citenamefont {Coles},
  \citenamefont {Bawa}, \citenamefont {Trenner},\ and\ \citenamefont
  {Dorazio}}]{coles2001introduction}%
  \BibitemOpen
  \bibfield  {author} {\bibinfo {author} {\bibfnamefont {Stuart}\ \bibnamefont
  {Coles}}, \bibinfo {author} {\bibfnamefont {Joanna}\ \bibnamefont {Bawa}},
  \bibinfo {author} {\bibfnamefont {Lesley}\ \bibnamefont {Trenner}}, \ and\
  \bibinfo {author} {\bibfnamefont {Pat}\ \bibnamefont {Dorazio}},\ }\href@noop
  {} {\emph {\bibinfo {title} {An introduction to statistical modeling of
  extreme values}}},\ Vol.\ \bibinfo {volume} {208}\ (\bibinfo  {publisher}
  {Springer},\ \bibinfo {year} {2001})\BibitemShut {NoStop}%
\bibitem [{\citenamefont {Hamze}\ and\ \citenamefont
  {de~Freitas}(2004)}]{hamze:04}%
  \BibitemOpen
  \bibfield  {author} {\bibinfo {author} {\bibfnamefont {Firas}\ \bibnamefont
  {Hamze}}\ and\ \bibinfo {author} {\bibfnamefont {Nando}\ \bibnamefont
  {de~Freitas}},\ }\bibfield  {title} {\enquote {\bibinfo {title} {From fields
  to trees},}\ }in\ \href {http://dl.acm.org/citation.cfm?id=1036873} {\emph
  {\bibinfo {booktitle} {UAI}}},\ \bibinfo {editor} {edited by\ \bibinfo
  {editor} {\bibfnamefont {David~Maxwell}\ \bibnamefont {Chickering}}\ and\
  \bibinfo {editor} {\bibfnamefont {Joseph~Y.}\ \bibnamefont {Halpern}}}\
  (\bibinfo  {publisher} {AUAI Press},\ \bibinfo {address} {Arlington,
  Virginia},\ \bibinfo {year} {2004})\ pp.\ \bibinfo {pages}
  {243--250}\BibitemShut {NoStop}%
\bibitem [{\citenamefont {Selby}(2014)}]{Selby:2014tx}%
  \BibitemOpen
  \bibfield  {author} {\bibinfo {author} {\bibfnamefont {Alex}\ \bibnamefont
  {Selby}},\ }\bibfield  {title} {\enquote {\bibinfo {title} {Efficient
  subgraph-based sampling of ising-type models with frustration},}\ }\href
  {http://arXiv.org/abs/1409.3934} {\bibfield  {journal} {\bibinfo  {journal}
  {arXiv:1409.3934}\ } (\bibinfo {year} {2014})}\BibitemShut {NoStop}%
\bibitem [{\citenamefont {Kadowaki}\ and\ \citenamefont
  {Nishimori}(1998)}]{kadowaki_quantum_1998}%
  \BibitemOpen
  \bibfield  {author} {\bibinfo {author} {\bibfnamefont {Tadashi}\ \bibnamefont
  {Kadowaki}}\ and\ \bibinfo {author} {\bibfnamefont {Hidetoshi}\ \bibnamefont
  {Nishimori}},\ }\bibfield  {title} {\enquote {\bibinfo {title} {Quantum
  annealing in the transverse \uppercase{I}sing model},}\ }\href {\doibase
  10.1103/PhysRevE.58.5355} {\bibfield  {journal} {\bibinfo  {journal} {Phys.
  Rev. E}\ }\textbf {\bibinfo {volume} {58}},\ \bibinfo {pages} {5355}
  (\bibinfo {year} {1998})}\BibitemShut {NoStop}%
\bibitem [{\citenamefont {Brooke}\ \emph {et~al.}(1999)\citenamefont {Brooke},
  \citenamefont {Bitko}, \citenamefont {F.}, \citenamefont {Rosenbaum},\ and\
  \citenamefont {Aeppli}}]{Brooke1999}%
  \BibitemOpen
  \bibfield  {author} {\bibinfo {author} {\bibfnamefont {J.}~\bibnamefont
  {Brooke}}, \bibinfo {author} {\bibfnamefont {D.}~\bibnamefont {Bitko}},
  \bibinfo {author} {\bibfnamefont {T.}~\bibnamefont {F.}}, \bibinfo {author}
  {\bibnamefont {Rosenbaum}}, \ and\ \bibinfo {author} {\bibfnamefont
  {G.}~\bibnamefont {Aeppli}},\ }\bibfield  {title} {\enquote {\bibinfo {title}
  {Quantum annealing of a disordered magnet},}\ }\href {\doibase
  10.1126/science.284.5415.779} {\bibfield  {journal} {\bibinfo  {journal}
  {Science}\ }\textbf {\bibinfo {volume} {284}},\ \bibinfo {pages} {779--781}
  (\bibinfo {year} {1999})}\BibitemShut {NoStop}%
\bibitem [{\citenamefont {Das}\ and\ \citenamefont
  {Chakrabarti}(2008)}]{RevModPhys.80.1061}%
  \BibitemOpen
  \bibfield  {author} {\bibinfo {author} {\bibfnamefont {Arnab}\ \bibnamefont
  {Das}}\ and\ \bibinfo {author} {\bibfnamefont {Bikas~K.}\ \bibnamefont
  {Chakrabarti}},\ }\bibfield  {title} {\enquote {\bibinfo {title}
  {\textit{Colloquium}: Quantum annealing and analog quantum computation},}\
  }\href {\doibase 10.1103/RevModPhys.80.1061} {\bibfield  {journal} {\bibinfo
  {journal} {Rev. Mod. Phys.}\ }\textbf {\bibinfo {volume} {80}},\ \bibinfo
  {pages} {1061--1081} (\bibinfo {year} {2008})}\BibitemShut {NoStop}%
\bibitem [{\citenamefont {Johnson}\ \emph {et~al.}(2011)\citenamefont
  {Johnson}, \citenamefont {Amin}, \citenamefont {Gildert}, \citenamefont
  {Lanting}, \citenamefont {Hamze}, \citenamefont {Dickson}, \citenamefont
  {Harris}, \citenamefont {Berkley}, \citenamefont {Johansson}, \citenamefont
  {Bunyk}, \citenamefont {Chapple}, \citenamefont {Enderud}, \citenamefont
  {Hilton}, \citenamefont {Karimi}, \citenamefont {Ladizinsky}, \citenamefont
  {Ladizinsky}, \citenamefont {Oh}, \citenamefont {Perminov}, \citenamefont
  {Rich}, \citenamefont {Thom}, \citenamefont {Tolkacheva}, \citenamefont
  {Truncik}, \citenamefont {Uchaikin}, \citenamefont {Wang}, \citenamefont
  {Wilson},\ and\ \citenamefont {Rose}}]{Dwave}%
  \BibitemOpen
  \bibfield  {author} {\bibinfo {author} {\bibfnamefont {M.~W.}\ \bibnamefont
  {Johnson}}, \bibinfo {author} {\bibfnamefont {M.~H.~S.}\ \bibnamefont
  {Amin}}, \bibinfo {author} {\bibfnamefont {S.}~\bibnamefont {Gildert}},
  \bibinfo {author} {\bibfnamefont {T.}~\bibnamefont {Lanting}}, \bibinfo
  {author} {\bibfnamefont {F.}~\bibnamefont {Hamze}}, \bibinfo {author}
  {\bibfnamefont {N.}~\bibnamefont {Dickson}}, \bibinfo {author} {\bibfnamefont
  {R.}~\bibnamefont {Harris}}, \bibinfo {author} {\bibfnamefont {A.~J.}\
  \bibnamefont {Berkley}}, \bibinfo {author} {\bibfnamefont {J.}~\bibnamefont
  {Johansson}}, \bibinfo {author} {\bibfnamefont {P.}~\bibnamefont {Bunyk}},
  \bibinfo {author} {\bibfnamefont {E.~M.}\ \bibnamefont {Chapple}}, \bibinfo
  {author} {\bibfnamefont {C.}~\bibnamefont {Enderud}}, \bibinfo {author}
  {\bibfnamefont {J.~P.}\ \bibnamefont {Hilton}}, \bibinfo {author}
  {\bibfnamefont {K.}~\bibnamefont {Karimi}}, \bibinfo {author} {\bibfnamefont
  {E.}~\bibnamefont {Ladizinsky}}, \bibinfo {author} {\bibfnamefont
  {N.}~\bibnamefont {Ladizinsky}}, \bibinfo {author} {\bibfnamefont
  {T.}~\bibnamefont {Oh}}, \bibinfo {author} {\bibfnamefont {I.}~\bibnamefont
  {Perminov}}, \bibinfo {author} {\bibfnamefont {C.}~\bibnamefont {Rich}},
  \bibinfo {author} {\bibfnamefont {M.~C.}\ \bibnamefont {Thom}}, \bibinfo
  {author} {\bibfnamefont {E.}~\bibnamefont {Tolkacheva}}, \bibinfo {author}
  {\bibfnamefont {C.~J.~S.}\ \bibnamefont {Truncik}}, \bibinfo {author}
  {\bibfnamefont {S.}~\bibnamefont {Uchaikin}}, \bibinfo {author}
  {\bibfnamefont {J.}~\bibnamefont {Wang}}, \bibinfo {author} {\bibfnamefont
  {B.}~\bibnamefont {Wilson}}, \ and\ \bibinfo {author} {\bibfnamefont
  {G.}~\bibnamefont {Rose}},\ }\bibfield  {title} {\enquote {\bibinfo {title}
  {Quantum annealing with manufactured spins},}\ }\href {\doibase
  10.1038/nature10012} {\bibfield  {journal} {\bibinfo  {journal} {Nature}\
  }\textbf {\bibinfo {volume} {473}},\ \bibinfo {pages} {194--198} (\bibinfo
  {year} {2011})}\BibitemShut {NoStop}%
\bibitem [{\citenamefont {Boixo}\ \emph {et~al.}(2014)\citenamefont {Boixo},
  \citenamefont {Ronnow}, \citenamefont {Isakov}, \citenamefont {Wang},
  \citenamefont {Wecker}, \citenamefont {Lidar}, \citenamefont {Martinis},\
  and\ \citenamefont {Troyer}}]{q108}%
  \BibitemOpen
  \bibfield  {author} {\bibinfo {author} {\bibfnamefont {Sergio}\ \bibnamefont
  {Boixo}}, \bibinfo {author} {\bibfnamefont {Troels~F.}\ \bibnamefont
  {Ronnow}}, \bibinfo {author} {\bibfnamefont {Sergei~V.}\ \bibnamefont
  {Isakov}}, \bibinfo {author} {\bibfnamefont {Zhihui}\ \bibnamefont {Wang}},
  \bibinfo {author} {\bibfnamefont {David}\ \bibnamefont {Wecker}}, \bibinfo
  {author} {\bibfnamefont {Daniel~A.}\ \bibnamefont {Lidar}}, \bibinfo {author}
  {\bibfnamefont {John~M.}\ \bibnamefont {Martinis}}, \ and\ \bibinfo {author}
  {\bibfnamefont {Matthias}\ \bibnamefont {Troyer}},\ }\bibfield  {title}
  {\enquote {\bibinfo {title} {Evidence for quantum annealing with more than
  one hundred qubits},}\ }\href {\doibase 10.1038/nphys2900} {\bibfield
  {journal} {\bibinfo  {journal} {Nat. Phys.}\ }\textbf {\bibinfo {volume}
  {10}},\ \bibinfo {pages} {218--224} (\bibinfo {year} {2014})}\BibitemShut
  {NoStop}%
\bibitem [{\citenamefont {Katzgraber}\ \emph {et~al.}(2014)\citenamefont
  {Katzgraber}, \citenamefont {Hamze},\ and\ \citenamefont
  {Andrist}}]{2014Katzgraber}%
  \BibitemOpen
  \bibfield  {author} {\bibinfo {author} {\bibfnamefont {Helmut~G.}\
  \bibnamefont {Katzgraber}}, \bibinfo {author} {\bibfnamefont {Firas}\
  \bibnamefont {Hamze}}, \ and\ \bibinfo {author} {\bibfnamefont {Ruben~S.}\
  \bibnamefont {Andrist}},\ }\bibfield  {title} {\enquote {\bibinfo {title}
  {Glassy chimeras could be blind to quantum speedup: Designing better
  benchmarks for quantum annealing machines},}\ }\href {\doibase
  10.1103/PhysRevX.4.021008} {\bibfield  {journal} {\bibinfo  {journal} {Phys.
  Rev. X}\ }\textbf {\bibinfo {volume} {4}},\ \bibinfo {pages} {021008--}
  (\bibinfo {year} {2014})}\BibitemShut {NoStop}%
\bibitem [{\citenamefont {Venturelli}\ \emph {et~al.}(2015)\citenamefont
  {Venturelli}, \citenamefont {Mandr{\`a}}, \citenamefont {Knysh},
  \citenamefont {O'Gorman}, \citenamefont {Biswas},\ and\ \citenamefont
  {Smelyanskiy}}]{Venturelli:2014nx}%
  \BibitemOpen
  \bibfield  {author} {\bibinfo {author} {\bibfnamefont {Davide}\ \bibnamefont
  {Venturelli}}, \bibinfo {author} {\bibfnamefont {Salvatore}\ \bibnamefont
  {Mandr{\`a}}}, \bibinfo {author} {\bibfnamefont {Sergey}\ \bibnamefont
  {Knysh}}, \bibinfo {author} {\bibfnamefont {Bryan}\ \bibnamefont {O'Gorman}},
  \bibinfo {author} {\bibfnamefont {Rupak}\ \bibnamefont {Biswas}}, \ and\
  \bibinfo {author} {\bibfnamefont {Vadim}\ \bibnamefont {Smelyanskiy}},\
  }\bibfield  {title} {\enquote {\bibinfo {title} {Quantum optimization of
  fully connected spin glasses},}\ }\href
  {http://link.aps.org/doi/10.1103/PhysRevX.5.031040} {\bibfield  {journal}
  {\bibinfo  {journal} {Phys. Rev. X}\ }\textbf {\bibinfo {volume} {5}},\
  \bibinfo {pages} {031040--} (\bibinfo {year} {2015})}\BibitemShut {NoStop}%
\bibitem [{\citenamefont {Boixo}\ \emph {et~al.}(2016)\citenamefont {Boixo},
  \citenamefont {Smelyanskiy}, \citenamefont {Shabani}, \citenamefont {Isakov},
  \citenamefont {Dykman}, \citenamefont {Denchev}, \citenamefont {Amin},
  \citenamefont {Smirnov}, \citenamefont {Mohseni},\ and\ \citenamefont
  {Neven}}]{Boixo:2014yu}%
  \BibitemOpen
  \bibfield  {author} {\bibinfo {author} {\bibfnamefont {Sergio}\ \bibnamefont
  {Boixo}}, \bibinfo {author} {\bibfnamefont {Vadim~N.}\ \bibnamefont
  {Smelyanskiy}}, \bibinfo {author} {\bibfnamefont {Alireza}\ \bibnamefont
  {Shabani}}, \bibinfo {author} {\bibfnamefont {Sergei~V.}\ \bibnamefont
  {Isakov}}, \bibinfo {author} {\bibfnamefont {Mark}\ \bibnamefont {Dykman}},
  \bibinfo {author} {\bibfnamefont {Vasil~S.}\ \bibnamefont {Denchev}},
  \bibinfo {author} {\bibfnamefont {Mohammad~H.}\ \bibnamefont {Amin}},
  \bibinfo {author} {\bibfnamefont {Anatoly~Yu}\ \bibnamefont {Smirnov}},
  \bibinfo {author} {\bibfnamefont {Masoud}\ \bibnamefont {Mohseni}}, \ and\
  \bibinfo {author} {\bibfnamefont {Hartmut}\ \bibnamefont {Neven}},\
  }\bibfield  {title} {\enquote {\bibinfo {title} {Computational multiqubit
  tunnelling in programmable quantum annealers},}\ }\href
  {http://dx.doi.org/10.1038/ncomms10327} {\bibfield  {journal} {\bibinfo
  {journal} {Nat Commun}\ }\textbf {\bibinfo {volume} {7}} (\bibinfo {year}
  {2016})}\BibitemShut {NoStop}%
\bibitem [{\citenamefont {King}\ \emph
  {et~al.}(2015{\natexlab{b}})\citenamefont {King}, \citenamefont {Lanting},\
  and\ \citenamefont {Harris}}]{King:2015zr}%
  \BibitemOpen
  \bibfield  {author} {\bibinfo {author} {\bibfnamefont {Andrew~D.}\
  \bibnamefont {King}}, \bibinfo {author} {\bibfnamefont {Trevor}\ \bibnamefont
  {Lanting}}, \ and\ \bibinfo {author} {\bibfnamefont {Richard}\ \bibnamefont
  {Harris}},\ }\bibfield  {title} {\enquote {\bibinfo {title} {Performance of a
  quantum annealer on range-limited constraint satisfaction problems},}\ }\href
  {http://arXiv.org/abs/1502.02098} {\bibfield  {journal} {\bibinfo  {journal}
  {arXiv:1502.02098}\ } (\bibinfo {year} {2015}{\natexlab{b}})}\BibitemShut
  {NoStop}%
\bibitem [{\citenamefont {Gershman}\ and\ \citenamefont
  {Blei}(2012)}]{gershman2012tutorial}%
  \BibitemOpen
  \bibfield  {author} {\bibinfo {author} {\bibfnamefont {Samuel~J}\
  \bibnamefont {Gershman}}\ and\ \bibinfo {author} {\bibfnamefont {David~M}\
  \bibnamefont {Blei}},\ }\bibfield  {title} {\enquote {\bibinfo {title} {A
  tutorial on bayesian nonparametric models},}\ }\href@noop {} {\bibfield
  {journal} {\bibinfo  {journal} {Journal of Mathematical Psychology}\ }\textbf
  {\bibinfo {volume} {56}},\ \bibinfo {pages} {1--12} (\bibinfo {year}
  {2012})}\BibitemShut {NoStop}%
\bibitem [{\citenamefont {Albash}\ \emph {et~al.}(2015)\citenamefont {Albash},
  \citenamefont {Vinci}, \citenamefont {Mishra}, \citenamefont {Warburton},\
  and\ \citenamefont {Lidar}}]{q-sig2}%
  \BibitemOpen
  \bibfield  {author} {\bibinfo {author} {\bibfnamefont {Tameem}\ \bibnamefont
  {Albash}}, \bibinfo {author} {\bibfnamefont {Walter}\ \bibnamefont {Vinci}},
  \bibinfo {author} {\bibfnamefont {Anurag}\ \bibnamefont {Mishra}}, \bibinfo
  {author} {\bibfnamefont {Paul~A.}\ \bibnamefont {Warburton}}, \ and\ \bibinfo
  {author} {\bibfnamefont {Daniel~A.}\ \bibnamefont {Lidar}},\ }\bibfield
  {title} {\enquote {\bibinfo {title} {Consistency tests of classical and
  quantum models for a quantum annealer},}\ }\href
  {http://link.aps.org/doi/10.1103/PhysRevA.91.042314} {\bibfield  {journal}
  {\bibinfo  {journal} {Phys. Rev. A}\ }\textbf {\bibinfo {volume} {91}},\
  \bibinfo {pages} {042314--} (\bibinfo {year} {2015})}\BibitemShut {NoStop}%
\end{thebibliography}%

\end{document}